\documentclass[fleqn,usenatbib]{mnras}

\usepackage{newtxtext,newtxmath}

\usepackage[T1]{fontenc}

\DeclareRobustCommand{\VAN}[3]{#2}
\let\VANthebibliography\thebibliography
\def\thebibliography{\DeclareRobustCommand{\VAN}[3]{##3}\VANthebibliography}

\usepackage{graphicx}	
\usepackage{amsmath}	
\usepackage{nccmath}
\usepackage{adjustbox}

\usepackage{physics}
\usepackage{siunitx}
\usepackage{booktabs}
\usepackage{adjustbox}
\usepackage{caption}
\usepackage{subcaption}
\usepackage{float}
\usepackage{makecell}

\title[Observations of the DSX meteor shower]{Characterizing the Daytime Sextantids Meteor Shower and Unveiling the Nature of the Phaethon-Geminid Stream Complex}

\author[Y. Kipreos et al.]{
Y. Kipreos,$^{1,2}$\thanks{E-mail: ykipreos@uwo.ca}
 Margaret Campbell-Brown,$^{1,2}$, P. Brown,$^{1,2}$, D. Vida$^{1,2}$
\\
$^{1}$Department of Physics and Astronomy, Western University, London, Ontario, N6A 3K7, Canada\\
$^{2}$Institute for Earth and Space Exploration, University of Western Ontario, N6A 5B7, Canada\\
}

\date{Accepted XXX. Received YYY; in original form ZZZ}

\pubyear{2021}

\begin{document}
\label{firstpage}
\pagerange{\pageref{firstpage}--\pageref{lastpage}}
\maketitle

\begin{abstract}
The Daytime Sextantids meteor shower, part of the Phaethon-Geminid Stream Complex (PGC), is closely related to the Geminids, currently the strongest meteor shower visible at the Earth. The DSX share a similar orbit to asteroid 2005 UD, but the nature of the association remains unclear. From optical data we find that DSX meteors ablate similarly to Geminids, suggesting that they are also high density and consistent with a common origin.  From radar data we have isolated 19,007 DSX orbits through application of a novel convex hull approach to determine stream membership. We find at the peak the mean semi-major axis is near 1 AU, eccentricity is 0.86 and that both decrease as a function of solar longitude. The inclination averages 25 degrees at the peak but increases over time.  Noticeable DSX activity extends from solar longitude 173-196$^{\circ}$ with a flux plateau between 186 - 189$^{\circ}$. The peak flux is $2 \pm 0.05 \times 10^{-3}$ km$^{-2}$ hr$^{-1}$, equivalent to a ZHR of 20. We estimate a true differential mass index for the shower of $s = 1.64 \pm 0.06$ at the time of peak and an average of $1.70 \pm 0.07$ for days surrounding the peak. The mass of the DSX stream is estimated to be $10^{16}$ g, the same order as 2005 UD, suggesting the stream is too massive to have been created by recent meteoroid production from 2005 UD. We propose that the DSX and 2005 UD were created in the same break-up event that created 3200 Phaethon.
\end{abstract}




\section{Introduction}

The strongest meteor shower currently visible on Earth is the Geminids. Occurring every December, the shower regularly produces in excess of 100 visual meteors per hour at its peak \citep{Rendtel2014}. The Geminids share a close orbital affinity with 3200 Phaethon, suggesting it is its immediate parent \citep{Gustafson1989, Williams1993, Ryabova2019}. Complicating this simple picture, however, is the stream's dynamical association with several other showers, notably the Daytime Sextantids and Canis Minorids \citep{Babadzhanov1991}. 

The discovery of 2005 UD \citep{ohtsuka2006} on a near-identical orbit to the Daytime Sextantids and the possible association of Phaethon and 2005 UD with 1999 YC has led to suggestions of the existence of a Phaethon-Geminid Stream complex (PGC) \citep{Jewitt2006a}. Additional puzzles concerning the origin of the PGC include the role of the unusually low perihelion orbit of Phaethon in stream formation, Phaethon's non-comet-like reflectance spectrum \citep{Licandro2007} and the comparatively high strength of Geminid meteoroids relative to other meteor showers \citep{Spurny1993, Borovicka2007}. 

The Geminid shower's unusual orbit (high eccentricity, high inclination, and low perihelion distance) is suggestive that Phaethon (or its progenitor) may have originated in the Main Belt \citep{DeLeon2010}. An asteroidal origin for Phaethon comes with its own problems, namely the difficulty in forming a stream as massive as the Geminids \citep{Ryabova2017} from a parent generating minimal ejected dust. While Phaethon has been observed to eject small dust as it undergoes thermal heating at its closest approach of the Sun, this debris is, by orders of magnitude, inadequate to explain the mass of the Geminids \citep{jewitt2015}. 

The Geminids and Phaethon have been studied extensively, but the debate over the cometary vs. asteroidal origin for the Geminids and the existence/origin of the PGC remains unsettled \citep{Ryabova2019}. Due to the recent close approach of Phaeton in 2017, there has been renewed focus on the physical characterization of both Phaethon \citep{Jewitt2010, Ye2018, Tabeshian2019, Huang2021} and 2005 UD \citep{Kinoshita2007,Huang2021, ishiguro2022polarimetric} as well as their relationship to each other \citep{MacLennan2020,Devogele2020, kareta2021}. However, a major component of the PGC, namely the Daytime Sextantids (DSX), is comparatively unstudied. Few modern measurements of the DSX exist, and no physical properties of the associated meteoroids have been published. The goal of this work is to provide measurements of the DSX orbital parameters, their variation with solar longitude, the shower flux, mass distribution, and physical properties of the DSX meteoroids. Our aim is to contribute constraints to future models of the PGC and estimate the mass of the stream.

The Daytime Sextantids is a moderate-strength annual meteor shower that occurs in late September and early October. This stream is linked to the Geminids through precession, being the pre-perihelion daylight twin of the Geminids \citep{Babadzhanov1991}. The two streams have similar unusual orbits with high eccentricities, inclinations, and small perihelion distances. As with the Geminids and Phaethon, the probable parent body of the Daytime Sextantids stream, 2005 UD, has an unclear origin \citep{ishiguro2022polarimetric}. From its discovery, 2005 UD has been linked to the Daytime Sextantids stream on the basis of their orbital similarities and a small orbital evolutionary time lag of $\sim100$ years \citep{ohtsuka2006}.

Some authors have suggested that 2005 UD is a split nucleus of Phaethon. One line of evidence supporting this association is through backward and forward orbital integrations performed by \citet{ohtsuka2006}, which demonstrated that Phaethon and 2005 UD have similar orbital evolution with a time lag of $\sim4600$ years.

Another line of evidence connecting the two bodies is their similar reflectance spectra \citep{Jewitt2006a}. Both Phaethon and 2005 UD have blue reflectance spectra in the visible, which is rarely found in asteroids. However, \citet{kareta2021} recently presented the first measurement of 2005 UD in the near-infrared and found that it was not similar to Pheathon's near-infrared spectrum. \citet{kareta2021} did not find a suitable alternative mechanism that could explain how the asteroids could have similar orbits and visible-light reflectance spectra and yet have dissimilar near-infrared spectra. More measurements of 2005 UD and Phaethon are needed to clarify the relationship between the two asteroids and their associated meteor streams.

The close orbital and physical relationship suggests that the nature of 2005 UD is intricately linked to Phaethon. Thus in studying 2005 UD via the Daytime Sextantids, we expand our understanding of the nature of Phaethon and the PGC more broadly.

While the Daytime Sextantids' physical characteristics have yet to be measured, the shower and its orbital properties have been observed for more than 60 years. The first detection of the Daytime Sextantids was made in 1957 by \citet{Weiss1960}, who conducted a two-year radio survey. Weiss originally called the DSX the Sextantids-Leonids shower. Unfortunately, only one part of the radar was in service during the main DSX shower activity period, so the DSX radiant measured by Weiss contained significant biases. 

The Daytime Sextantids were initially thought to be periodic in nature. There was no evidence of the DSX before Weiss' study, and a meteor survey in 1960 by \citet{kashcheyev1967} did not detect the shower. It was not until 1961 that the Daytime Sextantids were detected again by \citet{Nilsson1964} during a radio survey in the southern hemisphere. Nilsson suggested the seemingly periodic nature of the DSX, referencing the fact that it was not detected in earlier surveys made in both hemispheres. Nilsson was also the first to connect the Daytime Sextantids stream to the Geminids due to their orbital similarities. 

The stream was again detected during the Harvard Radio Meteor project survey \citep{Sekanina1976}, as one of a total of 275 streams reported during the 1968-1969 survey. The next DSX detections were not made until almost three decades later by the Advanced Meteor Orbit Radar (AMOR) \citep{Galligan2003}. More recently, radar measurements of the stream have also been reported by \citet{Brown2008, Brown2010} using the Canadian Meteor Orbit Radar (CMOR), by \citet{Younger2009} using Southern hemisphere meteor wind radars, and by \citet{Pokorny2017} using the Southern Argentina Agile Meteor Radar (SAAMER).

Despite being a daytime shower with a radiant only 30$^{\circ}$ from the Sun, some optical meteors of the shower are detectable in the hour before local sunrise near the peak time. Some optical DSX orbits have been reported by the Cameras for All-Sky Surveillance project (CAMS) \citep{Jenniskens2016a}, by the SonotaCo network \citep{Sonotaco2009}, and by the Global Meteor Network \citep{Vida2021}, among others. \citet{Rudawaska2015} identified ten DSX meteors in the EDMOND database using a D-criterion approach.

Here we examine the flux profile, radiant drift, and orbital element variation with solar longitude together with the mass index of the DSX using CMOR radar observations. Additionally, we measure the apparent bulk strength of DSX meteoroids relative to the Geminids using optical data collected by the GMN.

Table A1 and A2, found in Appendix A in the supplementary materials, summarize literature measurements of the radiant and orbital elements of the Daytime Sextantids meteor shower both from radar and optical instruments. Our measurements (described later) are also shown for comparison. 

 The present study of the DSX is particularly timely as the Japan Aerospace Exploration Agency (JAXA) will launch the DESTINY+ mission in 2024 to rendezvous with Phaethon and (possibly) 2005 UD. This mission aims to provide a deeper understanding of the Phaethon-Geminid complex. It is vitally important that all components of the complex: the Geminids, Phaethon, 2005 UD, and the Daytime Sextantids, are measured and as fully characterized as possible to aid in the analysis of the data measured by DESTINY+. Of these components, the Daytime Sextantid meteor shower is by far the least characterized and therefore the focus of our study.

\section{Equipment}
\subsection{The Canadian Meteor Orbit Radar (CMOR)}

Our radar data is derived from CMOR (the Canadian Meteor Orbit Radar), which is comprised of three five-element interferometric backscatter radars located in Tavistock, Ontario. The three radar systems are synchronized in transmission and receiving and operate using three different frequencies: 17.45 and 38.15 MHz with a peak power of 6 kW, and 29.85 MHz with a peak power of 15 kW. CMOR is an all-sky radar able to detected echoes in virtually all directions except for meteors located lower than 20$^{\circ}$ above the horizon, which are rejected due to higher measurement errors. A detailed description of the CMOR system is provided in \cite{Jones2005}, \cite{Brown2008}, and \cite{Brown2010}. Here we provide a brief overview of CMOR pertinent to the DSX data collection. 

CMOR uses SKiYMET software to calculate the time of occurrence, interferometric position in the sky, range, height, decay time, and maximum amplitude of meteor echoes \citep{Hocking2001}. Additionally, the 29.85 MHz radar system has five remote outlying stations. The interferometry from the main site, coupled with the measured time delays from the remote sites, allows the velocity vector to be calculated. This velocity calculation is referred to as the time-of-flight (TOF) method. CMOR's 29.85 MHz system uses the velocity vector to calculate the meteor's orbit, and this calculation can be applied to meteors with magnitudes as low as $\sim+8.0$, which roughly corresponds to a mass of $10^{-7}$ kg at DSX in-atmosphere speeds of 35 km/s.

This study calculates the radiant and orbital parameters of the Daytime Sextantids shower using data collected from the multi-station 29.85 MHz system. The DSX mass index and flux calculations require the inclusion of low amplitude meteors, making the use of the single-station station meteor echoes more appropriate. For single station echoes velocities cannot be computed; instead we use the pre-t0 speed technique \citep{Mazur2020}, where the meteor echoes rate of change of phase is used to calculate the speed to help select shower members.

\subsection{Optical Instruments}

For optical measurements of the DSX we make use of the distributed stations of the Global Meteor Network, described in \citet{Vida2021}. The GMN is a citizen science project made up of a collection of CMOS video meteor cameras owned and operated by members of the public. Meteor trajectories are computed by correlating common observations and applying a strict set of quality filters which allow this procedure to be automated, and orbits are computed using the method of \citet{vida2020a}.

The GMN uses consumer-grade Internet Protocol (IP) cameras based on inexpensive low-light Starvis Sony CMOS sensors paired with Raspberry Pi single-board computers which perform data collection and processing. Typical GMN cameras have fields of view of order 50$^{\circ}$-90$^{\circ}$, limiting stellar magnitudes around +6 and operate at 25 frames per second. Currently, the GMN has over 600 cameras in 35 countries which allow 24/7 meteor monitoring in both hemispheres.

\section{Observational Results : Radar}

\subsection{Wavelet Measurement of the DSX Radiant}

To isolate the DSX radiant from the background of meteor radiants detected by CMOR, we first use a three-dimensional wavelet transform, following the process in \cite{Brown2008}. In practice, the radiant of a meteor shower at a particular time is an overdensity in the local number of individual radiants in three dimensions: Sun-centered geocentric ecliptic longitude ($\lambda - \lambda_{\odot}$), latitude $\beta$, and the geocentric speed of the shower $V_g$. The use of this coordinate system minimizes the drift in the radiant as it accounts for the movement of the Earth around the Sun. 

A three-dimensional wavelet transform applied in one-degree time windows of solar longitude on CMOR-measured meteor radiants has been found to be effective at discerning meteor shower radiants from the sporadic background \citep{Brown2010}. When the wavelet is applied across the radiant distribution, regions where there is a higher concentration of meteor radiants are easily isolated as given by the location of the maximum wavelet coefficient \citep{Galligan2000}. 

The mother wavelet function that is best suited for meteor shower radiant searches is the Ricker wavelet:

\begin{multline}
 W(x_0,y_0,v_{g0}) = \frac{1}{(2\pi)^{2/3} \sigma_{v}{1/2} a} \int^{v_{gmax}}_{v_{gmin}} \int^{+\infty}_{-\infty} \int^{+\infty}_{-\infty}\\ f(x,y,v_g) \times (3 - \frac{(x-x_0)^2 + (y-y_0)^2}{a^2} - \frac{(v_{g0} - v_g)^2}{\sigma^2_v}) \\
 \times \exp(-0.5[\frac{(x-x_0)^2 + (y-y_0)^2}{a^2} - \frac{(v_{g0} - v_g)^2}{\sigma^2_v}])
\end{multline}

\noindent where $W(x_0,y_0,v_{g0})$ is the wavelet coefficient, $\sigma_v$ is the velocity probe size measured in km $\:$ s$^{-1}$, $a$ is the angular probe size measured in degrees, $v_g$ is the geocentric velocity, and $x$ and $y$ are the spatial coordinates, $(\lambda - \lambda_\odot)$ and $\beta$, respectively.

The raw wavelet coefficient indicates the relative local concentration of radiants, but this is affected by the relative density of the local sporadic meteor radiant background. To more robustly define the strength of the shower relative to the general sporadic background, the excursion of the wavelet coefficient in units of standard deviations ($\sigma$) above the median average wavelet value at the same sun-centred radiant location across the entire year is used \citep{Brown2008}, a value termed xsig or $\sigma_x$.

The signal strength produced by the wavelet transform is dependent on defining the characteristic scales, $\sigma_v$ and $a_v$, for each dimension of the wavelet. The size of the probes should reflect both the natural spread of the actual radiant distribution and the spread caused by the instrumental uncertainty in the radiant measurements. Hence, to optimize the wavelet analysis requires some a priori knowledge of these characteristic scales. For this reason, we perform the wavelet analysis in a two step process. This is done by first establishing the approximate radiant location for the shower and then computing the optimal probe sizes for the shower of interest at the time of maximum.  
	
\subsection{Optimizing the Velocity and Angular Probe Size}

The CMOR radiant measurements being used in this study were collected between the years 2011 and 2019 and comprise 19.6M orbits. For this application, we aim to maximize the shower signal, so following the original approach of \citet{Galligan2000}, all years are combined and binned temporally into one-degree solar longitude windows, producing a stacked "virtual year." 

In the first stage, an initial run of the three-dimensional wavelet transform using the Ricker wavelet equation is applied to the stacked virtual year. The wavelet probe sizes were fixed to a default velocity probe size of 10\% of the velocity and an angular separation probe size of 4$^{\circ}$. These probe sizes have been empirically found to provide reasonable results for most meteor shower radiants as determined in \citet{Brown2008}.

From this initial wavelet analysis, the radiant location of the shower and its approximate time of peak activity is determined, following the approach described by \citet{Bruzzone2015}. In the next stage, the wavelet probe sizes are optimized specifically for the DSX. To accomplish this optimization, the velocity and angular probes sizes at the radiant location at the time of shower maximum are incremented in small steps to determine which combination of probe sizes produces the best signal-to-noise ratio for the shower as defined by the maximum xsig value. 

\begin{figure}
	\centering
	\includegraphics[width=0.9\linewidth]{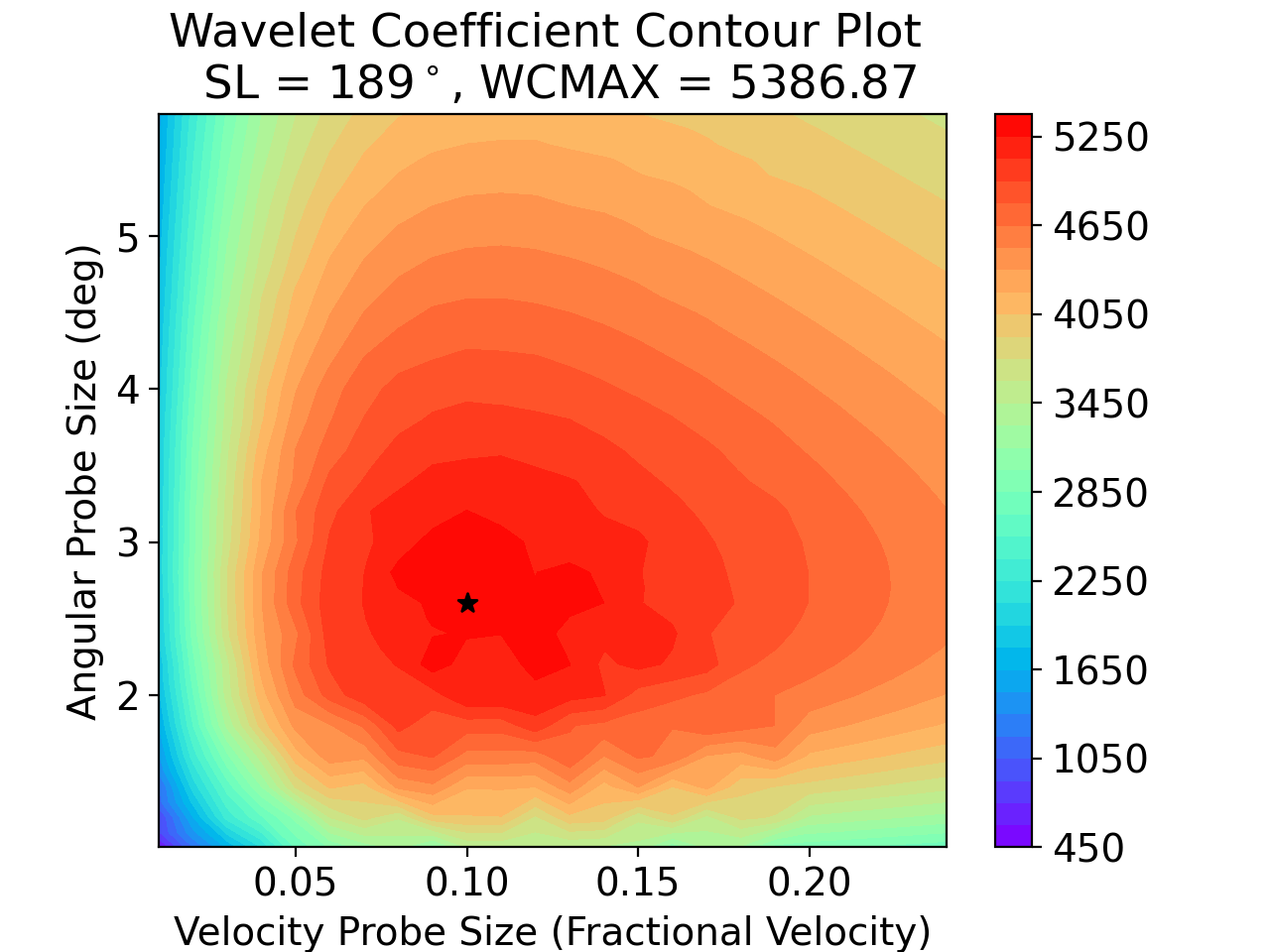}
	\caption{The wavelet coefficient (color bar), WCMAX, at Solar longitude (SL)=189$^{\circ}$ as a function of  probe size. WCMAX is a measure of the concentration of meteor radiants, where larger WCMAX values indicate regions of higher concentration. The angular and velocity probe sizes that maximize WCMAX are $2.60^\circ$ and $10 \%$, respectively, and is represented by a black star in the figure.
	The wavelet coefficient is being computed at $\lambda-\lambda_\odot$=329.8$^{\circ}$, $\beta$=-10.8$^{\circ}$, V$_g$=31.7 km/s, the nominal position of the DSX radiant for SL=189$^{\circ}$ as found in the first stage analysis where a probe size of 4$^{\circ}$ and 10\% in velocity was used .}
	\label{fig:189WCMAX}
\end{figure}

\begin{figure}
	\centering
	\includegraphics[width=0.9\linewidth]{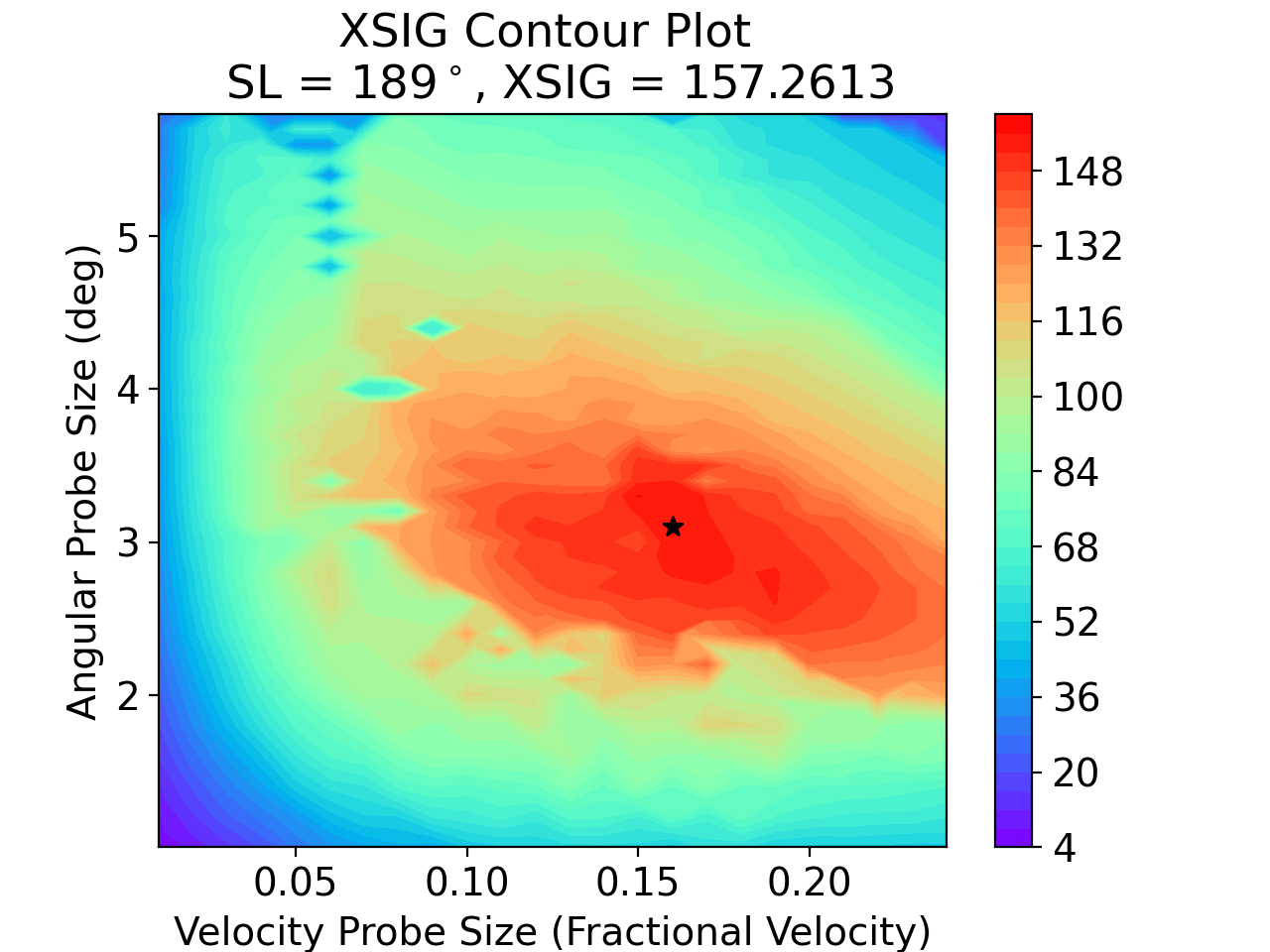}
	\caption{Variation of XSIG (color coded) at SL=189, near the peak of the  DSX shower, as a function of probe size. XSIG is a measure of the number of standard deviations that the shower is above the median wavelet coefficient of the background . The angular and velocity probe sizes that maximize XSIG are $3.10^\circ$ and $16 \%$, respectively, and is represented by a black star in the figure. The xsig coefficient is being computed at $\lambda-\lambda_\odot$=329.8$^{\circ}$, $\beta$=-10.8$^{\circ}$, V$_g$=31.7 km/s.}
	\label{fig:189XSIG}
\end{figure}

Figure \ref{fig:189WCMAX} and Figure  \ref{fig:189XSIG} show the variation in wavelet coefficient and xsig as a function of probe size, respectively. From Figure \ref{fig:189XSIG} a best estimate for the optimal angular probe size for the DSX is 3.10$^{\circ}$ while the optimal velocity probe size is 0.16$\ \times \ $V$_g$. We use these values for all later wavelet analysis.

\begin{figure}
	\centering
	\begin{subfigure}{.5\textwidth}
	\includegraphics[width=0.9\linewidth]{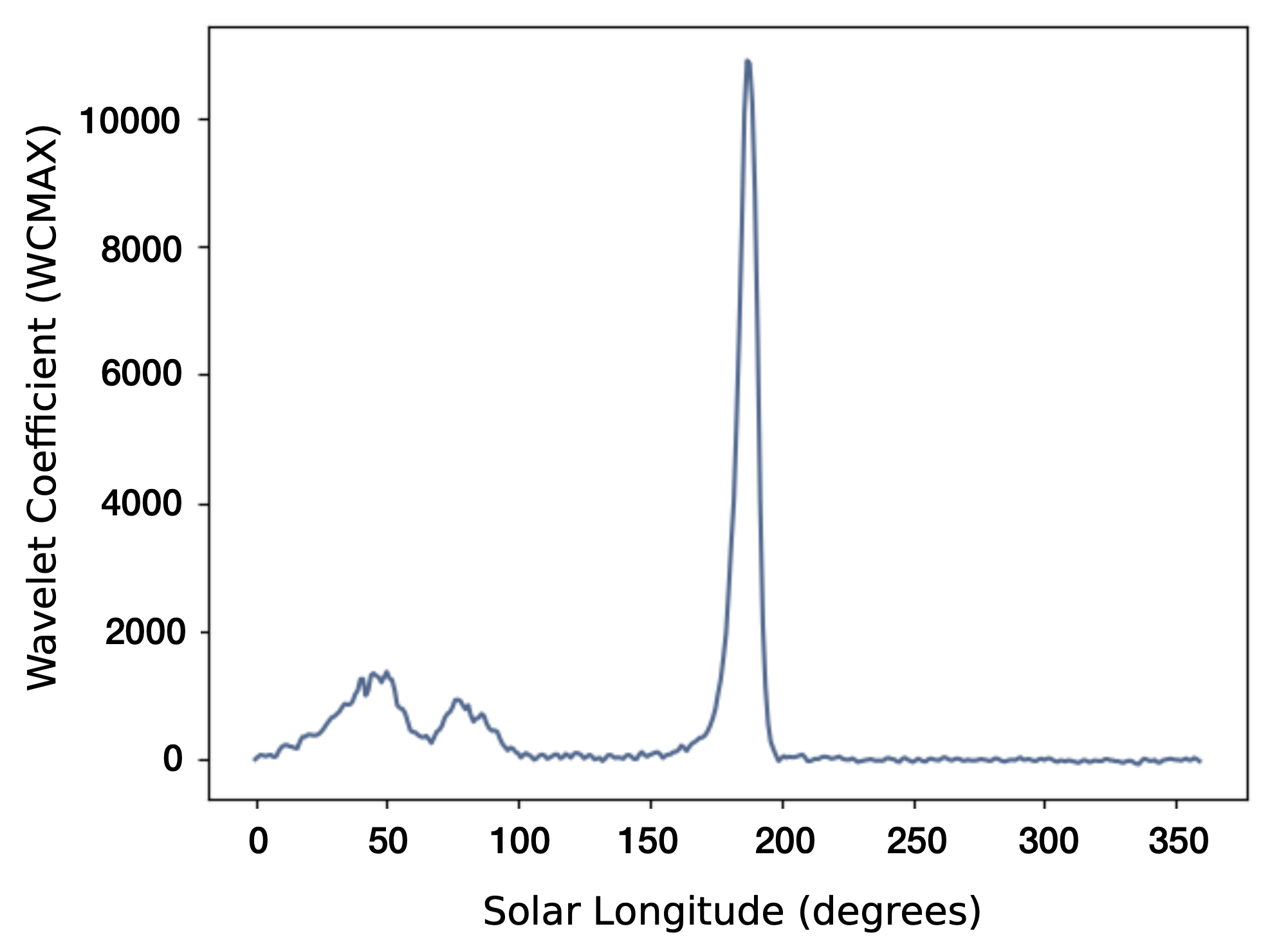}
	\end{subfigure}
	\begin{subfigure}{.5\textwidth}
	    \includegraphics[width=0.9\linewidth]{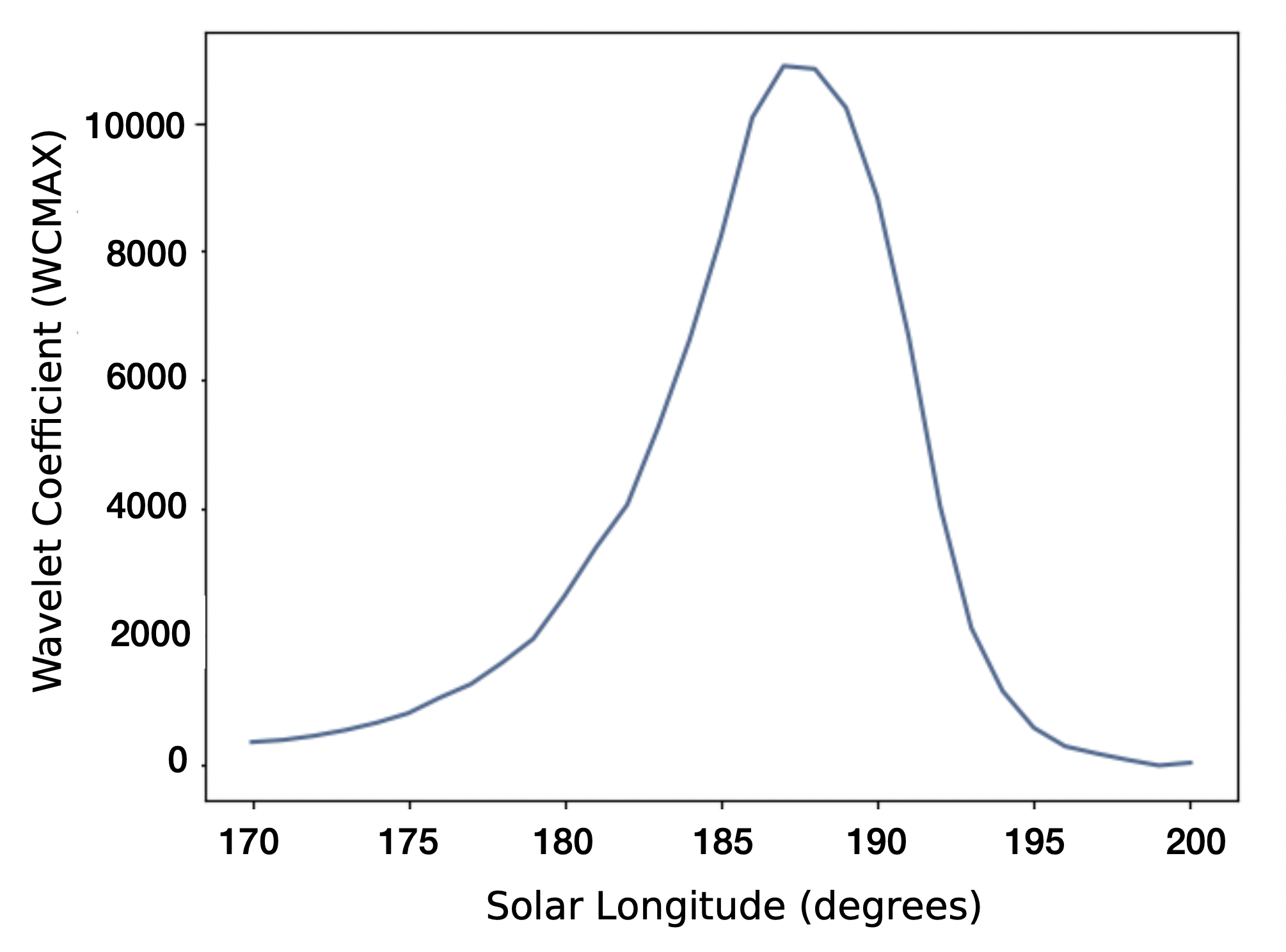}
	\end{subfigure}

	\caption{Wavelet coefficient, WCMAX, plotted over the full year (top) and over the solar longitudes around the DSX shower (bottom). Here the wavelet coefficient is being computed at $\lambda-\lambda_\odot$=329.8$^{\circ}$, $\beta$=-10.8$^{\circ}$, V$_g$=31.7 km/s.}
	\label{fig:wcmax}
\end{figure}

\begin{figure}
	\centering
	\begin{subfigure}{.5\textwidth}
	\includegraphics[width=0.9\linewidth]{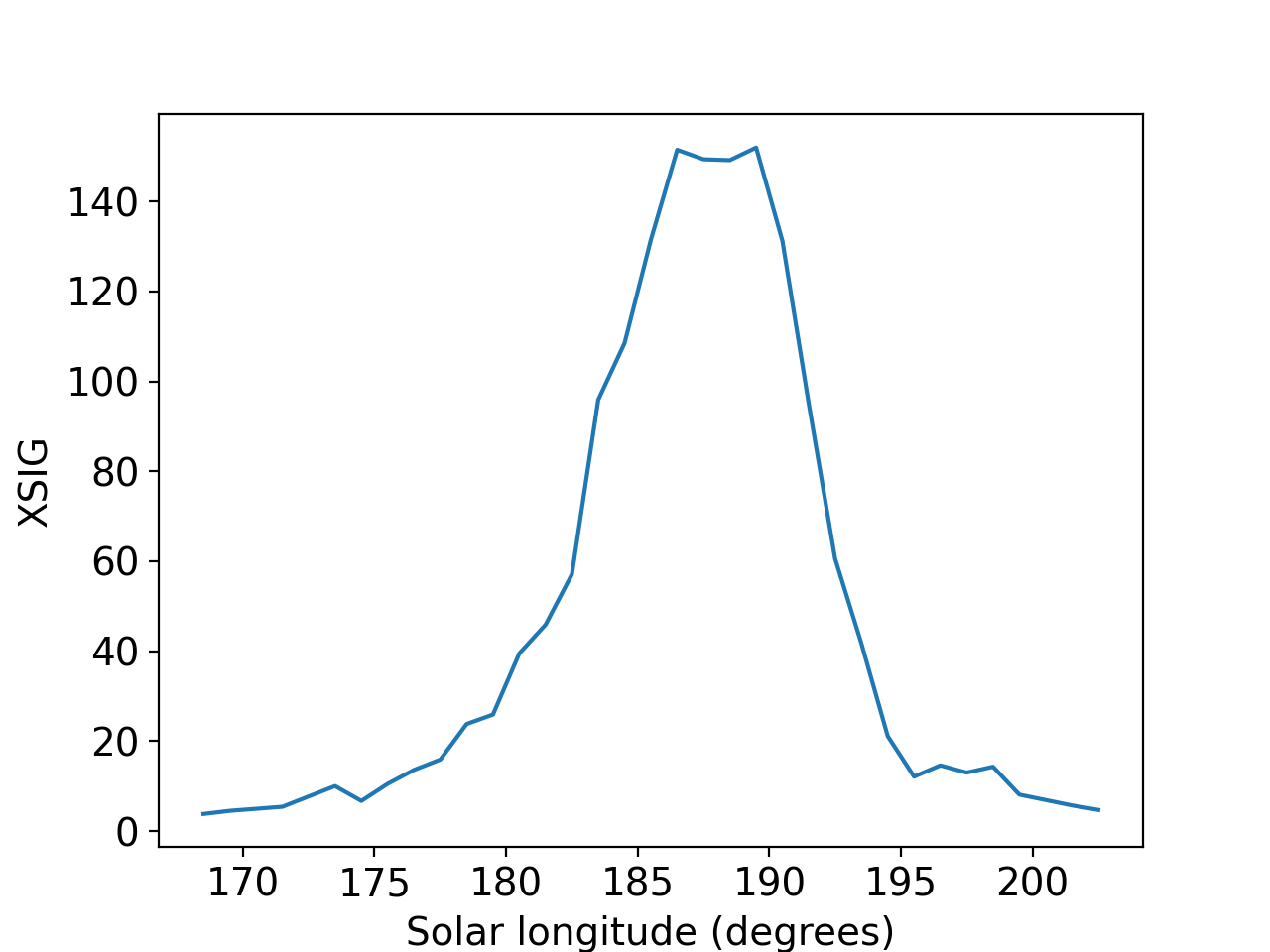}
	\end{subfigure}

	\caption{The XSIG values for the solar longitudes for the duration of the DSX shower. XSIG values are a measure of the excursion of the wavelet coefficient over the average wavelet value calculated at the same radiant location during the year, as measured in standard deviations. The XSIG values are calculated using the best-estimate optimal probe sizes, which is an angular probe size of 3$^{\circ}$ and a velocity probe size of 0.16$\ \times \ $V$_g$, as determined from Figure \ref{fig:189XSIG}. Here the wavelet coefficient is being computed at $\lambda-\lambda_\odot$=329.8$^{\circ}$, $\beta$=-10.8$^{\circ}$, V$_g$=31.7 km/s.}
	\label{fig:xsig}
\end{figure}

Using these optimal probe sizes, the wavelet coefficients vs. solar longitude at the sun-centred radiant location of the DSX at its time of maximum activity is shown in Figure \ref{fig:wcmax} while Figure \ref{fig:xsig} shows a plot of the corresponding xsig values.  
	
\subsection{Wavelet Radiant-drift}

Using our optimal probe sizes, we applied the wavelet transform on the subset of CMOR radiants measured between 2011-2020 near the time of DSX activity in a large region around the sun-centred radiant. Specifically, we examined the solar longitude interval from 159-205$^{\circ}$ to search for localized radiant maxima in the radiant interval $\SI{320}{\degree} \leq \lambda-\lambda_\odot \leq\SI{350}{\degree}$, $\SI{-15}{\degree} \leq \beta \leq\SI{0}{\degree}$ and in the velocity range $\SI{26}{km/s} \leq V_g \leq\SI{38}{km/s}$.  

\begin{figure*}

\begin{tabular}{@{}c@{}}
    \includegraphics[width=0.45\textwidth]{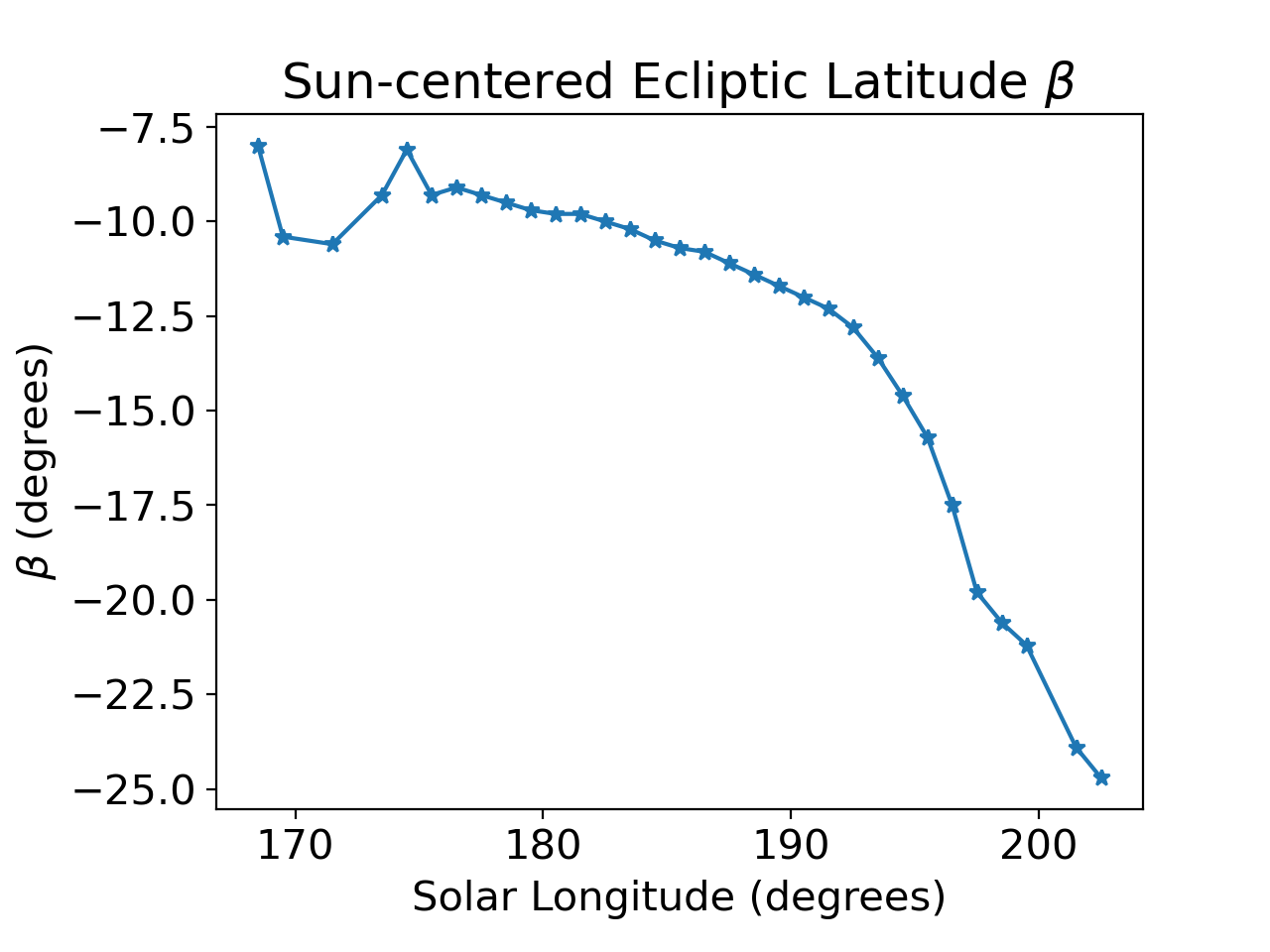}
\end{tabular}
\begin{tabular}{@{}c@{}}
    \includegraphics[width=0.45\textwidth]{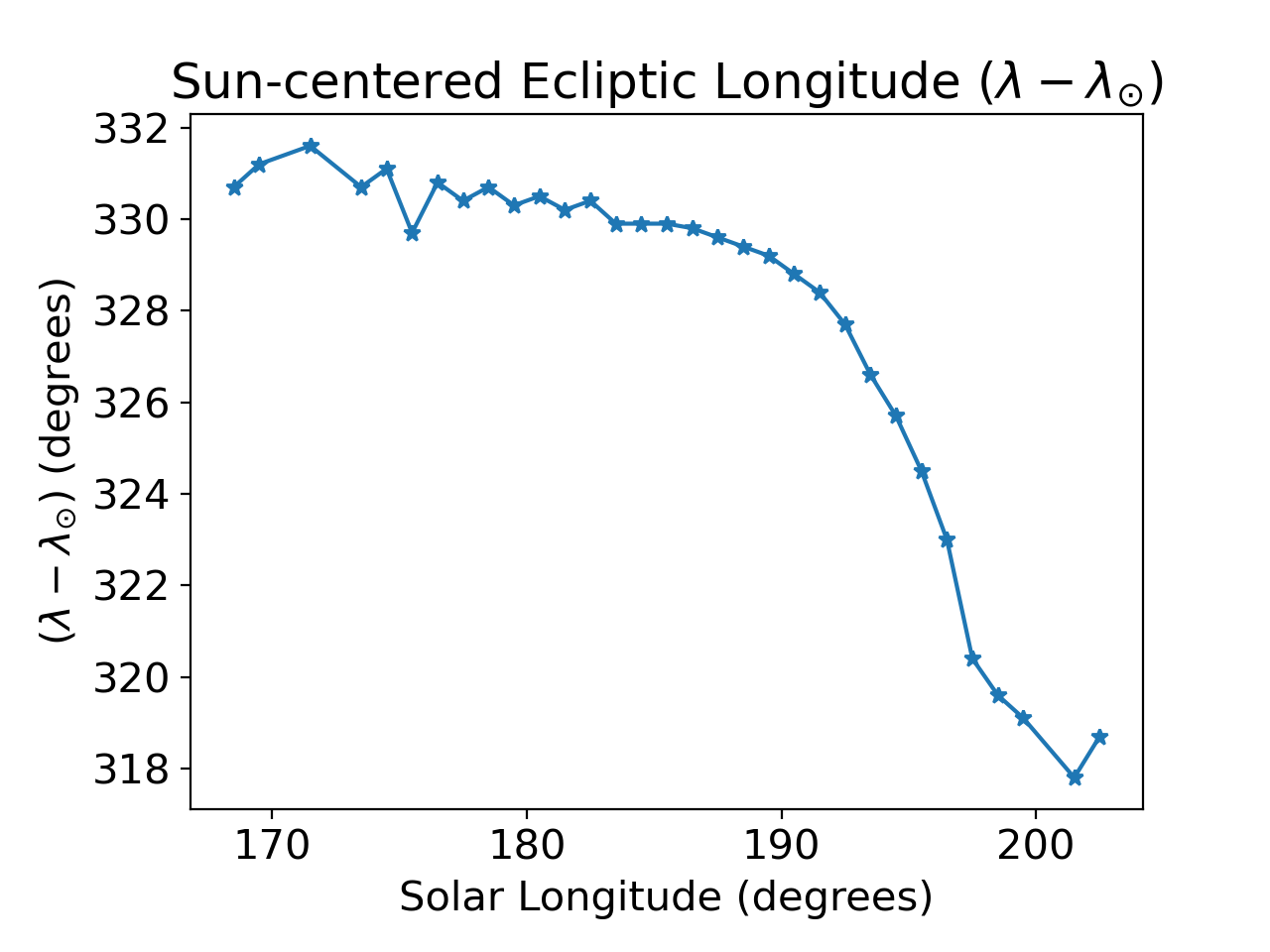}
\end{tabular}
\begin{tabular}{@{}c@{}}
    \includegraphics[width=0.45\textwidth]{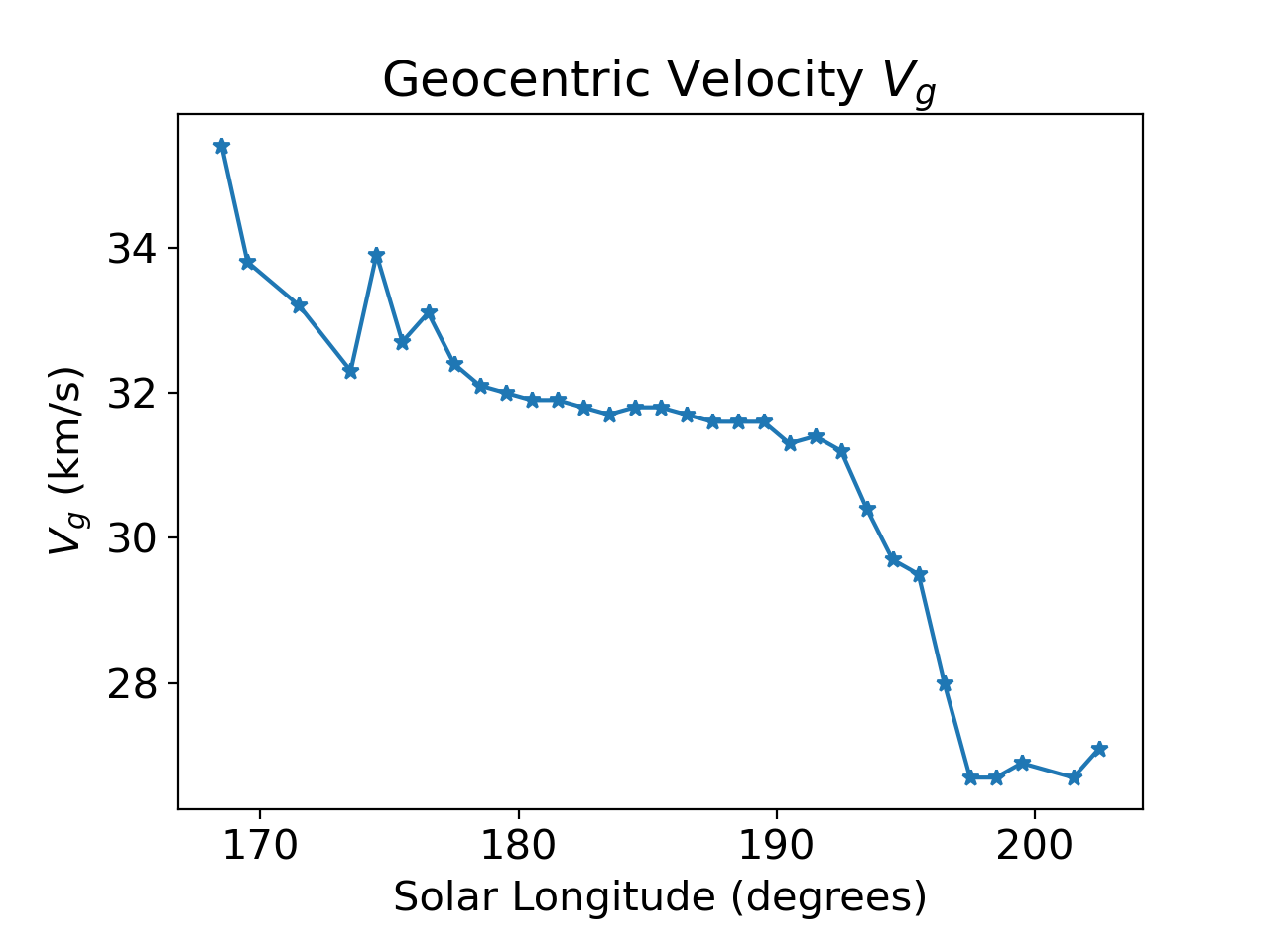}
\end{tabular}

\caption{Sun-centered ecliptic longitude, latitude, and geocentric velocity of the radiant of the Daytime Sextantids meteor shower per solar longitude degree, calculated by the wavelet transformation. The DSX radiant exhibits moderate drift during the duration of the shower. The precision in radiant coordinates is 0.1$^{\circ}$ and 0.03 km/s in speed as set by the step-size used in the wavelet transform search.}
\label{fig:wavelet_radiant}
\end{figure*}

We applied the three-dimensional wavelet transform on the DSX meteor distribution and found the local maximum per degree solar longitude in ecliptic longitude, latitude, and geocentric velocity. We were then able to link these maxima together and associate them with the DSX shower, based on the radiant ephemeris as given in \cite{Brown2010}. The final wavelet-based radiant ephemeris is shown in Figure \ref{fig:wavelet_radiant}. There is a distinct inflection in the drift near $\lambda_\odot$=193$^\circ$, suggesting either possible confusion with a background source or significant change in the stream properties in the latter part of its activity window.

\subsection{A New Meteor Radiant Selection Methodology applied to the DSX}

The wavelet approach is a good statistical estimate of the mean shower radiant, but it does not permit direct association of any single meteor radiant to a shower.

As a result, contamination from sporadic meteors complicates the wavelet-based analysis of meteor showers. The difficulty in removing the sporadic meteors from the meteor shower radiant distribution lies in how well the meteor shower radiant space is defined. Ideally, such a radiant space should objectively define which individual meteors are members of the meteor shower and which are not at some confidence level. 

A common technique is to use an orbital dissimilarity criterion, such as the D-criterion, to serially associate members of a stream \citep{Williams2019}. This approach is dependent on defining an acceptance threshold, which often is subjective and frequently does not account for the local background population near a shower radiant \citep{Neslusan2017}.

Another approach to defining a meteor shower is to select an acceptable range in each of the observed radiant parameters ($\lambda - \lambda_{\odot}$), $\beta$, and $V_g$. These ranges are used to enclose the shower radiant such that the meteors contained within all three parameter ranges are considered to share the same radiant with the meteor shower. In practice, this traditional method assumes that the radiant space of a meteor shower can be roughly modeled as a rectangular prism, as defined by the acceptable parameter ranges, which are usually chosen subjectively. Ideally, we would define this radiant space as being statistically more populated than the background. 

The radiant space of a shower is generally ellipsoid-shaped, as can be seen in the clustering of meteor radiants in Figure \ref{fig:rad_distribution}. The issue with the traditional method is that it assumes that the dimensions of the parameter space are linearly independent. However, the geocentric velocity directly depends on the radiant angular distance from the apex \citep{vida2020}, a measure derived from the other two angular coordinates. Furthermore, by defining the cuts independently, the chosen shower radiant space may not match the actual clustering of the radiant distribution and may cause the introduction of sporadic contamination by radiants located in the edges of the rectangular-prism-shaped radiant space. The measurement of the radiant space of a meteor shower can be improved to better select only the shower meteors if no assumptions about its shape and dimensional independence are made. Instead, a threshold for radiant density as set by the background levels is a better measure of shower spread. We can then define the radiant space with varying degrees of statistical significance.
	
\subsubsection{The Convex Hull Radiant Space Method}
\label{convex_hull_section}

To identify specific radiants for a particular shower, we present a new meteor selection method to objectively define the radiant space of a meteor shower. This method (termed the convex hull approach) finds radiants within a shower radiant space determined to be members with a confidence level of 95\% relative to the background population. This new method relies on a sufficiently large dataset to define the sporadic background contamination, improving the accuracy of the meteor shower radiant identification. The convex hull approach uses a series of statistical steps to remove the contamination of sporadic meteors to better isolate the shower population.

\begin{figure}
	\centering
	\includegraphics[width=0.9\linewidth]{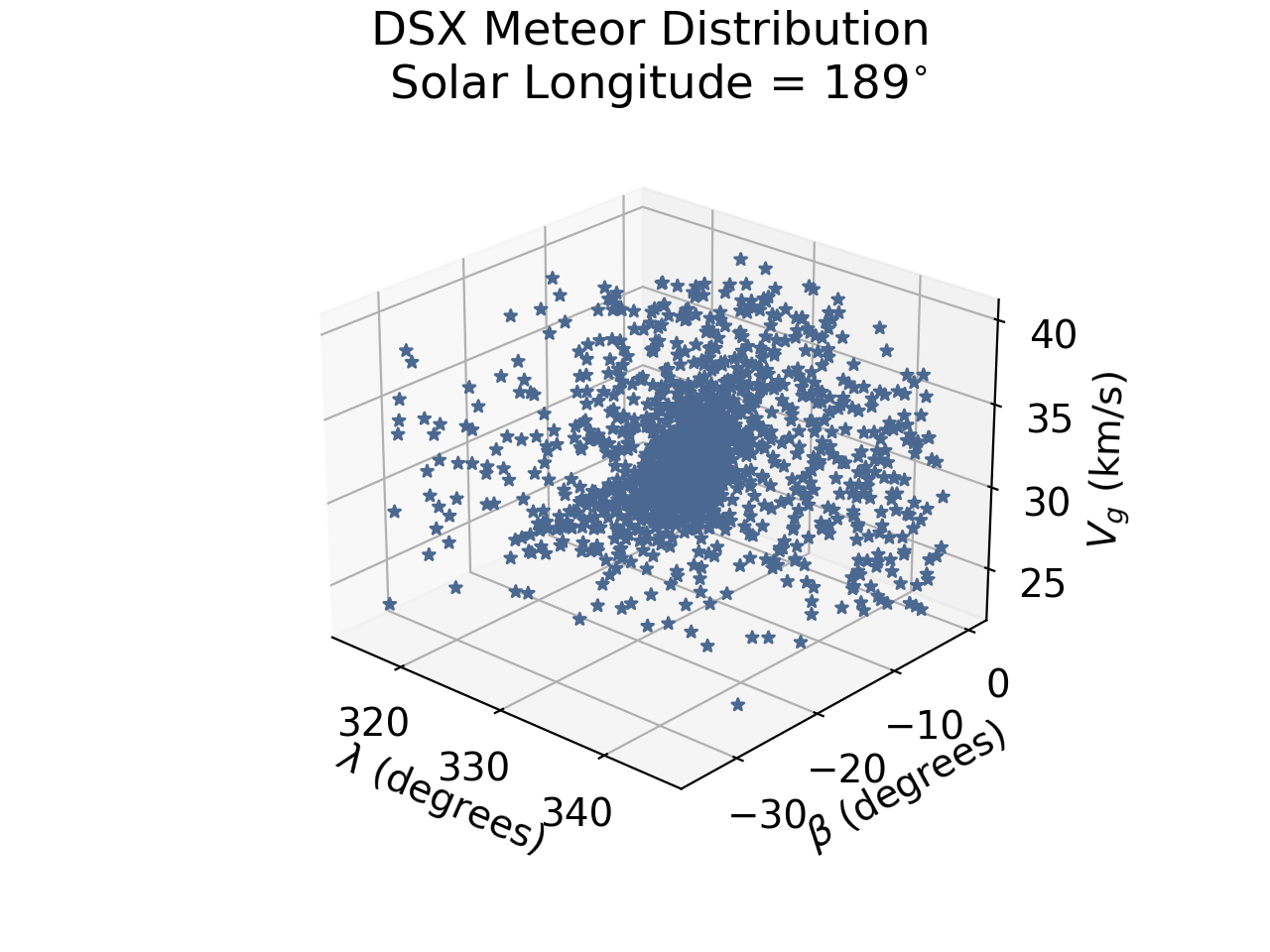}
	\caption{The 3D distribution of individual meteor radiants measured by CMOR between 2011-2020. Here we show radiants recorded at $\lambda_\odot$=189$^\circ$ in the area near the expected DSX radiant based on the maximum location provided by the wavelet approach. The clustering of radiants represents the core of the shower.}
	\label{fig:rad_distribution}
\end{figure}

The meteor radiant distribution is defined in the three dimensions of ($\lambda - \lambda_{\odot}$), $\beta$, and $V_g$. An example of such a radiant distribution for solar longitude 189$^{\circ}$ (near the DSX activity peak) is shown in Figure \ref{fig:rad_distribution}. Each point in the figure represents the radiant of an individual meteor echo detected by CMOR. Lower-quality meteor echo observations, with radiant uncertainties greater than one degree, have been removed from the distribution. This distribution is then converted into a 3D number density matrix by dividing the three-dimensional space into 8$\times$8$\times$8 cubes called voxels. Each voxel approximately covers a range in sun-centered ecliptic longitude and latitude of 3.75$^{\circ}$ and 4.38$^{\circ}$, respectively, and a geocentric velocity range of 2.88 km/s. The number of meteor radiants contained within each cube of radiant space is then measured. 

\begin{figure}
	\centering
	\includegraphics[width=0.9\linewidth]{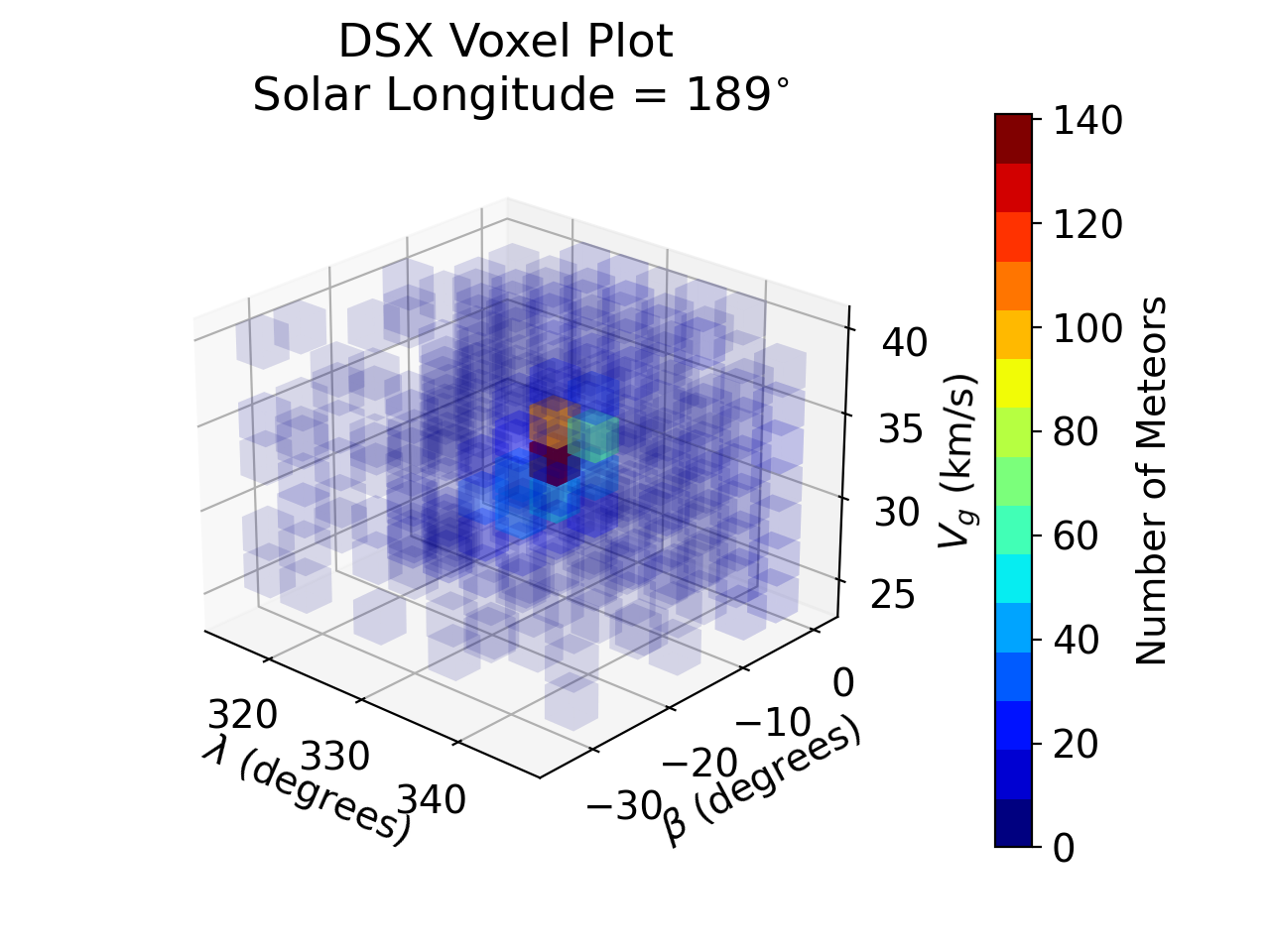}
	\caption{The number density matrix of all radiants recorded for $\lambda_\odot$=189$^\circ$ in the stacked virtual year. Each cube is referred to as a "voxel". Both the color and transparency of the voxel represent the number of meteors with radiants located within each voxel of radiant space. Voxels that contain no meteors are invisible.}
	\label{fig:raw_voxels}
\end{figure}

An example voxel plot for the DSX is shown in Figure \ref{fig:raw_voxels} for solar longitude 189$^{\circ}$. Each voxel represents one of the cubes in the divided 8$\times$8$\times$8 meteor distribution, and both the color and transparency of each voxel represent the number of meteors contained within it. Voxels that contain no meteors are transparent. Note that the goal of the Convex Hull approach is to isolate the edges of the shower radiant, not to precisely identify the core location (which is done well with the wavelet approach). As a result relatively large voxels are used to improve statistics. 

The average sporadic background density matrix must first be calculated before the DSX meteors can be separated from the sporadic meteors. The average of multiple background days is used to provide good number statistics. The background meteor radiant distribution is created by combining all meteor observations made in five solar longitude bins before the beginning of the DSX shower (160$^{\circ}$-164$^{\circ}$) and five solar longitude bins after the end of the DSX shower (210$^{\circ}$-214$^{\circ}$). We assume the sporadic radiant distribution is constant within our chosen radiant space window in sun-centred coordinates over the entire 54 day interval $\SI{160}{\degree} \leq \lambda-\lambda_\odot \leq\SI{214}{\degree}$. 

\begin{figure}
	\centering
	\includegraphics[width=0.9\linewidth]{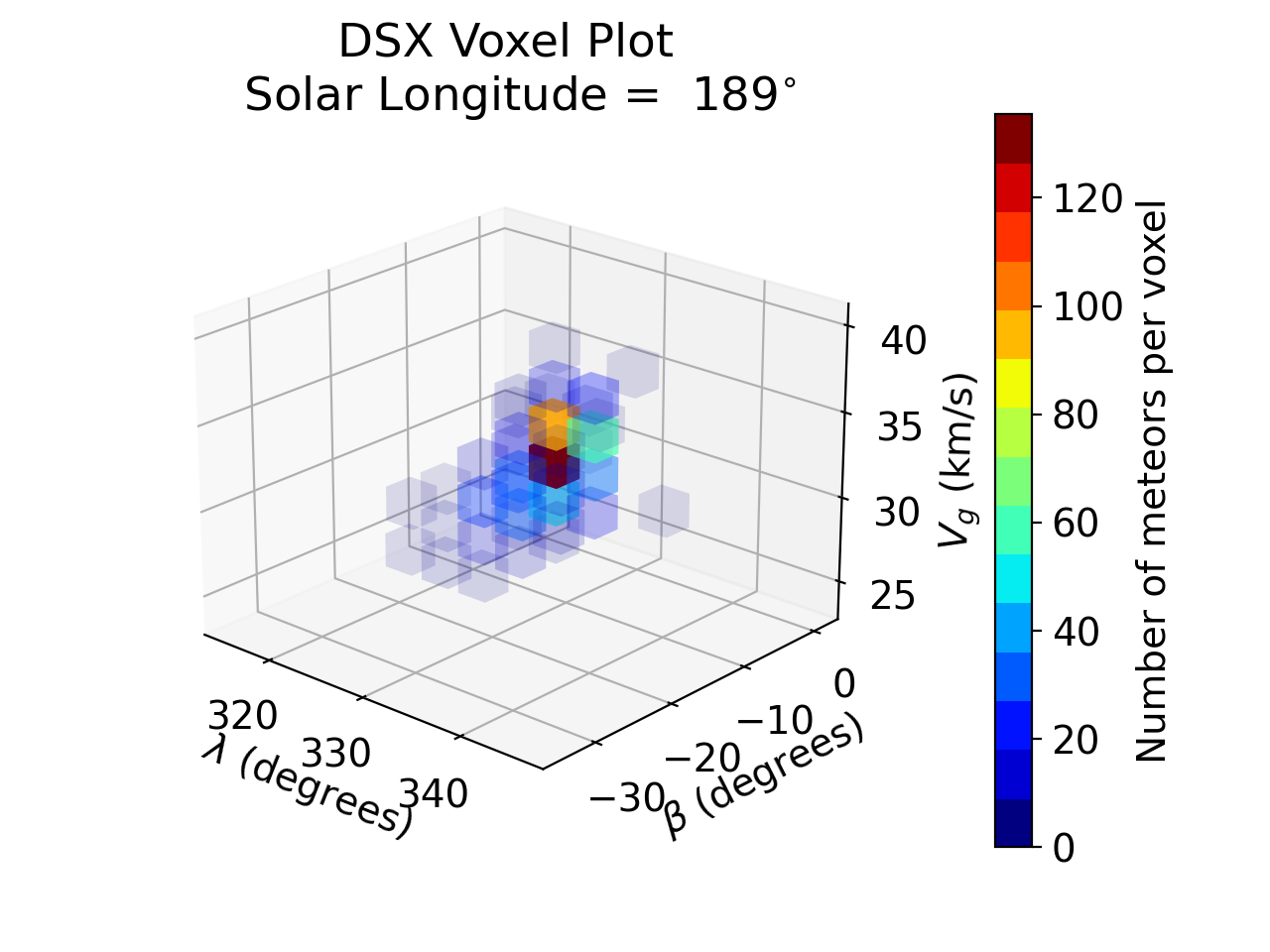}
	\caption{Starting from the distribution in Figure \ref{fig:raw_voxels} shown here are the remaining voxels that contain at least four meteors after the DSX number density matrix and the average background number density matrix are subtracted. Voxels that contain no meteors are invisible.}
	\label{fig:core_voxels}
\end{figure}

In the first step of sporadic contamination removal, we subtract the DSX number density matrix from the average background number density matrix. After subtraction, we take the background matrix's standard deviation and remove any DSX voxel less than three standard deviations above the number of meteors in the corresponding background voxel. Only voxels that contain more than four meteors are kept. This additional requirement was chosen to improve the isolation of the shower core. After this removal step, the voxels which remain are shown in Figure \ref{fig:core_voxels} for solar longitude 189$^{\circ}$.

To further separate the DSX meteors from the sporadic background meteors, we take the remaining voxels and extract the set of meteors with radiants contained within them. Next, we compare those meteor radiants with the wavelet-generated radiant. For each meteor to remain as a possible DSX member, its radiant must be within three standard deviations of each of the ($\lambda - \lambda_{\odot}$), $\beta$, and $V_g$ distributions, centered on the wavelet-generated radiant, where the standard deviation is calculated from the total meteor radiant distribution. The remaining radiants are considered members of the Daytime Sextantids, with a confidence level of 95\%.

\begin{figure}
	\centering
	\includegraphics[width=0.9\linewidth]{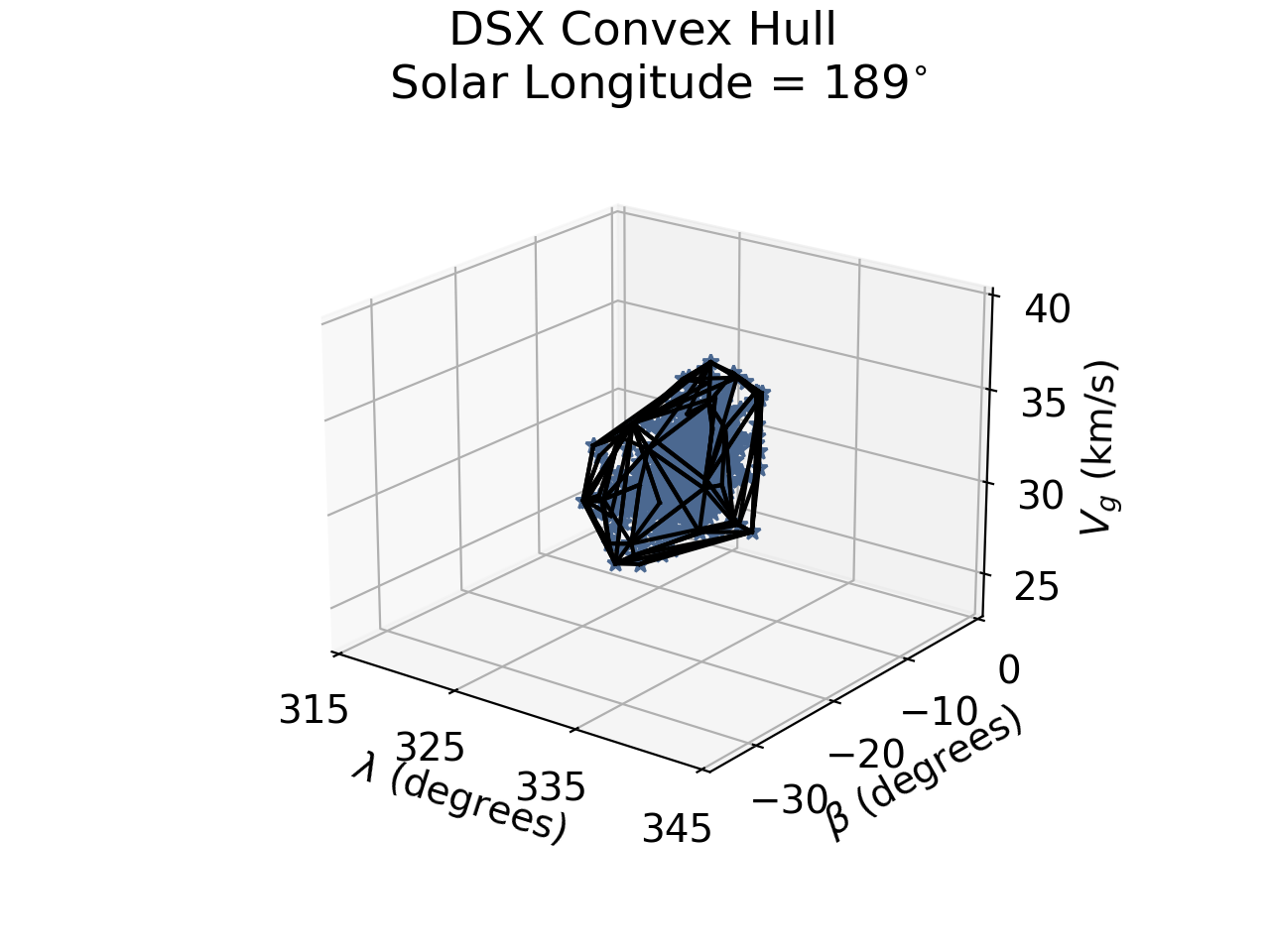}
	\caption{The convex hull applied to the DSX shower. The convex hull has been calculated so that any meteor with a radiant located within the 3D shape is considered to be a member of the DSX shower. This meteor radiant selection method has a confidence level of 95\%.}
	\label{fig:convex_hull}
\end{figure}

\begin{figure}
	\centering
	\includegraphics[width=0.9\linewidth]{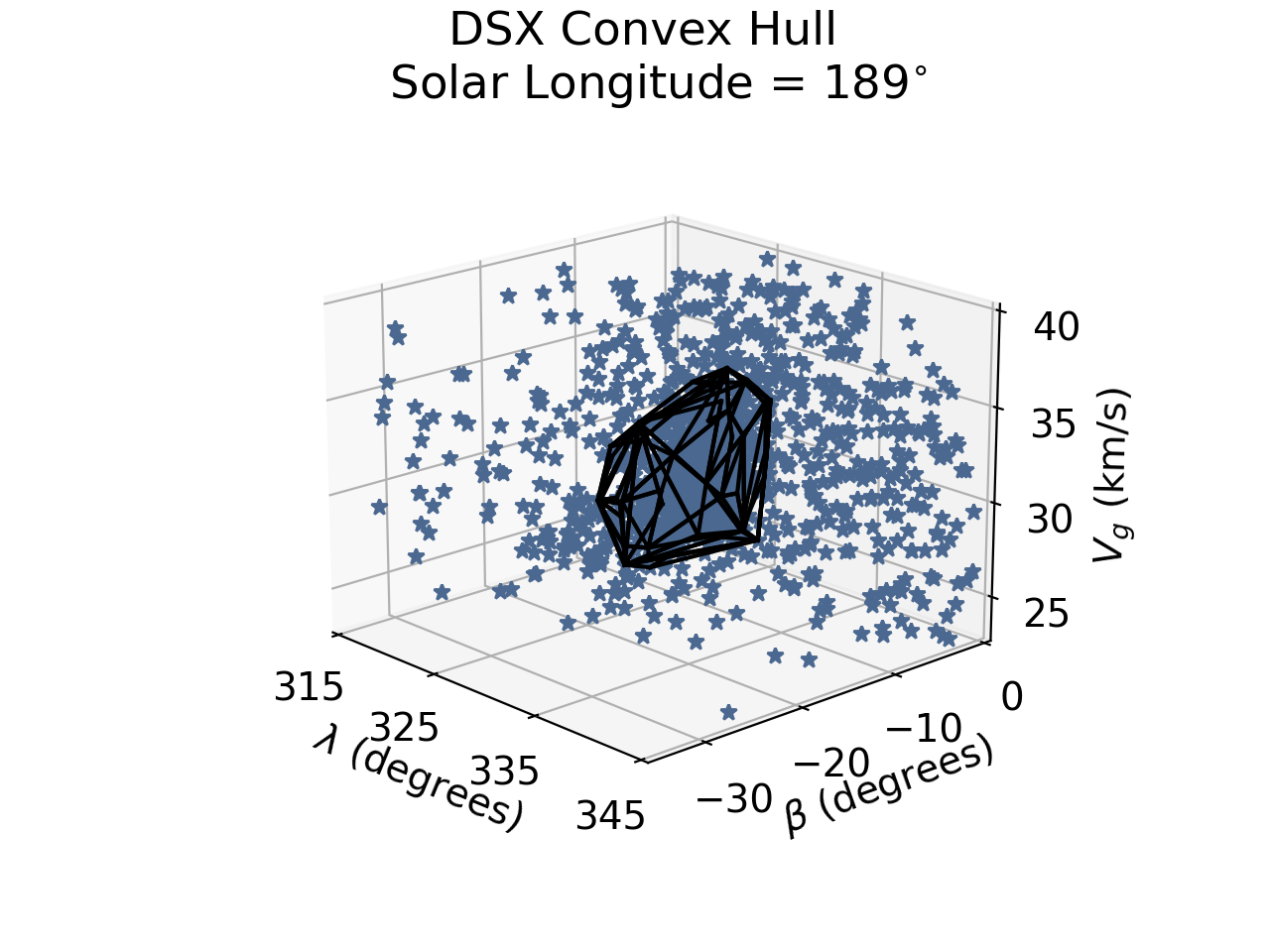}
	\caption{Convex hull of the DSX shower is overlaid on top of the original radiant distribution shown in Figure \ref{fig:rad_distribution} to demonstrate how the convex hull captures the core of the shower.} 
	\label{fig:convex_hull_all}
\end{figure}

The convex hull approach defines the DSX radiant in three-dimensional space, properly accounting for the dimensional dependence of the parameters. We calculate the smallest convex shape that contains all the DSX meteor radiants in the three-dimensional space to accomplish this. This shape, called the convex hull, can be used to identify DSX meteor radiants in other surveys which have similar measurement accuracy. The convex hull for the DSX shower for solar longitude 189$^{\circ}$ is shown in Figure \ref{fig:convex_hull}. Figure \ref{fig:convex_hull_all} shows the convex hull overlaid on top of the total meteor radiant distribution to demonstrate how the convex hull captures the core of the shower. The convex hull method detected the DSX shower from solar longitude 173$^{\circ}$ to 196$^{\circ}$. A complete set of convex hull diagrams as a function of solar longitude for the duration of the DSX shower is given in Appendix B in supplementary materials. 

The 8$\times$8$\times$8 number density size was chosen empirically by repeating this process using different number density sizes. This size was found to produce the most stable convex hull volumes per solar longitude and so was chosen as a suitable size for our dataset. 

It is important to note that the convex hull method fails when the strength of the shower approaches the relative strength of the average sporadic background, which typically occurs near the wings of the meteor shower. This breakdown occurs because there are not enough meteors that pass the 95\% confidence limit when the shower strength is weak. The convex hull method performs best during the peak of the meteor shower when the strength of the shower is significantly greater than the strength of the average sporadic background. 

We also investigated an alternative, more robust convex hull calculation. The convex hull method explored in this section models meteor radiants as discrete points in radiant space. However, radiants contain measurement uncertainties and are more realistically modeled as three-dimensional probability distributions in radiant space. We explored the effect that this change makes on the convex hull results and found no significant difference in the results. We conclude that the more computationally simple convex hull method models the meteors well, and the alternative convex hull method is not a necessary change. A more in-depth discussion of the alternative convex hull method and a comparison of the results can be found in the Supplementary Materials.  
 
\subsubsection{Comparing the convex hull to the 3D wavelet approach}
\label{comparing_convex_wavelet}
	This section will explore the differences between how the wavelet and convex hull approaches define the radiant of a meteor shower and how these differences affect the kind of information we can extract. 
	
	The wavelet transformation searches through a complicated meteor distribution and identifies the region of the highest meteor density. The radiant of a meteor shower is defined as a single set of numbers in ecliptic longitude, latitude, and geocentric velocity space. However, in representing the radiant as a single set of numbers, much of the information about the clustering of radiants that make up the shower is lost. 
	
	The convex hull approach aims to capture as much information about the shower's radiant as possible. The convex hull uses the wavelet approach first to identify the location of the shower. Then it expands on the information gained by the wavelet approach by identifying the three-dimensional geometric extent of the shower radiant in (($\lambda - \lambda_{\odot}$), $\beta$, $V_g$) space. In doing so, the convex hull can determine the shower association for individual meteor radiants, a feature unique to the convex hull approach. This radiant identification isolates the shower meteors and reduces the amount of sporadic contamination in subsequent calculations. 

\begin{figure*}
\centering

    \includegraphics[width=0.9\columnwidth]{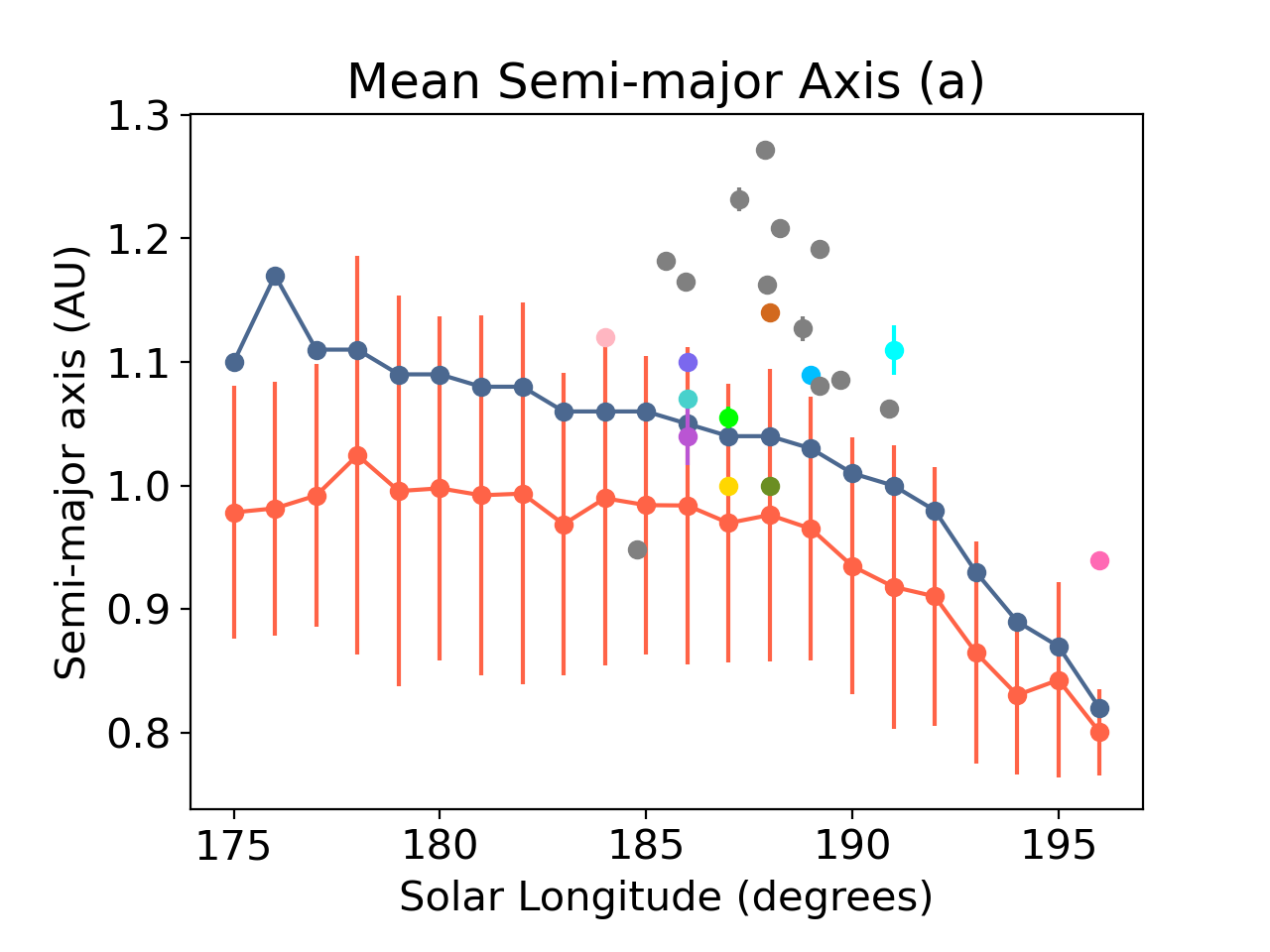}
    \includegraphics[width=0.9\columnwidth]{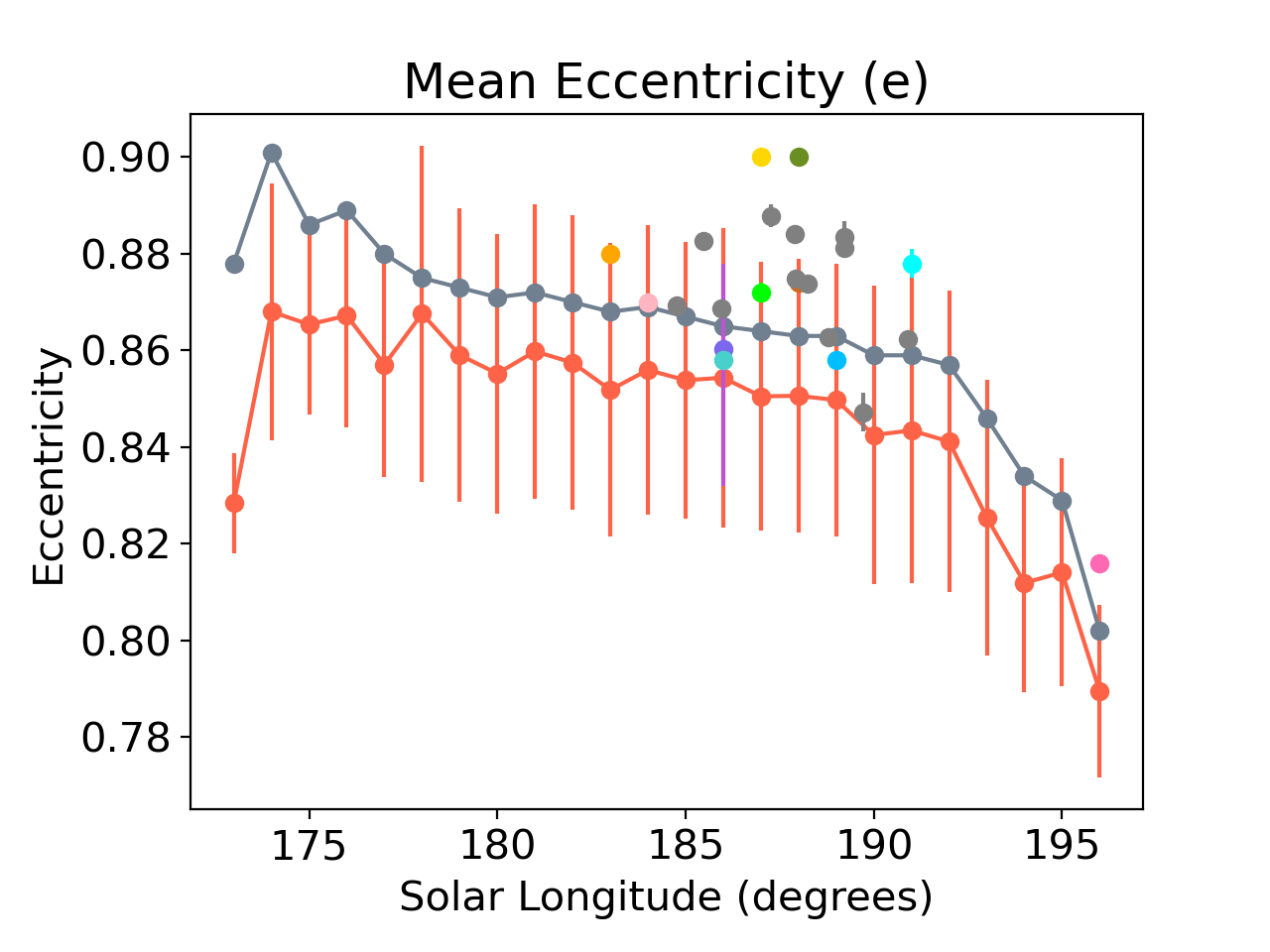}
    
    \includegraphics[width=0.9\columnwidth]{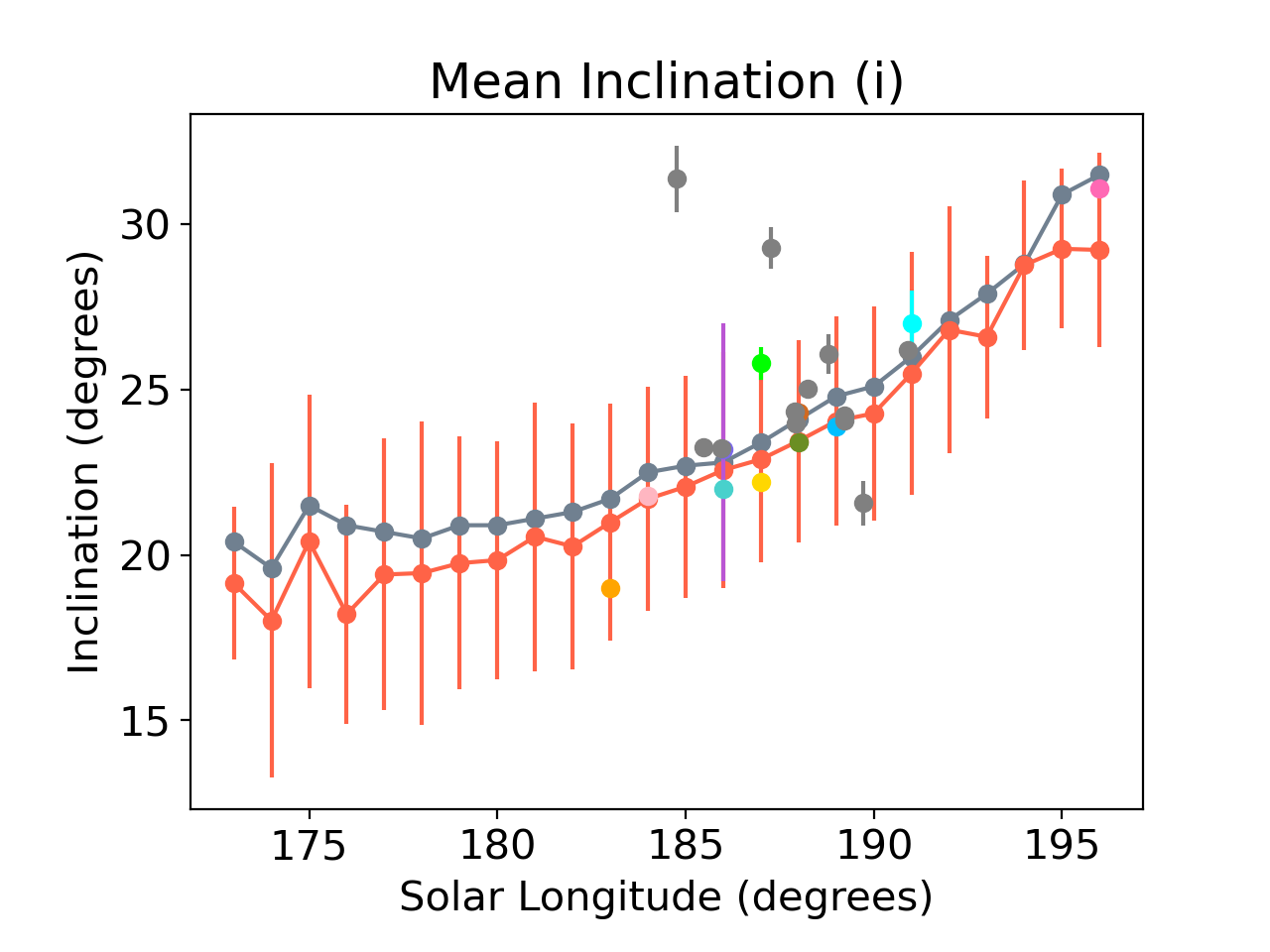}
    \includegraphics[width=0.9\columnwidth]{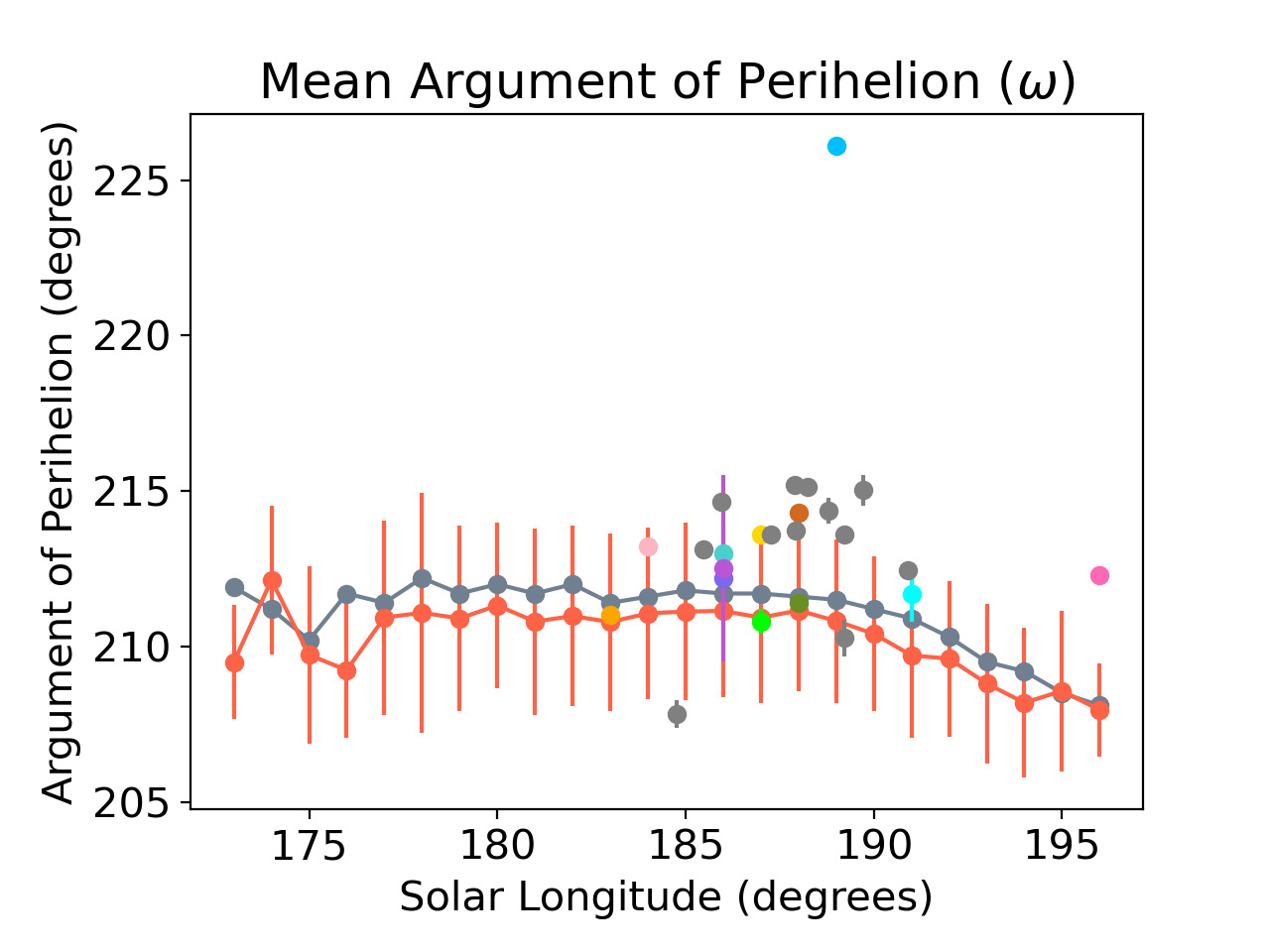}
    
    \includegraphics[width=0.9\columnwidth]{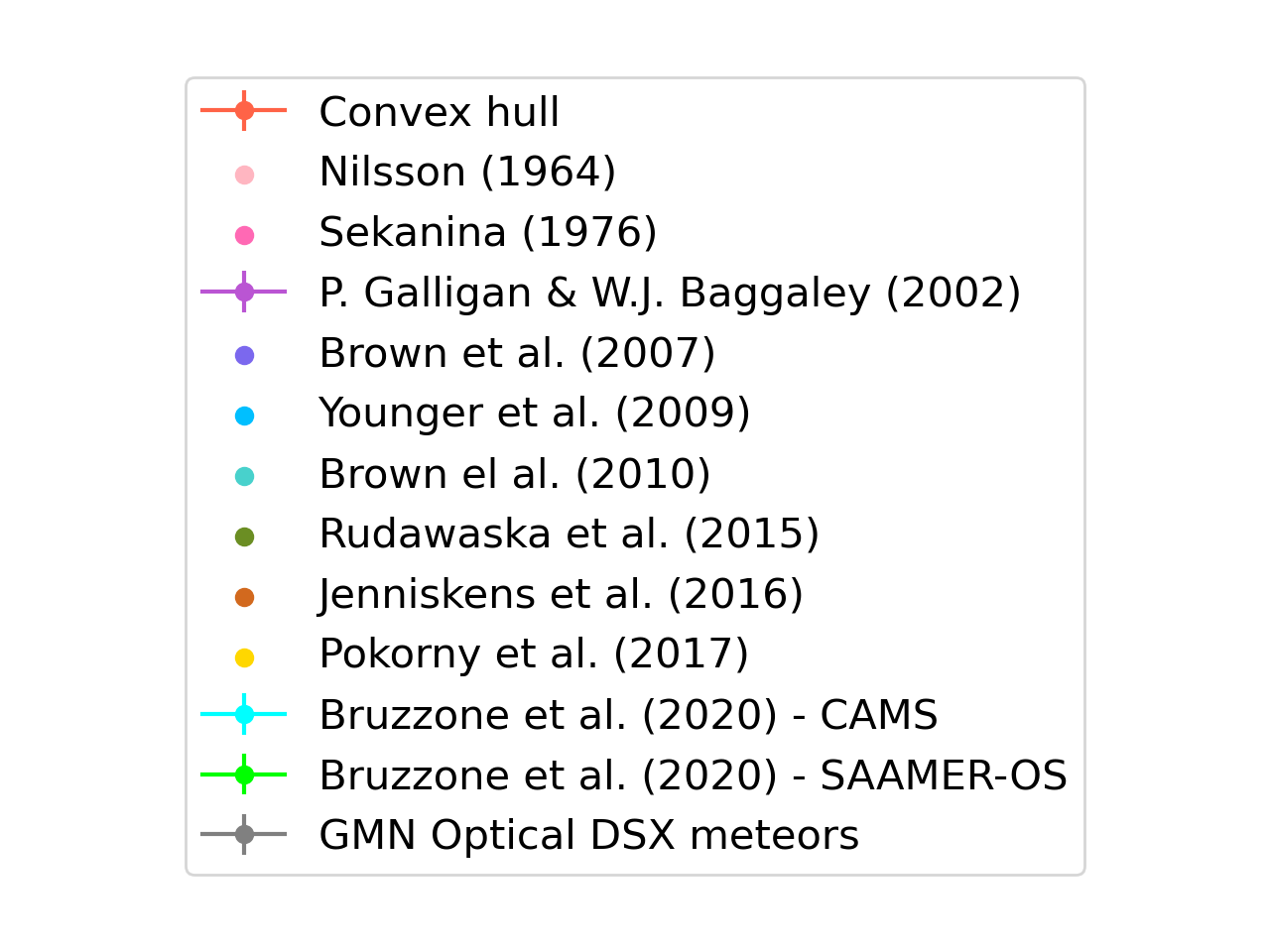}
\caption[short]{Variation of mean orbital elements as a function of solar longitude for the DSX shower calculated using the computationally simple convex hull method (red line and symbols) and using a 3D wavelet (grey symbols and lines). The convex hull is used to extract the set of DSX meteors with a confidence level of 95\%. The mean orbital elements have been calculated following \citet{Jopek2006}. The uncertainty bars represent one standard deviation.The DSX orbital elements from other literature sources (described in Appendix A of the supplementary materials) are shown in each figure. Note that these figures contain the solar longitude range from 173$^{\circ}$ to 196$^{\circ}$.}
\label{fig:orbital_elements}
\end{figure*}

\begin{figure*}
\centering
\begin{tabular}{@{}c@{}}
    \includegraphics[width=0.45\textwidth]{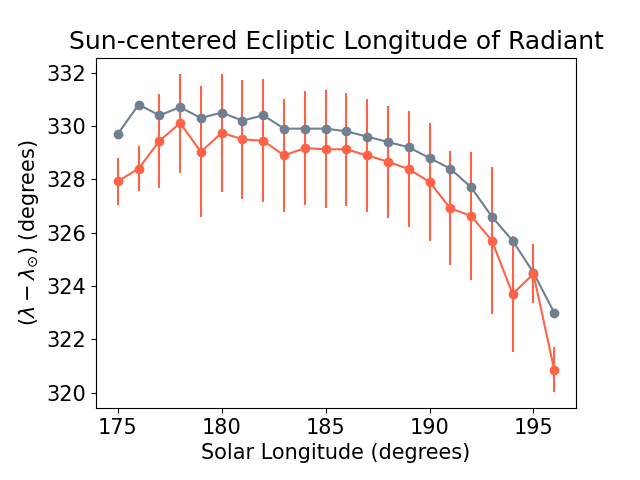}
\end{tabular}
\begin{tabular}{@{}c@{}}
    \includegraphics[width=0.45\textwidth]{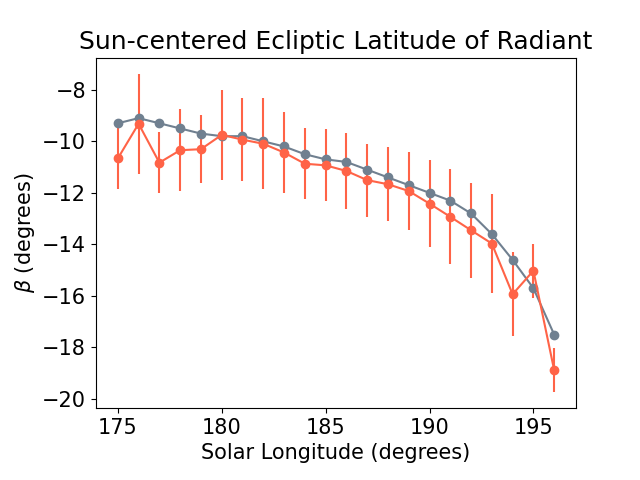}
\end{tabular}
\begin{tabular}{@{}c@{}}
    \includegraphics[width=0.45\textwidth]{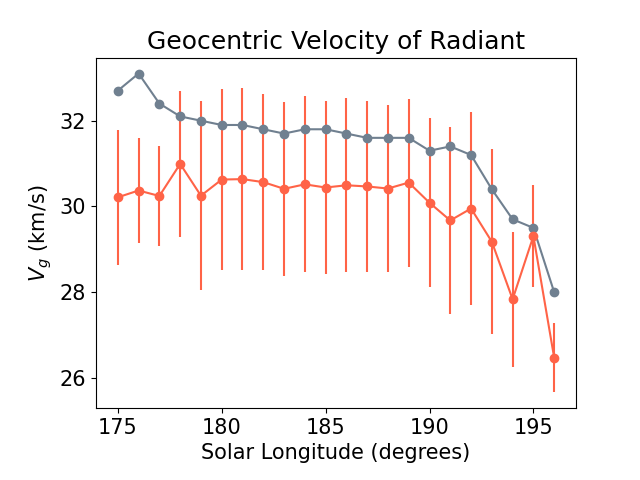}
\end{tabular}
\begin{tabular}{@{}c@{}}
    \includegraphics[width=0.45\textwidth]{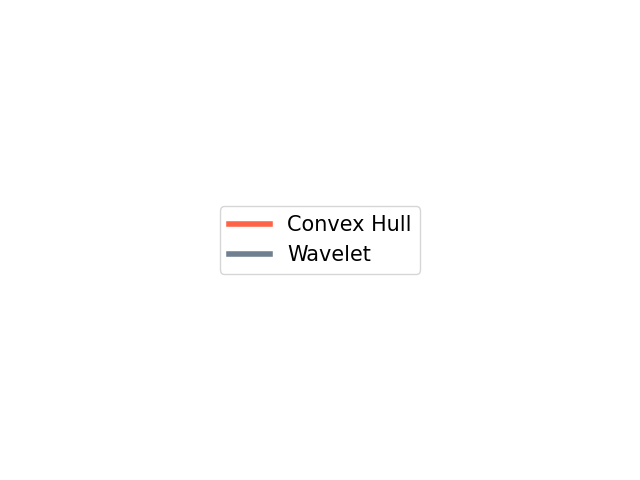}
\end{tabular}
\caption[short]{Radiant drift over the duration of the DSX shower calculated using the computationally simple convex hull method, which models the meteor radiants as points in radiant space. The convex hull is used to extract the set of DSX meteors with a confidence level of 95\%. The mean DSX radiant from the convex hull distribution has been calculated with the method described in \citet{Jopek2006}} 
\label{fig:DSX radiant}
\end{figure*}

Both methods produce similar orbital elements and radiants of the meteor shower stream, but the convex hull approach can take the 3D spread of the shower radiant into consideration when calculating the uncertainty of the average stream orbit, unlike the wavelet approach. Figures \ref{fig:orbital_elements} and \ref{fig:DSX radiant} show a comparison between the average stream orbits calculated by the wavelet and convex hull for each solar longitude of the meteor shower. Note that part of the difference in the mean elements is a small difference in the deceleration correction used between the wavelet data and the single radiants in the convex hull approach, though this systematic difference is within the uncertainty of the convex hull distribution.

\subsection{The DSX Mass Index}
\label{mass_index_section}

In addition to a shower's variation of orbital elements with solar longitude, another diagnostic shower characteristics which is important in understanding a stream's origin and evolution is its mass distribution. 

As the Daytime Sextantids meteor shower occurs primarily during the day, observations are made primarily using radar rather than optical cameras. An indirect measurement of the size distribution of meteoroids in the DSX stream can be made using radar by calculating the mass index of the shower from the distribution of shower echo amplitudes. This quantity is also needed to calculate other fundamental shower properties, such as flux.

The physical significance of the mass index can be illustrated by comparing the mass index of the shower and sporadic meteor populations. The mass contained in meteor shower streams tends to be more concentrated in larger particles, which corresponds to shower populations having a differential mass index below 2. In contrast, the mass distribution of the sporadic meteor population tends to be distributed more heavily in smaller particles rather than larger particles. This size distribution corresponds to a mass index greater than 2. For example, \citet{Blaauw2011} found that the average mass index for sporadic meteors is 2.17 $\pm$ 0.07. Meteor populations with a mass index of exactly 2 would have equal proportions of mass distributed in both large and small particles.

Meteor observations made with radar cannot be used to measure a meteoroid's mass directly. However, it is possible to use the amplitude of the returned meteor echo as a proxy for meteoroid mass, as discussed in \citet{Blaauw2011}. With this approximation we can compute the mass index of the Daytime Sextantids. Here we follow the process introduced in \citet{Pokorny2016}.

\begin{figure}
	\centering
	\includegraphics[width=0.9\linewidth]{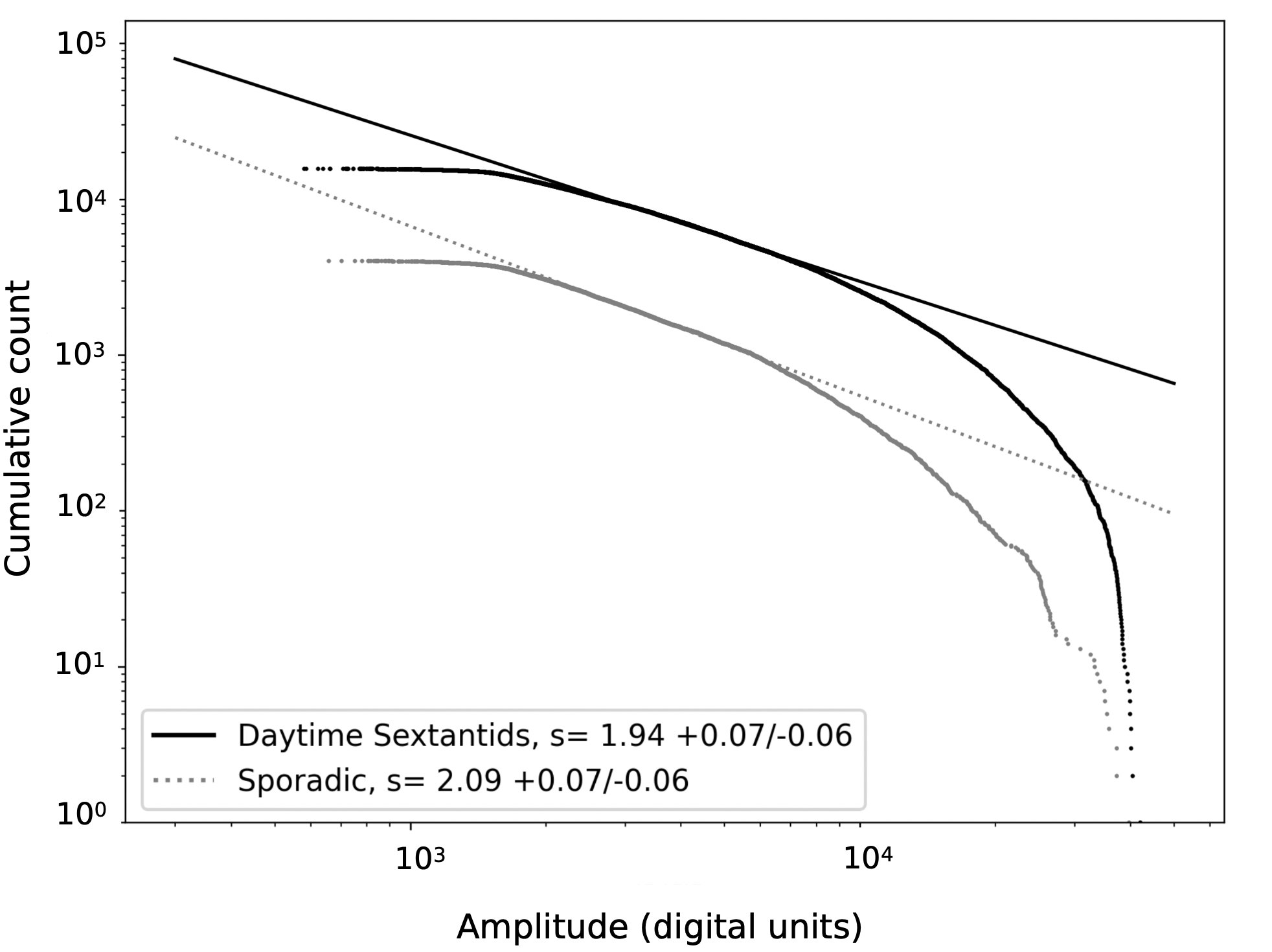}
	\caption{Multinest fit of the cumulative number of single station echoes which meet our filter criteria for the stream above a given peak amplitude value over the entire duration of the Daytime Sextantids meteor shower. Shown for comparison is the corresponding fit of the average sporadic population, representing echoes meeting the same filtering criteria but for time periods before and after the end of the DSX shower.}
	\label{fig:multinest_all}
\end{figure}

The mass distribution of a meteoroid population is normally assumed to follow a power law of the form: 

\begin{equation}
N_c \propto M^{(1-s)}
\end{equation}

\noindent where $N_c$ is the cumulative number of meteoroids with mass greater than $M$, and $s$ is the differential mass index. The power-law nature of the mass distribution allows us to use the logarithmic plot of the cumulative number distribution vs. amplitude to determine the mass index of the meteor population. For the underdense region of the plot, the slope is equal to $1-s$ and can be identified in Figure \ref{fig:multinest_all} as the straight portion of the plotted values. 

For underdense meteors, the amplitude of the returned echo is proportional to the electron line density of the meteor trail, which is in turn proportional to the mass of the meteor \citep[$A \propto q \propto m$;][]{Ceplecha1998e}. For overdense meteors the power received is proportional to q$^{0.5}$. In Figure \ref{fig:multinest_all}, the region of overdense echoes is located to the right of the underdense region where the amplitude distribution shows a steep roll-off since $A \propto q^{0.25}$, with a transition region between the underdense and overdense regions. The left side of the plot shows a rollover from the straight underdense region as the limiting sensitivity of the radar is approached and echoes missed.   

In practice, measuring the mass index of a shower from radar amplitude distributions involves isolating individual echoes belonging to the shower and then applying an algorithm to measure the straight-line portion of the resulting cumulative number - amplitude plot. However, how the boundaries of this straight portion are chosen can significantly affect the calculated slope value. For example, if the start and endpoints of the straight region are chosen by hand, different slopes are found by different analysts for the same data. 

A demonstrated solution to this issue is to use the Multinest algorithm \citep{Pokorny2016}. Multinest is a Bayesian inference tool that determines the ideal boundaries of the straight region of such a plot. It can be used to objectively calculate mass index and uncertainty. Although, it should be noted that an assumption incorporated into Multinest is that the sporadic population before and after the shower can be used to approximate the sporadic background throughout the shower duration. The mass index calculations which follow for the  Daytime Sextantids use the implementation of the Multinest program as discussed in \citet{dewsnap2021}.

A challenge, even using Multinest, in calculating the mass index of a meteor shower is that there must be a good straight line in the logarithmic plot of the cumulative count vs. amplitude, as shown in Figure \ref{fig:multinest_all}. However, CMOR's  multi-station orbital data system requires a good signal on several remote stations and, for this reason, is biased against detecting the weakest and smallest echoes. This filtering causes the range in mass between the limiting mass of the orbital data observations and the overdense roll-off to be too small to obtain a meaningful straight line fit.

Unlike multi-station data, single station 29.85 MHz CMOR data contains observations of weaker echoes, although these detections cannot be used for orbital data. The single station CMOR data also contains a larger number of echoes. For example, the number of single-station echoes made between 2011-2020 at solar longitude 189$^{\circ}$ is 143,576, but the number of these with multi-station observations is only 43,554.

Single-station CMOR meteor echo observations do not contain unique radiant measurements, but do contain the direction to each echo obtained through interferometry. They also contain speed estimates through measurement of the phase change prior to the specular point, termed pre-t0 speeds. Here we use the algorithm described in \citet{Mazur2020} to estimate pre-t0 speed.  Because of the specular reflection condition for backscatter echoes, the radiant is known to be located in a band of sky 90$^{\circ}$ from the observed echo direction \citep{Jones2006}. Because of this, the shower meteors cannot be filtered using the new meteor selection method described in detail earlier, which requires exact radiant information. Instead, we determine which meteors are potential members of a meteor shower using echo location and selecting those which are at right angles to the apparent radiant as estimated from the wavelet-generated ephemeris and which fall within a speed window about the nominal shower value.

The velocity filter used to calculate the DSX mass index is 8\%, meaning that only meteors with velocities within  8\% of the wavelet-generated velocity of the peak shower day pass through this filter. The angular separation filter is 3.5$^{\circ}$, meaning that only meteors with observed directions within 3.5$^{\circ}$ of right angles to the shower radiant pass through this filter. The latter value is chosen to account for measurement error in the interferometry for weaker echoes. Individual meteors must pass through both the velocity and angular separation filter to be considered members of the meteor shower. Section \ref{filter_investigation} provides an in-depth discussion on how these filters have been chosen. 

\begin{figure}
	\centering
	\includegraphics[width=0.9\linewidth]{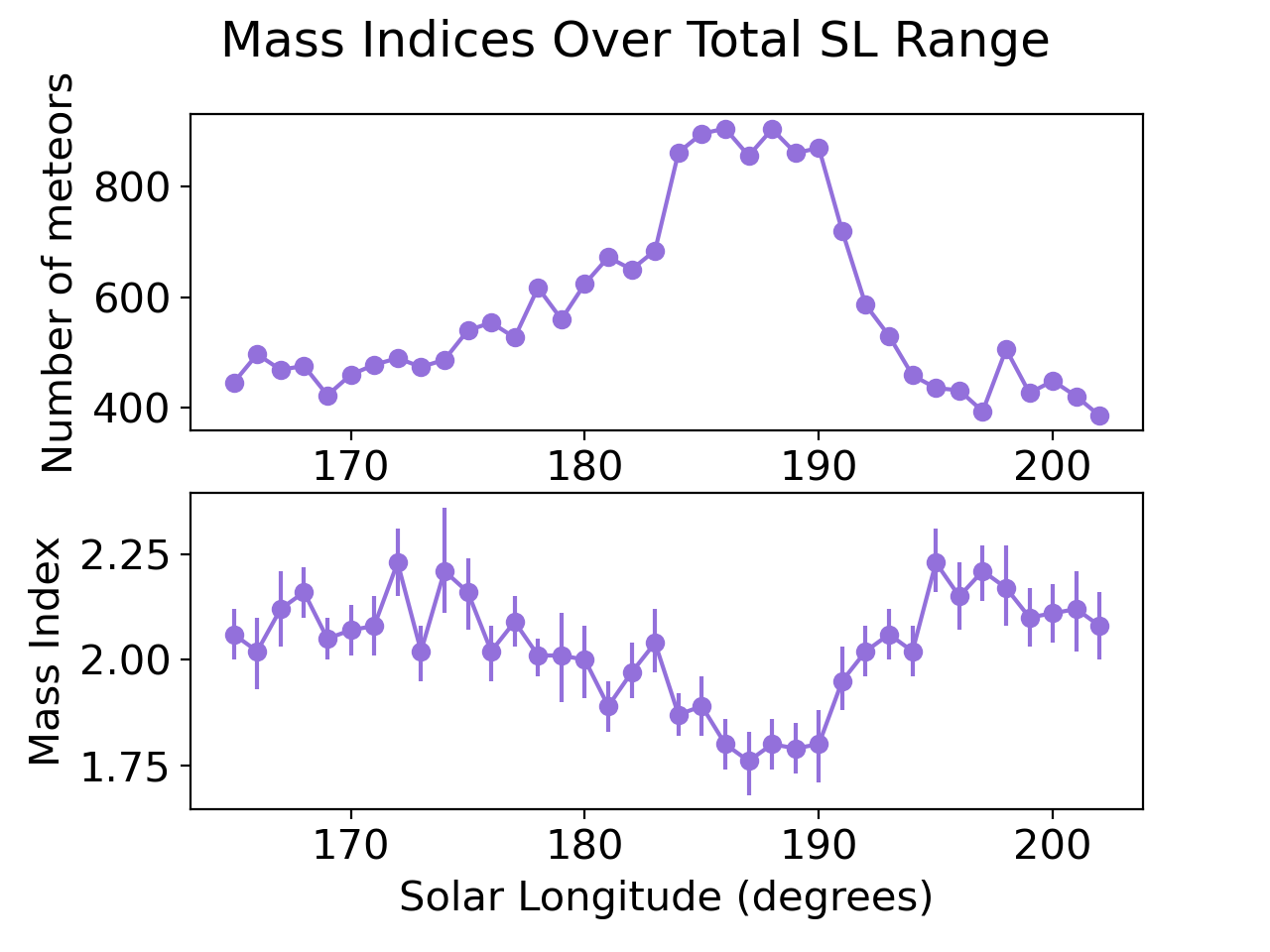}
	\caption{Mass index of the DSX shower as a function of solar longitude, calculated by Multinest, along with the number of DSX single station meteor echoes  that have passed through the filter. The peak of the shower, shown in the upper diagram, is associated with a lower mass index as can be seen in the lower plot. Uncertainties are from the Multinest fits.}
	\label{fig:mass_index_total_range}
\end{figure}

\begin{figure}
	\centering
	\includegraphics[width=0.9\linewidth]{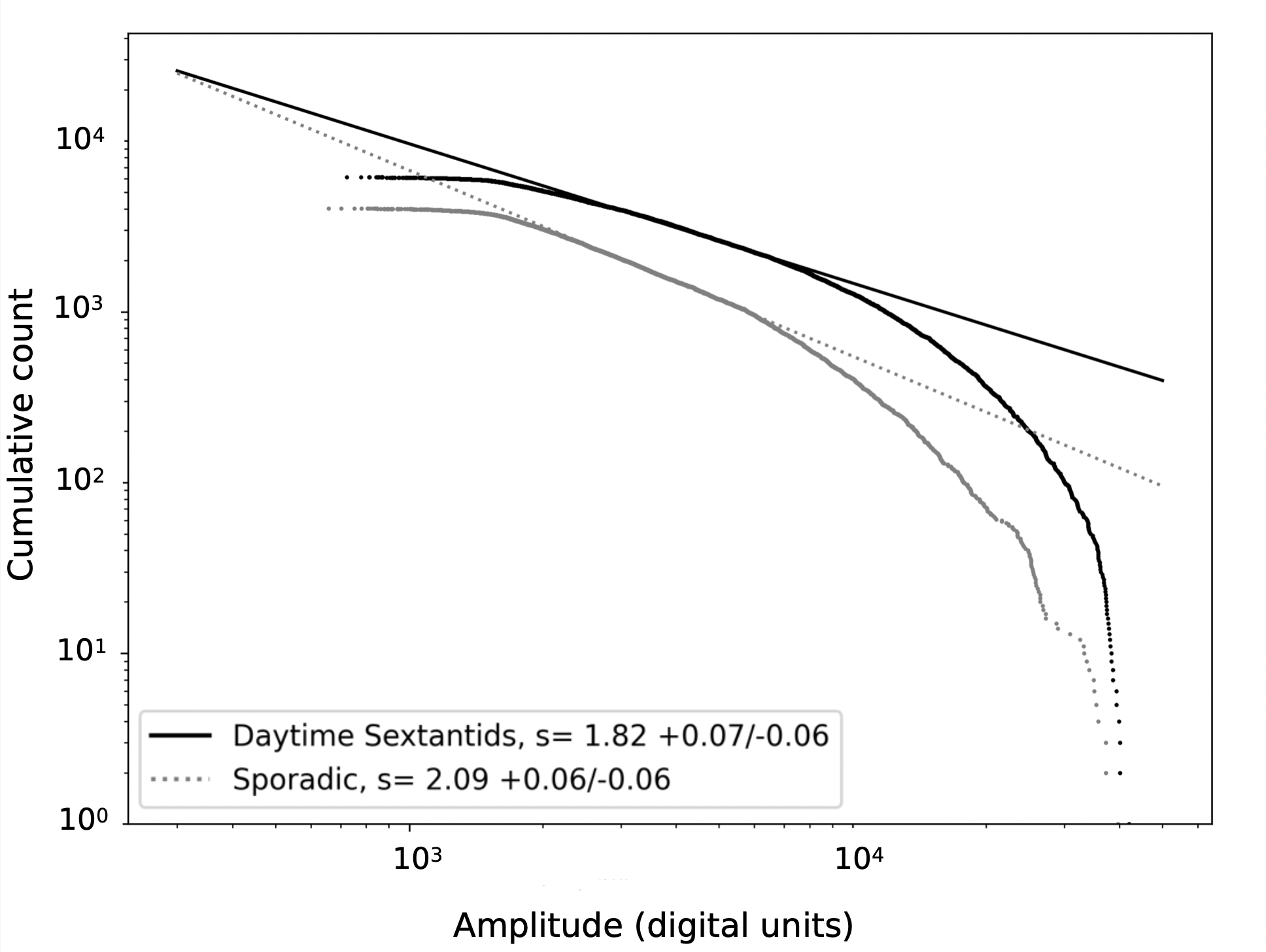}
	\caption{Multinest fit during the peak of the Daytime Sextantids meteor shower (solar longitude 184$^{\circ}$ to 190$^{\circ}$) and the fit of the average sporadic population.}
	\label{fig:multinest_peak}
\end{figure}

At this stage, the single station meteor echo dataset contains only the possible shower meteors determined by the velocity and angular separation filters. Next, we analyze this dataset using the Multinest implementation described in \citep{dewsnap2021}. We use only the solar longitude range 185$^{\circ}$ - 190$^{\circ}$ surrounding the peak of the DSX due to the higher level of sporadic contamination in the wings of the shower. This contamination effect can be seen in Figure \ref{fig:mass_index_total_range}, where the outer wings of the shower have higher mass indices, characteristic of the sporadic meteor population. The outer wings also have lower number of shower meteors. The Multinest fit around the peak is shown in Figure \ref{fig:multinest_peak} and the Multinest fit over the entire duration of the shower is shown in Figure \ref{fig:multinest_all}.

\subsubsection{Mass Index Refinement Using a Mixing Model}
\label{mixing model}
The meteor filtering method used to separate shower meteors from sporadic meteors is not stringent, and thus it allows a considerable number of sporadic meteors to contaminate the final mass index calculation of the meteor shower. This contamination tends to cause the calculated mass index of the shower to be artificially raised from its actual value because sporadic meteors tend to have higher mass indices than showers. To address this problem, we have developed a mixing-model method to minimize the effect of sporadic contamination using the Multinest mass index calculation approach.

Prior to the Multinest analysis, the sporadic meteor population undergoes the same filtering process as the shower. For the DSX analysis, the average number of single station sporadic meteor echoes that pass the filtering per solar longitude day is 285 meteors between solar longitudes 165$^{\circ}$-170$^{\circ}$ and 200$^{\circ}$-202$^{\circ}$. Assuming no time variation of the sporadic sources, we suppose that a similar number of sporadic meteors per day are "mixed" with the filtered DSX meteor population. 

\begin{figure*}
\centering
    \begin{tabular}{@{}c@{}}
        \includegraphics[width=0.45\textwidth]{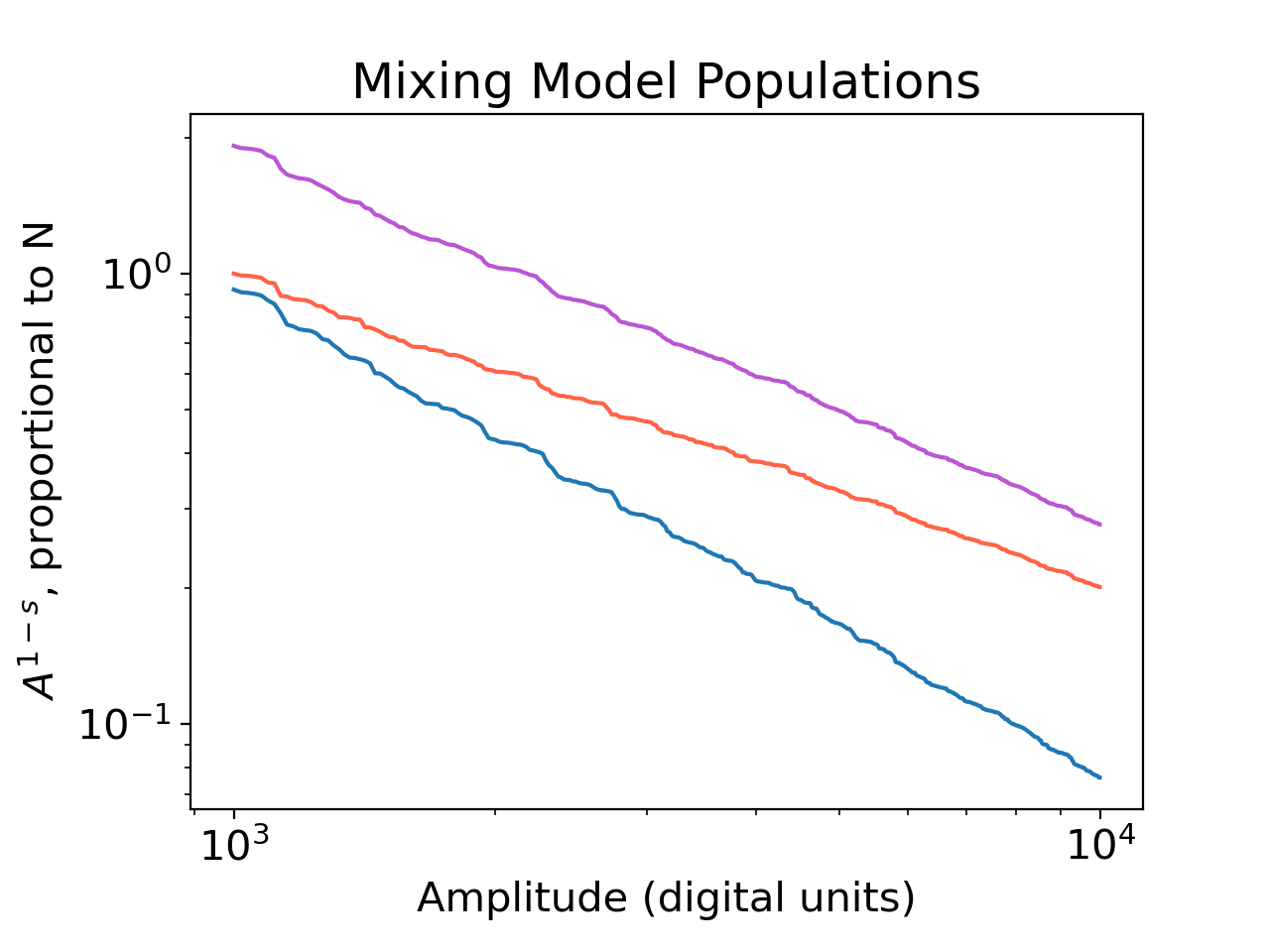}
    \end{tabular}
	\begin{tabular}{@{}@{}c}
	\includegraphics[width=0.40\linewidth]{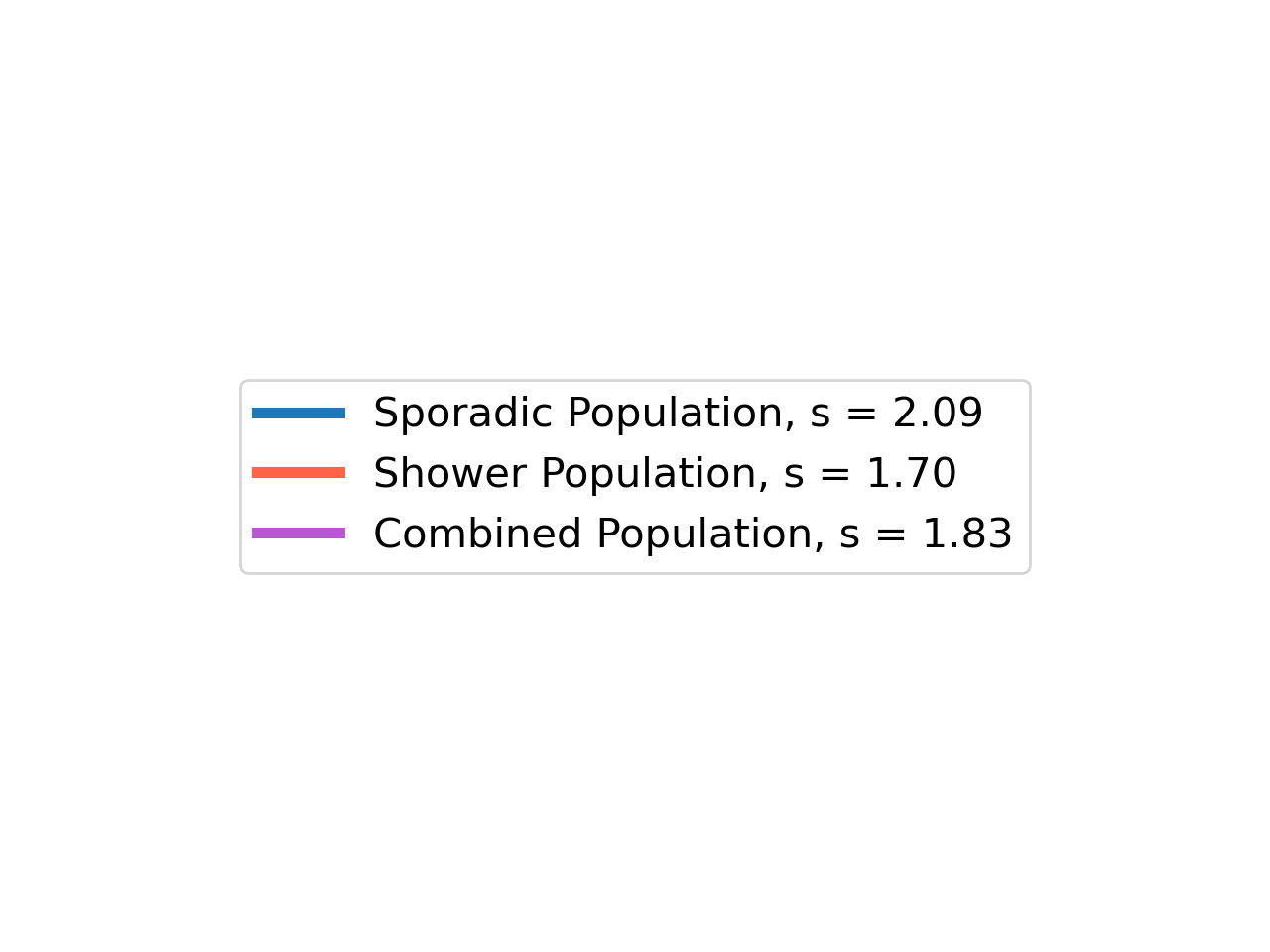}
	\end{tabular}
	\caption{An example of the mass index of the  shower and sporadic synthetic populations that together result in the expected observed (combined) population.}
	\label{fig:mixingmodel}
\end{figure*}

The mixing model simulated two meteor populations using the equation $N \propto A^{1-s}$. The first simulated population represents the sporadic meteor population, which contains an average of 285 meteors per solar longitude day. The simulated sporadic population is given a mass index equal to the sporadic mass index calculated by Multinest, which approximates the sporadic background using the solar longitudes 165$^{\circ}$-170$^{\circ}$ and 200$^{\circ}$-202$^{\circ}$. The second meteor population is an ideal shower population, containing only shower meteors and with no sporadic contamination. The number of meteors in the idealized meteor population is equal to the number of shower meteors that pass the filters in the Multinest analysis minus the average number of sporadic meteors. The mass index of the simulated shower population is iterated from 1.5 to 2.5 in steps of 0.1. For each iteration, this shower population is combined with the simulated sporadic population, and Multinest is used to calculate the mass index of the combined population. An example of the populations used in the mixing model is shown in Figure \ref{fig:mixingmodel}, where the shower population has a mass index of 1.70, as an example. 

Next, we take the observed mass index during the DSX shower, calculated by Multinest, and we determine which mass index of the idealized simulated meteor shower produces the measured DSX mass index when mixed with the known sporadic mass index. The mass index of the best fit shower population is then taken to be the true mass index of the DSX shower, corrected for sporadic contamination. Using the mixing model, we calculated a contamination-corrected mass index of $1.64 \pm 0.06$ at the peak of the DSX, with an average of $1.70 \pm 0.07$ measured over solar longitudes $184^\circ - 190^\circ$.

\subsubsection{Mixing Model Uncertainty}
While multinest outputs formal uncertainty values for fits of the mass index, application of the mixing model introduces another source of uncertainty. To account for this  we calculate the associated uncertainty using a Monte Carlo approach. 

Our method is to first generate a simulated shower meteor population modeled as a Gaussian distribution in $s$, using the Multinest-calculated mass index as the mean of the distribution and the averaged uncertainty of the Multinest output as the standard deviation. The same process is done to generate a sporadic population. These Gaussian distributions represent the spread of the individual populations extracted from the Multinest calculation uncertainty. 

Next, 1000 realizations are performed where a random value is sampled from both the shower and sporadic Gaussian distributions. These are then normalized to the number of shower and average number of sporadic meteors that pass filtering. The final aggregate distribution then has a combined Multinest-calculated mass index and uncertainty produced. The standard deviation of the resulting set of 1000 contamination-corrected mass indices is taken as the total uncertainty in the mixing model mass index calculation.

\subsubsection{Defining When the Mixing Model Can Be Applied}

The mixing model is sensitive to the ratio of shower to sporadic meteors. It is most useful when there are many more shower meteors than sporadics as is generally the case around a shower's peak. The mixing model assumes that we can approximate the sporadic background by averaging the meteor population before and after the shower, when shower contamination to the sporadic background is minimal. This assumption is valid only when the strength of the shower is much less than the strength of the sporadic background. Additionally, we expect small variations in the sporadic mass index during the time of the shower so any mixing model estimate that is very sensitive to the absolute mass index of the sporadic background is more prone to systematic uncertainty.  

To quantify the validity range of the assumption of constant sporadic mass index, a sensitivity analysis was performed across the activity period of the DSX. The goal of the sensitivity test is to determine time intervals when the corrected mass index is particularly sensitive to small changes in the sporadic background. We repeated the mixing model Monte Carlo runs using different mass indices for the sporadic background, ranging from the largest observed value, 2.23, to the lowest observed value, 2.10, of the sporadic background days before and after the shower. DSX solar longitude days that do not show much variation in the final resulting shower mass index values are chosen as appropriate candidates for the mixing model. 

\begin{figure*}
\centering
    \begin{tabular}{@{}c@{}}
        \includegraphics[width=0.5\linewidth]{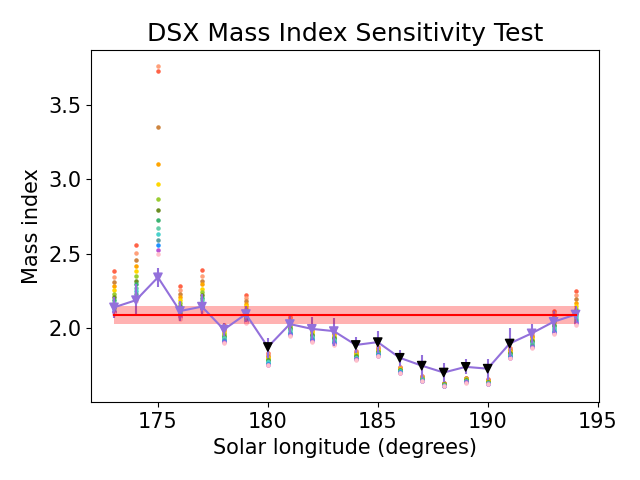}
    \end{tabular}
	\begin{tabular}{@{}@{}c}
    	\includegraphics[width=0.30\linewidth]{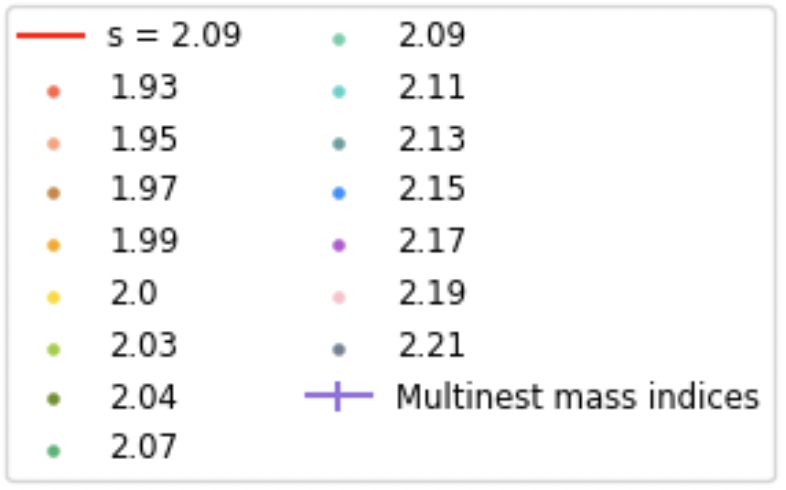}
	\end{tabular}
	\caption{Sensitivity analysis used to determine the period of validity of the mass index mixing model. The observed (combined) mass index measured by Multinest for filtered data (which includes shower and sporadic meteor echoes) is shown by purple inverted triangles. The red horizontal line represents the mass index of the average sporadic background and the red shaded zone its variance from solar longitudes 165$^{\circ}$-170$^{\circ}$ and 200$^{\circ}$-202$^{\circ}$. The colored dots represent the mixing model corrected shower mass index for a given synthetic value of the sporadic mass index (shown in legend). The black inverted triangles represent those solar longitude days that had mass index variations less than 0.1.}
	\label{fig:sensitivity}
\end{figure*}

We chose a variation limit of 0.1, as this is the size of the variance in mass index observed in the sporadic background by CMOR over the time interval of the shower \citep{Pokorny2016}. Any day showing more variation than 0.1 in the mixing model interpreted shower mass index is considered an unsuitable candidate for the mixing model. For those days, the mixing model  cannot be reliably corrected for sporadic contamination. A plot of the sensitivity test results is shown in Figure \ref{fig:sensitivity}. We found the solar longitude days from 184 - 191 produced reliable mass indices for the Daytime Sextantids based on this criteria and using the mixing model corrections.

\subsection{Optimizing Single Station Echo Filters for the DSX Shower}
\label{filter_investigation}

The usual shower filtering for CMOR uses a default value of $\sim4^{\circ}$ for the radiant filter and 10\% for the velocity filter in selecting single station echoes associated with a meteor shower \citep{Brown2008}. We now investigate how different choices of these filters affect the mass index of the DSX and introduce a more robust method for determining which filter is appropriate for a given shower. 
	
Choosing an appropriate filter is challenging. Small filters are more stringent and remove a significant number of sporadics but at the expense of many meteors that are members of the shower.  Larger filters increase the total number of meteors, leading to good Multinest fits, but at the expense of increased sporadic contamination.  The optimal filter choice should maximize the shower/sporadic ratio.  

To define this optimal filter, we first investigated how different filters affect the calculated mass index of the shower. To do this, we varied the radiant filter from 1$^{\circ}$ to 8$^{\circ}$ and the velocity filter from 5\% to 20\%. Velocity filters below 5\% resulted in too few echoes and caused an error in the Multinest code and were therefore not used in this investigation. The mass index, corrected using the mixing model described in Section \ref{mixing model}, is calculated for each possible filter combination. Additionally, the same filters were used to calculate the mass index of the sporadic background, and the solar longitudes before and after the shower were used. 

\begin{figure}
	\centering
	\includegraphics[width=0.9\linewidth]{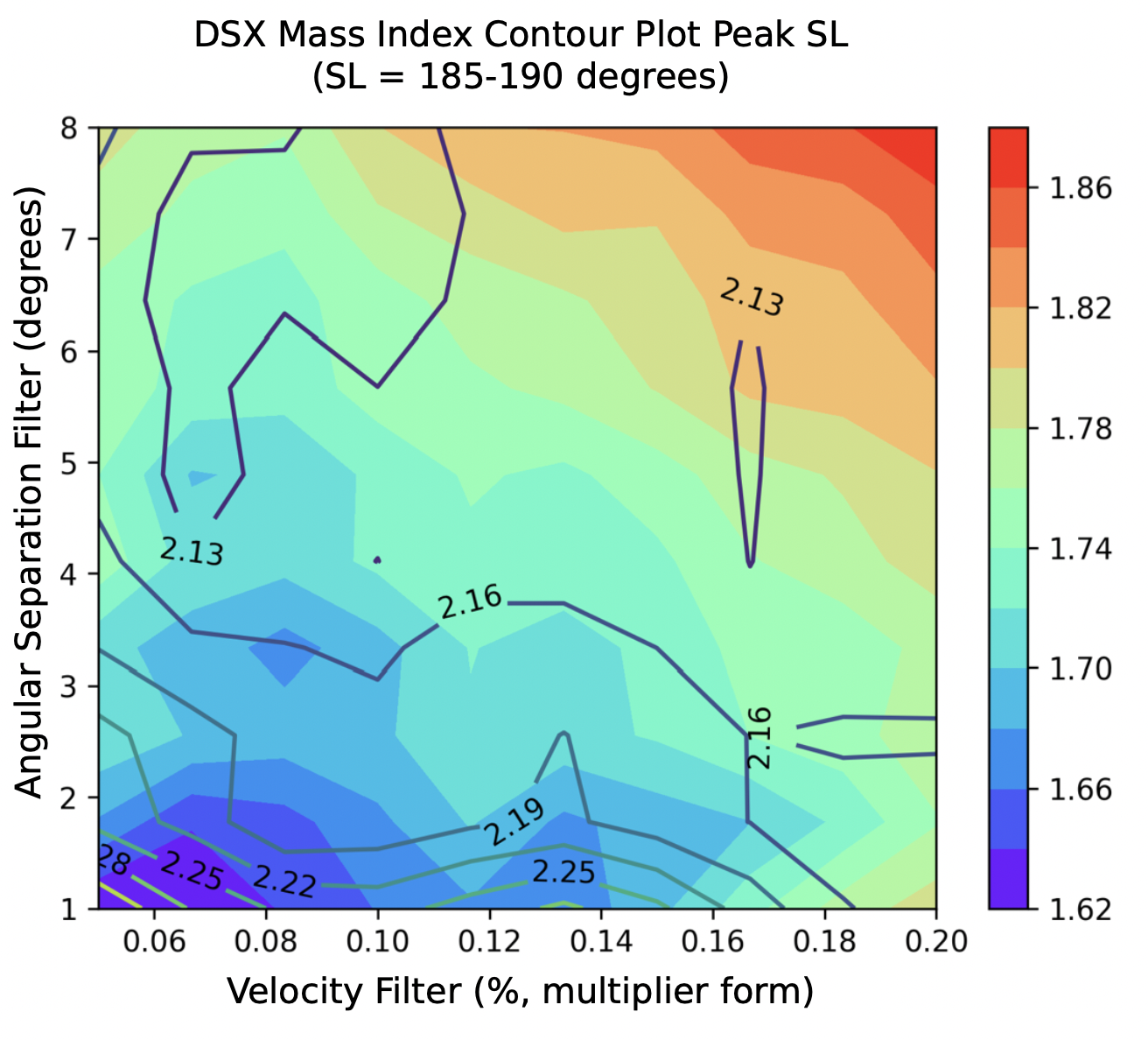}
	\caption{The mixing model estimate of the mass index of the DSX shower calculated using different radiant and velocity filters (contour colors) for the aggregate of all single station echoes from solar longitude 185$^{\circ}$ to 200$^{\circ}$ for all data collected between 2011-2020. The contour lines are the mass index of the sporadic population measured using the same radiant and velocity filters and also using MultiNest.}
	\label{fig:mass index contour}
\end{figure}

The mass index of the sporadic background is taken from mass index fits of all echoes. Figure \ref{fig:mass index contour} shows the mass index resulting from different radiant and velocity filters. The colored regions in the contour plot represent the calculated shower mass index value using the mixing model. The labelled dark contour lines represent the mass index of the sporadic background computed using the same radiant/velocity filters. 

From Figure \ref{fig:mass index contour}, the mass index of the DSX shower is lower for smaller filters and higher for less stringent filters, as expected, since the larger filters allow more sporadic meteors to contaminate the sample and raise the mass index. At radiant filters less than 3$^{\circ}$, the mass index value of the sporadic background shows large variances and has larger mass index values. This is likely due to the small number statistics and poor Multinest fits. 

\begin{figure}
	\centering
	\includegraphics[width=0.9\linewidth]{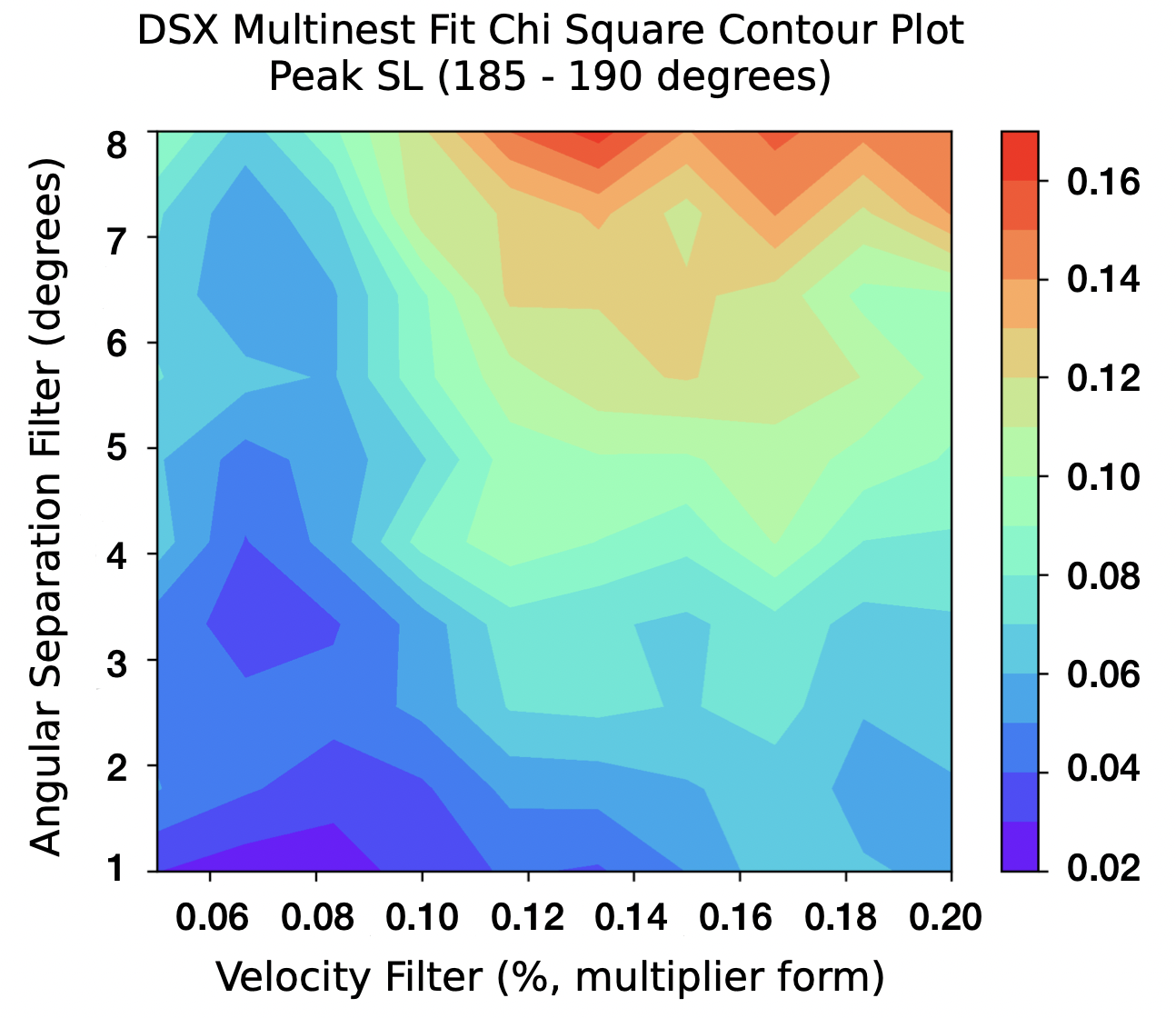}
	\caption{Chi square result (color bar) of the Multinest fits calculated using different filters. The Multinest amplitude ranges are used as the boundaries to perform the chi square test.} 
	\label{fig:chi square contour}
\end{figure}

\begin{figure}
	\centering
	\includegraphics[width=0.9\linewidth]{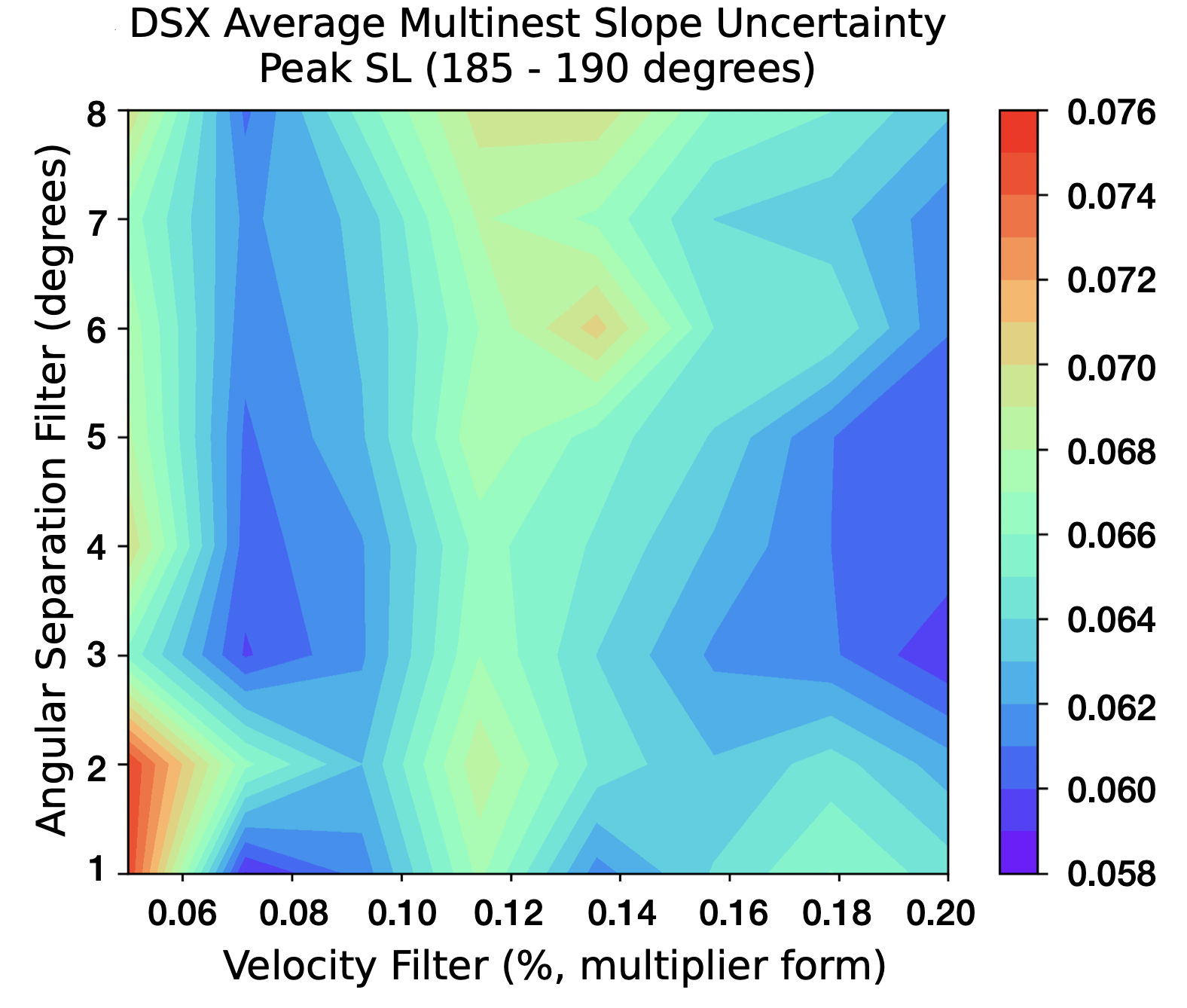}
	\caption{Uncertainties in the Multinest fits using different radiant and speed filters.}
	\label{fig:uncertainty contour}
\end{figure}

The filters in the lower-left corner of Figure \ref{fig:mass index contour} produce the smallest DSX mass indices. However, these filters are so stringent that many DSX meteors are rejected along with the sporadic meteors, leading to poor Multinest fits. To evaluate which filters reject the most sporadic meteors without removing too many DSX meteors, we do a chi-square test on the Multinest fits of each radiant and velocity filter combination. For each filter combination, the Multinest-fitted straight line is compared to the actual data and the fit is evaluated using a chi-square test. A contour plot of the chi-square test results is shown in Figure \ref{fig:chi square contour}. Figure \ref{fig:uncertainty contour} shows a contour plot of the Multinest uncertainties, which can be used as a rough measure of the quality of the Multinest fit.

From Figures \ref{fig:mass index contour} we see that the shower mass index changes relatively slowly as a function of the filter size at mid range values (3$^{\circ}$-6$^{\circ}$ and 0.08-0.15 \%) where the measured mass index difference is of order 0.1 or less. The DSX shower radiant has some extent, and therefore the radiant filter must accommodate that natural spread. 

To estimate this spread, Figure \ref{fig:DSX radiant} shows the ecliptic latitude and longitude values per solar longitude calculated using the convex hull method. The uncertainty in the radiant values is around 2$^{\circ}$, including observational uncertainty and physical spread, so the radiant filter best suited for the DSX shower should also be larger than 2$^{\circ}$. Looking at Figures~\ref{fig:mass index contour}, \ref{fig:chi square contour}, and \ref{fig:uncertainty contour}, the filter which combines the lowest mass index, Multinest chi-square result, and Multinest uncertainty, and which is also above the uncertainty limits from the convex hull and interferometry error, is a velocity filter of 8\% and a radiant filter of 3.5$^{\circ}$. This filter combination rejects the most sporadic meteors without rejecting so many DSX meteors that number statistics are compromised. 

It is important to note that although this filter combination selection method provides a more robust approach, the final mass index values for each solar longitude does not vary significantly when slightly different filters are used. Figure \ref{fig:mass index contour} demonstrates the mass index changes for each filter combination; note the variation is within the uncertainty ranges.  
	
Our initial filter choice of 5$^{\circ}$ and a velocity filter of 10\% in fact produces very similar values to our optimal filters. After determining which filter best suits the DSX meteor shower, we redid the mass index and flux calculations using this new filter of 3.5$^{\circ}$ for the radiant filter and 8\% for the velocity filter. All results discussed in this paper have been made with this new filter.

\subsection{Average DSX Flux}
\label{flux_section}

Accurate measurement of the mass distribution of a meteor shower is a critical first step to estimating the flux of shower meteors. The flux calculation requires the inclusion of low-amplitude meteors, so we use the single station observations from the 38 MHz system instead of the orbital observations from 29 MHz. CMOR's 38 MHz system is more stable than the other systems, as determined by \citet{campbell2019solar}, because the equipment has not undergone any upgrades, and it is not subject to terrestrial noise which affects the 17 MHz system. To estimate an average DSX flux we include single station meteor observations recorded by the 38 MHz system between the years 2002 to 2021.

We calculated the flux of the Daytime Sextantids using the best estimate of the mass index of the DSX from our mixing model approach as discussed in Section \ref{mixing model}. This produces an average mass index of $1.64 \pm 0.06$ for the DSX peak between solar longitudes 180$^{\circ}$-190$^{\circ}$. We follow the process described in \citet{dewsnap2021} to calculate the average flux and activity profile of the DSX shower. 

\begin{figure}
	\centering
	\includegraphics[width=0.9\linewidth]{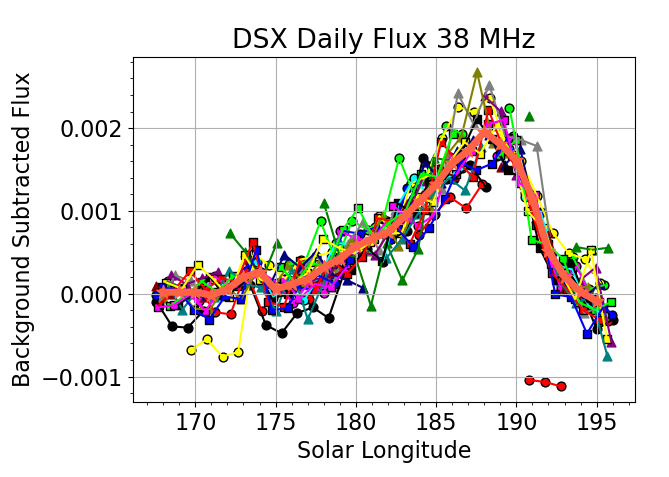}
	
	\includegraphics[width=0.5\linewidth]{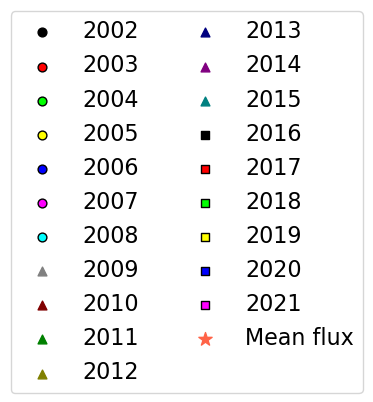}
	\caption{The flux of the Daytime Sextantids using single station echoes as detected by the CMOR 38.15 MHz radar in one degree solar longitude time bins for all years from 2002-2021. The orange solid line corresponds to the average flux profile. Here the background subtracted flux is in units of meteors km$^{-2}$hour$^{-1}$ to a limiting radar magnitude of +6.5.}
	    
	\label{fig:flux38}
\end{figure}

However, we have modified one aspect of the process of \citet{dewsnap2021} to allow us to incorporate the results of the convex hull meteor selection method, described in Section \ref{convex_hull_section}. The radiant of a meteor shower drifts throughout the duration of the shower. To account for this, the flux code used in \citet{dewsnap2021} updates the radiant location but applies a linear radiant drift model, calculated by fitting the wavelet radiant values. The convex hull method calculates the new radiant of the DSX for each solar longitude of the shower. Instead of updating the radiant location with the wavelet-calculated radiant drift, we use the convex hull radiant positions, which have a confidence level of 95\%, for each solar longitude day. The flux profile for the Daytime Sextantids is shown in Figure \ref{fig:flux38}, using data from the 38.15 MHz radar.

Note that  our resulting DSX fluxes have the sporadic background subtracted. The peak flux of 2$\pm$0.05$\times$10$^{-3}$ meteoroids km$^{-2}$hour$^{-1}$ is reached at solar longitude 187$^{\circ}$-188$^{\circ}$ and results in an equivalent ZHR of 20$\pm$5 for our measured mass index. This is to a limiting radar magnitude of +6.5 which, at DSX speeds using the mass-magnitude-velocity relation of \citet{Verniani1973}, is 2$\times$10$^{-7}$ kg. This corresponding to meteoroids with diameter of approximately 0.5~mm. A density of $3000 \ kg/m^3$ was used to estimate the diameter and a spherical shape was assumed.

Adopting a logarithmic model for the flux profile fit to the shower activity as first proposed by \citep{Jenniskens1994}, the DSX activity curve without background subtraction has a B+ slope of 0.06 before the peak and a B- slope of 0.11 after the peak. For comparison, the B+ and B- slopes measured in \citet{moorhead2019} are 0.063 and 0.167 respectively.

\section{Results: Optical}
As the radiant of the Daytime Sextantids is located near the sporadic Helion source, the vast majority of DSX meteors occur in daytime. However, a small number of DSX meteors have been observed just before sunrise and thus may be detected with optical instruments.

As the main motivation of our study is to better understand the broader Phaethon-Geminid Stream Complex through characterization of its least studied member, the Daytime Sextantids, comparison to the associated but much better studied Geminid meteor shower is desirable. Optical DSX meteor observations allow us to directly compare the GEM and DSX optical ablation parameters and make comparisons about the relative physical characteristics of the two meteor showers, such as their bulk strength. This comparison can be used to better define the relationship between the GEM and DSX meteor showers in a way that has not been done before.

The optical DSX meteor data we use were recorded by cameras belonging to the Global Meteor Network (GMN). From GMN observations between 2019-2021, we identified a total of 38 DSX meteors. The initial association with the DSX shower was made following the shower association procedure described in \citet{Vida2021}. To confirm that these meteors are members of the DSX shower, we took the radiant of every meteor in the GMN database during the DSX shower duration and checked if it was located within the radiant space inside the convex hull for the corresponding solar longitude day as determined earlier from CMOR measurements. Of the original 38 meteors identified as DSX by the GMN, two additional shower members were added, and seven were rejected, leaving 33 that were taken to be real DSX members. Of these 33 DSX meteors captured by the GMN, 13 meteors qualify as good candidates for optical analysis as they had complete lightcurves. These 13 fully observed DSX meteors are used for our subsequent analysis of ablation characteristics.

To perform the optical analysis each multi-station meteor observation was manually analyzed using the SkyFit program \citep{vida2018}. SkyFit allowed astrometric and photometric calibrations to be done for all cameras. After each calibration we extracted the apparent light curve and position of the leading edge of the meteor from the video frames. The trajectories were computed using the method of \citet{vida2020}.

\subsection[The KB and KC parameters]{The $k_B$ and $k_c$ parameters}
\label{kb_section}

Meteor light curves depend on the physical properties of each meteoroid as well as meteoroid entry speed and angle. In general, a meteor is optically visible only after it has absorbed enough energy to begin thermal ablation in the atmosphere. The height at which ablation begins depends on factors such as the atmospheric density, speed, entry angle, and the composition of the meteoroid. Typically, higher atmospheric density, higher speed, shallower entry angle, and a weaker/more volatile composition lead to higher begin heights. 

Using this information a value can be defined that characterizes the amount of energy required to begin ablation. Two such parameters have been proposed in the past, namely the $k_c$ and the $k_B$ parameter.

\citet{ceplecha1967} first proposed the $k_B$ parameter as:

\begin{align} \label{kb_equation}
k_B = log\rho_B \ + \ \frac{5}{2} log \ v_\infty \ - \ \frac{1}{2} log \ cos \ z_R \,,
\end{align}

\noindent where $\rho_B$ is the air mass density at the begin height in g~cm$^{-3}$, $v_\infty$ is the initial velocity in km/s, and $z_R$ is the zenith angle. The equation implies that the meteoroid surface temperature has a $v^{2.5}$ dependence. 

The $k_B$ parameter can be used to associate different meteoroid populations with stronger/weaker material properties \citep{Ceplecha1998e}. Unlike the $k_B$ parameter, the $k_c$ criterion \citep{jenniskens2016b} follows a $v^2$ meteoroid surface temperature dependence and is parameterized as:

\begin{align} \label{kc_equation}
k_c = H_b \ + \ (2.86 - 2.00 \ log \ v_\infty)/0.0612 \,,
\end{align}

\noindent where $H_b$ is the beginning height in km and $v_\infty$ is the initial velocity in km/s.

A complication to the interpretation of meteoroid properties using these parameters is that the observed beginning height depends on specific camera system sensitivity. This difference in detection height made by different systems affects the absolute value of calculated $k_c$ and $k_B$ values. However, we may use the relative value of these parameters to compare the bulk strength between meteoroids from different showers, provided the same camera systems observe both.

For this reason, we chose to compare the Geminid (GEM) and DSX showers using meteor observations made by the same type of camera system, the GMN. It has been proposed, based on dynamical arguments, that the GEM and DSX are related, and if this is the case we expect their associated meteoroids to have similar ablation behaviour.

The radiant of the Daytime Sextantid meteor shower is located near the Helion source, so the optical meteors that occur before sunrise enter the atmosphere with shallow entry angles. From the GMN database, we analyzed a random sample of 13 GEM meteors with shallow entry angles within one standard deviation of the mean optical DSX entry angle to maximize similarity in entry geometry between the two comparison datasets.

For the DSX meteors, we find an average $k_B$ value of 7.44 $\pm$ 0.17 and a $k_c$ value of 97.31 $\pm$ 1.81. Table \ref{optical_table_kb_kc} contains detailed information about each of the DSX optical meteors in our sample. For comparison the GEM meteor dataset have an average $k_B$ of 7.45 $\pm$ 0.21 and a $k_c$ of  96.71 $\pm$ 2.66. 

This direct comparison using both the $k_B$ and $k_c$ parameters demonstrates that the GEM and DSX optical meteors have the same relative bulk strength within uncertainties and are consistent with a common origin and/or similar evolution.

\begin{table*}
\begin{tabular}{|c|c|c|c|c|}
    \hline
    \textbf{Time (UTC)} & \textbf{Shower} & \textbf{$k_B$} & \textbf{$k_c$} \\
    \hline
    \textbf{2019-09-28 11:14:28} & DSX & 7.54 & 98.29 \\
    \hline
    \textbf{2019-10-03 11:58:37} & DSX &  7.65  &  93.77 \\
    \hline
    \textbf{2021-09-28 16:50:33} & DSX &  7.15  &  95.84 \\
    \hline
    \textbf{2021-09-29 04:35:04} & DSX &  7.65  &  98.82 \\
    \hline
    \textbf{2021-09-30 12:22:06} & DSX &  7.22  &  97.44 \\
    \hline
    \textbf{2021-10-01 03:48:48} & DSX &  7.56  &  95.71 \\
    \hline
    \textbf{2021-10-01 04:41:16} & DSX &  7.63  &  99.15 \\
    \hline
    \textbf{2021-10-01 12:16:03} & DSX &  7.27  &  98.44 \\
    \hline
    \textbf{2021-10-02 01:41:41} & DSX &  7.29  &  97.91\\
    \hline
    \textbf{2021-10-02 11:43:31} & DSX &  7.47  &  95.42 \\
    \hline
    \textbf{2021-10-02 11:55:25} & DSX &  7.37  &  99.03 \\
    \hline
    \textbf{2021-10-04 05:03:19} & DSX &  7.36  & 95.28 \\
    \hline
    \textbf{2021-10-02 11:59:52} & DSX &  7.37  &  99.87 \\
    \Xhline{5\arrayrulewidth}
    \textbf{2020-12-14 02:54:44} & GEM & 7.33  &  97.36 \\
    \hline
    \textbf{2020-12-13 17:31:10} & GEM & 7.36  &  99.23 \\
    \hline
    \textbf{2020-12-14 02:33:39} & GEM & 7.28  &  98.64 \\
    \hline
    \textbf{2019-12-14 17:06:07} & GEM & 7.40  &  98.29 \\
    \hline
    \textbf{2019-12-14 17:50:47} & GEM & 7.07  &  101.07 \\
    \hline
    \textbf{2019-12-14 18:08:00} & GEM & 7.54  &  94.50 \\
    \hline
    \textbf{2019-12-15 02:20:12} & GEM & 7.60  &  95.28 \\
    \hline
    \textbf{2020-12-12 02:29:53} & GEM & 7.61  &  94.50 \\
    \hline
    \textbf{2020-12-14 02:25:50} & GEM & 7.37  &  97.95 \\
    \hline
    \textbf{2019-12-14 17:41:47} & GEM & 7.33  &  98.00 \\
    \hline
    \textbf{2020-12-11 17:46:30} & GEM & 7.43  &  96.91 \\
    \hline
    \textbf{2020-12-13 17:45:54} & GEM & 7.22  &  94.46 \\
    \hline
    \textbf{2020-12-14 02:58:56} & GEM & 7.49  &  99.59 \\
    \hline
\end{tabular}
\caption{Summary of 13 optical Daytime Sextantid meteors analyzed from GMN data and a control population of 13 optical Geminid meteors used to compare with the DSX. The Geminid meteors chosen are low-entry angle GEM meteors with entry angles within one standard deviation of the mean DSX optical entry angles. More information about these optical meteors can be found in Appendix E.}
\label{optical_table_kb_kc}
\end{table*}

\section{Discussion}
\label{discussion_section}
Our radar analysis of the Daytime Sextantids shows them to have a roughly two week period of significant activity, with peak ZHRs of order 20, peak flux of 2$\pm$0.05$\times$10$^{-3}$ meteoroids km$^{-2}$hour$^{-1}$ at a peak solar longitude of 187-188 and a low mass index of $1.64 \pm 0.06$. The very low mass index indicates a predominance of larger meteoroids at the time of maximum. From our activity profile and mass index we can compute an approximate value for the total mass of the stream and compare to the estimated total mass of the parent, 2005UD.
\subsection{The mass of the DSX stream}
Following the meteor stream mass calculation methodology of \citet{hughes1989}, we calculated the mass of the Daytime Sextantids stream to be $1.11 \times 10^{16}$ grams. We have verified our calculation by reproducing the \citet{hughes1989}'s estimation of the mass in the GEM stream. However, there are several sources of uncertainty in this meteor stream mass calculation. Namely, the upper and lower mass limits of the DSX shower are unknown. Because the GEM and the DSX are related showers, we used the upper and lower mass limits for the GEM that were used in \citet{blaauw2017}'s calculation of the mass of the GEM meteor stream: $10^3$ grams and $10^{-6}$ grams respectively. 

The assumption that the DSX and GEM have similar upper and lower mass limits introduces considerable uncertainty in our calculation. For example, when the upper mass limit is increased to $10^5$ grams the estimated mass of the DSX stream is $4.44 \times 10^{16}$ grams and when it is decreased to $10^2$ grams the estimated mass of the DSX stream is $5.58 \times 10^{15}$ grams. Additionally, an assumption built-in to the \citet{hughes1989}'s stream mass calculation methodology is the estimated unknown area of the meteoroid stream that the Earth does not pass through. \citet{hughes1989} uses an f-factor (filling factor) of 10 to compensate for this unknown. 

Keeping the sources of uncertainty in mind, we can now compare the mass of the Daytime Sextantids stream with that of its parent body, 2005 UD. \citet{jewitt2006} observed 2005 UD at a distance of 1.6 AU from the Sun and  estimated the mass of 2005 UD to be $1 \times 10^{15}$ grams with a sublimation rate of $1.7 \times 10^{-4} kg \ m^{-2} \ s^{-1}$. 

This would suggest that the estimated mass of the DSX stream is larger than the mass of its potential parent body, 2005 UD. Two scenarios can explain this difference in mass. 

We believe the more likely scenario is that 2005 UD is not the immediate parent body of the DSX. Rather, the DSX stream and 2005 UD could be part of a break-up event followed by a decay-chain. Potentially, 2005 UD could be just the largest member of the DSX. This origin would explain the amount of mass of the DSX stream. If this decay-chain model is true, than there would likely be large  objects in the DSX stream, perhaps as large as tens of meters, as opposed to the predicted smaller objects if produced by the asteroid 2005 UD. Additionally, if the DSX stream was created in a decay-chain, this may imply that the GEM stream was also created in a decay-chain, rather than by 3200 Phaethon. This origin would also explain the apparent excess of mass in the GEM stream that cannot be accounted for by the low rate of mass-loss exhibited by Phaethon. 

An alternative scenario is that 2005 UD was much larger and more active in the recent past. If this was the case, then the origin of the DSX could be linked to an earlier epoch of profound mass loss from 2005 UD which might be reflected in its surface characteristics today. 

However, a further possibility, proposed by \citet{babadzhanov1987evolution} and \citet{jakubik2015meteor}, cannot be ruled out. \citet{jakubik2015meteor} simulated the dynamical evolution of various streams released by Phaethon and found that the DSX filament could be reproduced if the age of the DSX stream significantly exceeds 10 millennia. Additionally, another possible indication that the DSX and the GEM are related is that their radiants are symmetrical with respect to the Earth's apex. This scenario could also explain why the DSX stream is larger than 2005 UD.

\subsection{Comparisons to literature values}
The Geminid meteor shower and the Daytime Sextantids are thought to be related as part of the Phaethon Geminid Complex (PGC). We now put our results from the characterization of the DSX into context with other literature values of the Geminid meteor shower to further examine the relationship between the two showers.

Multiple studies have measured the mass index of the Geminids. \citep{reddy2008} measured the Geminids in 2003 and 2005 and found that the mass index at the peak of the GEM was 1.64 +/- 0.05 in 2003 and 1.65 +/- 0.04 in 2005. \citep{Blaauw2011} measured the mass index at the peak of the shower in 2007, 2008, and 2009 to be 1.63 +/- 0.04, 1.58 +/- 0.04, 1.62 +/- 0.04, respectively.

By comparison, for the Daytime Sextantids, we measured a mass index of 1.64 +/- 0.06. The DSX mass index, corrected for sporadic contamination using the mixing model method, is remarkably similar to the mass index of the Geminids. Similar mass indices indicate that the mass distribution of the two meteor streams is comparable, with similar relative amounts of small and large particles in the stream.

As discussed in Section \ref{kb_section}, we found that the Geminids and the Daytime Sextantids have similar $k_c$ and $k_B$ parameters. We calculated a $k_B$ value of $11.46 \pm 0.23$ for the GEM and $11.41 \pm 0.16$ for the DSX and a $k_c$ value of $96.63 \pm 2.98$ for the GEM and $97.59 \pm 1.59$ for the DSX. This similarity indicates that the relative strength of the meteoroids are comparable between the two showers. The DSX meteoroids are expected therefore to be of high density and comparatively refractory compared to other showers. 

The parent body of the Daytime Sextantids meteor shower has been proposed to be the asteroid 2005 UD. This parent relationship is mainly supported by the orbital similarity of the Daytime Sextantids and 2005 UD with a $\Delta t$ of 100 years, as noted by \citep{ohtsuka2006}. For completeness, we will compare the orbital elements calculated in this study with the known orbital elements of 2005 UD. From \citep{ohtsuka2006}, the orbital elements of 2005 UD are: $a = 1.27 AU$, $e = 0.87$, $i = 28.75^{\circ}$, and $\omega = 207.47^{\circ}$. The mean orbital elements of the DSX, as discussed in Section \ref{comparing_convex_wavelet}, are $a = 0.97 \pm 0.11 AU$, $e = 0.85 \pm 0.03$, $i = 24.05 \pm 3.16^{\circ}$, $\omega = 210.82 \pm 2.63^{\circ}$. From these orbital elements, we confirm that 2005 UD and the DSX have similar orbital elements, but are slightly offset. For example, while the orbital elements are similar, only the eccentricity of the DSX overlaps with 2005 UD's value. 
	
Our characterization of the Daytime Sextantids indicates that the DSX and GEM meteor showers share similar physical characteristics and are consistent with a relationship. However, while these findings may support the existence of the Phaethon-Geminid Complex, it is important to note that there are some separate lines of evidence that are in tension with this conclusion. 

Most notably, \citet{kareta2021} found that the near-infrared spectrum of 2005 UD was more red-sloped than Phaethon’s near-infrared spectrum and concluded that the two parent bodies might not be related. Additionally, \citet{lisse2022} found that the spectral reflectance spectrum of asteroids with small perihelion distances may develop a blue color due to thermal alteration and differential sublimation. If true, this would indicate that Phaethon and 2005 UD’s unusual blue reflectance spectrum would not be evidence of a relationship between the two bodies. If Phaethon and 2005 UD are not related, the unusually high strengths of the Geminid and Daytime Sextantid meteoroids may be due to the two streams' low perihelia, rather than a common parentage. The upcoming DESTINY+ mission will shed new light on the nature of Phaethon and 2005 UD’s relationship and the existence or non-existence of the Phaethon-Geminid Complex.

\section{Conclusions}
In this work, we use 9 years of orbital radar observations of the Daytime Sextantid meteor shower and 3 years of optical DSX observations to characterize the stream. 

From radar observations, we determined characteristics of the DSX such as its mass index, flux profile and radiant extent and drift. Using the optical observations we compared the relative strength and composition of the GEM meteors and DSX meteors. As discussed in Section \ref{discussion_section}, we found that the Geminids and Daytime Sextantids share many similar physical characteristics, and we conclude that the two showers are likely related. 

Additionally, we have developed two new meteor analysis techniques for this study. We developed a convex hull approach for defining the set of meteor shower members with a confidence limit of 95\% and used it to define the radiant, duration, and mean orbital elements of the DSX. We also developed a mixing model method that removes the effect of sporadic contamination from the mass index calculation. Using this new method, we calculated the contamination-corrected mass index along with the flux and activity profile of the DSX.

The major conclusions of this study are:
\begin{enumerate}
 \item The radiant and radiant drift of the Daytime Sextantids has been calculated as a function of solar longitude throughout the shower's duration. During the time of the peak of the DSX, from $180$ - $190^\circ$, the radiant drifted $0.70^\circ$ in right ascension and $-0.53^\circ$ in declination per degree solar longitude. Results are shown in Figure \ref{fig:DSX radiant} and compared to literature values in Appendix A in the supplementary materials.
 \item We find that the duration of the Daytime Sextantids shower is between 173$^{\circ}$ - 196$^{\circ}$ in solar longitude. 
 \item The mean orbital elements of the DSX have been calculated as a function of solar longitude. Results are shown in Figure \ref{fig:orbital_elements}, along with the corresponding orbital elements from the literature for comparison purposes.
 \item The mass index of the Daytime Sextantids has been calculated as a function of solar longitude. The mass index at the peak of the DSX is $1.64 \pm 0.06$ while the average mass index surrounding the peak is $1.70 \pm 0.07$, measured over solar longitudes $184^\circ - 190^\circ$. The results are shown in Figure \ref{fig:mass_index_total_range} and discussed in Section \ref{mass_index_section}.
 \item The peak flux of the DSX is found to be 2$\pm$0.05$\times$10$^{-3}$ meteoroids km$^{-2}$hour$^{-1}$ at a peak solar longitude of 187-188. Results are shown in Figure \ref{fig:flux38} and is discussed in Section \ref{flux_section}.
 \item Using the optical DSX observations and some GEM optical observations, we find that the DSX and GEM showers $k_c$ and $k_B$ parameters are identical within uncertainty. This suggests that the relative meteoroid strengths and composition at mm-sizes is similar for each shower. 
 \item We calculated the mass of the DSX stream to be $10^{16}$ grams. Comparing this value to the mass of 2005 UD, estimated to be $10^{15}$ grams, we conclude that it is unlikely that 2005 UD is the parent body of the DSX stream. We proposed two possible scenarios. The more likely scenario is that 2005 UD and the DSX stream are related via a decay chain, meaning that a break-up event of a precursor body created both the DSX stream and 2005 UD. An alternative scenario is that 2005 UD could have been much larger in the past and could have experienced much more extensive mass loss than it is currently exhibiting today. 
 
 We suggest that the origin of the DSX and 2005 UD is most likely formed via a decay chain and that their relationship is one of sibling rather than parent-child association. This suggests that origin of the GEM and Phaethon may be as part of a break-up forming the PGC more broadly. If so, it also suggests that 3200 Phaethon would not be the parent of the GEM stream, but rather simply the largest remnant of the original precursor body.
\end{enumerate}

\section*{Acknowledgements}

Funding for this work was provided in part through NASA co-operative agreement 80NSSC21M0073, by the Natural Sciences and Engineering Research Council of Canada Discovery Grants program (Grants no. RGPIN-2016-04433 \& RGPIN-2018-05659), and the Canada Research Chairs program. The authors thank J. Borovi\v{c}ka for helpful comments related to this study. 


\bibliographystyle{mnras}
\bibliography{bibliography} 

\begin{thebibliography}{}
\makeatletter
\relax
\def\mn@urlcharsother{\let\do\@makeother \do\$\do\&\do\#\do\^\do\_\do\%\do\~}
\def\mn@doi{\begingroup\mn@urlcharsother \@ifnextchar [ {\mn@doi@}
  {\mn@doi@[]}}
\def\mn@doi@[#1]#2{\def\@tempa{#1}\ifx\@tempa\@empty \href
  {http://dx.doi.org/#2} {doi:#2}\else \href {http://dx.doi.org/#2} {#1}\fi
  \endgroup}
\def\mn@eprint#1#2{\mn@eprint@#1:#2::\@nil}
\def\mn@eprint@arXiv#1{\href {http://arxiv.org/abs/#1} {{\tt arXiv:#1}}}
\def\mn@eprint@dblp#1{\href {http://dblp.uni-trier.de/rec/bibtex/#1.xml}
  {dblp:#1}}
\def\mn@eprint@#1:#2:#3:#4\@nil{\def\@tempa {#1}\def\@tempb {#2}\def\@tempc
  {#3}\ifx \@tempc \@empty \let \@tempc \@tempb \let \@tempb \@tempa \fi \ifx
  \@tempb \@empty \def\@tempb {arXiv}\fi \@ifundefined
  {mn@eprint@\@tempb}{\@tempb:\@tempc}{\expandafter \expandafter \csname
  mn@eprint@\@tempb\endcsname \expandafter{\@tempc}}}

\bibitem[\protect\citeauthoryear{Babadzhanov}{Babadzhanov}{1991}]{Babadzhanov1991}
Babadzhanov P.~B.,  1991, Soviet Astronomy, 35, 538

\bibitem[\protect\citeauthoryear{Babadzhanov \& Obrubov}{Babadzhanov \&
  Obrubov}{1987}]{babadzhanov1987evolution}
Babadzhanov P.,  Obrubov I.~V.,  1987, Publications of the Astronomical
  Institute of the Czechoslovak Academy of Sciences, 2, 141

\bibitem[\protect\citeauthoryear{{Blaauw}}{{Blaauw}}{2017}]{blaauw2017}
{Blaauw} R.~C.,  2017, \mn@doi [Planetary and Space Science]
  {10.1016/j.pss.2017.04.007}, \href
  {https://ui.adsabs.harvard.edu/abs/2017P&SS..143...83B} {143, 83}

\bibitem[\protect\citeauthoryear{{Blaauw}, {Campbell-Brown}  \&
  {Weryk}}{{Blaauw} et~al.}{2011}]{Blaauw2011}
{Blaauw} R.~C.,  {Campbell-Brown} M.~D.,   {Weryk} R.~J.,  2011, \mn@doi
  [Monthly Notices of the Royal Astronomical Society]
  {10.1111/j.1365-2966.2010.18038.x}, 412, 2033

\bibitem[\protect\citeauthoryear{Borovi{\v{c}}ka}{Borovi{\v{c}}ka}{2006}]{Borovicka2007}
Borovi{\v{c}}ka J.,  2006, Proceedings of the International Astronomical Union,
  2, 107

\bibitem[\protect\citeauthoryear{{Brown}, {Weryk}, {Wong}  \& {Jones}}{{Brown}
  et~al.}{2008}]{Brown2008}
{Brown} P.,  {Weryk} R.~J.,  {Wong} D.~K.,   {Jones} J.,  2008, \mn@doi
  [Icarus] {10.1016/j.icarus.2007.12.002}, 195, 317

\bibitem[\protect\citeauthoryear{{Brown}, {Wong}, {Weryk}  \&
  {Wiegert}}{{Brown} et~al.}{2010}]{Brown2010}
{Brown} P.,  {Wong} D.~K.,  {Weryk} R.~J.,   {Wiegert} P.,  2010, \mn@doi
  [Icarus] {10.1016/j.icarus.2009.11.015}, 207, 66

\bibitem[\protect\citeauthoryear{{Bruzzone}, {Brown}, {Weryk}  \&
  {Campbell-Brown}}{{Bruzzone} et~al.}{2015}]{Bruzzone2015}
{Bruzzone} J.~S.,  {Brown} P.,  {Weryk} R.~J.,   {Campbell-Brown} M.~D.,  2015,
  \mn@doi [Monthly Notices of the Royal Astronomical Society]
  {10.1093/mnras/stu2200}, 446, 1625

\bibitem[\protect\citeauthoryear{Campbell-Brown}{Campbell-Brown}{2019}]{campbell2019solar}
Campbell-Brown M.,  2019, Monthly Notices of the Royal Astronomical Society,
  485, 4446

\bibitem[\protect\citeauthoryear{Ceplecha}{Ceplecha}{1967}]{ceplecha1967}
Ceplecha Z.,  1967, Smithsonian Contr. Astrophys, 11, 35

\bibitem[\protect\citeauthoryear{Ceplecha, Borovi{\v{c}}ka, Elford, ReVelle,
  Hawkes, Porub{\v{c}}an  \& {\v{S}}imek}{Ceplecha
  et~al.}{1998}]{Ceplecha1998e}
Ceplecha Z.,  Borovi{\v{c}}ka J.,  Elford W.~G.,  ReVelle D.,  Hawkes R.,
  Porub{\v{c}}an V.,   {\v{S}}imek M.,  1998, \mn@doi [Space Science Reviews]
  {10.1023/A:1005069928850}, 84, 327

\bibitem[\protect\citeauthoryear{{De Leon}, Campins, Tsiganis, Morbidelli  \&
  Licandro}{{De Leon} et~al.}{2010}]{DeLeon2010}
{De Leon} J.,  Campins H.,  Tsiganis K.,  Morbidelli a.,   Licandro J.,  2010,
  \mn@doi [Astronomy and Astrophysics] {10.1051/0004-6361/200913609}, 513, A26

\bibitem[\protect\citeauthoryear{Devog{\`{e}}le et~al.,}{Devog{\`{e}}le
  et~al.}{2020}]{Devogele2020}
Devog{\`{e}}le M.,  et~al., 2020, \mn@doi [The Planetary Science Journal]
  {10.3847/psj/ab8e45}, 1, 15

\bibitem[\protect\citeauthoryear{Dewsnap \& Campbell-Brown}{Dewsnap \&
  Campbell-Brown}{2021}]{dewsnap2021}
Dewsnap R.~L.,  Campbell-Brown M.,  2021, Monthly Notices of the Royal
  Astronomical Society, 507, 4521

\bibitem[\protect\citeauthoryear{Galligan}{Galligan}{2000}]{Galligan2000}
Galligan D.~P.,  2000, Doctoral, Canterbury

\bibitem[\protect\citeauthoryear{Galligan}{Galligan}{2003}]{Galligan2003}
Galligan D.~P.,  2003, Monthly Notices of the Royal Astronomical Society, 898,
  893

\bibitem[\protect\citeauthoryear{Gustafson}{Gustafson}{1989}]{Gustafson1989}
Gustafson B. {\AA}.~S.,  1989, Astronomy and Astrophysics, 225, 533

\bibitem[\protect\citeauthoryear{Hocking, Fuller  \& Vandepeer}{Hocking
  et~al.}{2001}]{Hocking2001}
Hocking W.,  Fuller B.,   Vandepeer B.,  2001, Journal of Atmospheric and
  Solar-Terrestrial Physics, 63, 155

\bibitem[\protect\citeauthoryear{Huang, Muinonen, Chen  \& Wang}{Huang
  et~al.}{2021}]{Huang2021}
Huang J.-N.,  Muinonen K.,  Chen T.,   Wang X.-B.,  2021, Planetary and Space
  Science, 195, 105120

\bibitem[\protect\citeauthoryear{Hughes \& McBride}{Hughes \&
  McBride}{1989}]{hughes1989}
Hughes D.~W.,  McBride N.,  1989, Monthly Notices of the Royal Astronomical
  Society, 240, 73

\bibitem[\protect\citeauthoryear{Ishiguro et~al.,}{Ishiguro
  et~al.}{2022}]{ishiguro2022polarimetric}
Ishiguro M.,  et~al., 2022, Monthly Notices of the Royal Astronomical Society,
  509, 4128

\bibitem[\protect\citeauthoryear{Jakub{\'\i}k \& Neslu{\v{s}}an}{Jakub{\'\i}k
  \& Neslu{\v{s}}an}{2015}]{jakubik2015meteor}
Jakub{\'\i}k M.,  Neslu{\v{s}}an L.,  2015, Monthly Notices of the Royal
  Astronomical Society, 453, 1186

\bibitem[\protect\citeauthoryear{Jenniskens}{Jenniskens}{1994}]{Jenniskens1994}
Jenniskens P.,  1994, Astronomy and Astrophysics

\bibitem[\protect\citeauthoryear{Jenniskens et~al.,}{Jenniskens
  et~al.}{2016a}]{jenniskens2016b}
Jenniskens P.,  et~al., 2016a, Icarus, 266, 384

\bibitem[\protect\citeauthoryear{Jenniskens et~al.,}{Jenniskens
  et~al.}{2016b}]{Jenniskens2016a}
Jenniskens P.,  et~al., 2016b, \mn@doi [Icarus] {10.1016/j.icarus.2015.09.013},
  266, 331

\bibitem[\protect\citeauthoryear{Jewitt \& Hsieh}{Jewitt \&
  Hsieh}{2006a}]{Jewitt2006a}
Jewitt D.,  Hsieh H.,  2006a, \mn@doi [The Astronomical Journal]
  {10.1086/507483}, 132, 1624

\bibitem[\protect\citeauthoryear{Jewitt \& Hsieh}{Jewitt \&
  Hsieh}{2006b}]{jewitt2006}
Jewitt D.,  Hsieh H.,  2006b, The Astronomical Journal, 132, 1624

\bibitem[\protect\citeauthoryear{Jewitt \& Li}{Jewitt \& Li}{2010}]{Jewitt2010}
Jewitt D.,  Li J.,  2010, The Astronomical Journal, 140, 1519

\bibitem[\protect\citeauthoryear{Jewitt, Hsieh, Agarwal  et~al.}{Jewitt
  et~al.}{2015}]{jewitt2015}
Jewitt D.,  Hsieh H.,  Agarwal J.,   et~al., 2015, Asteroids IV, pp 221--241

\bibitem[\protect\citeauthoryear{Jones \& Jones}{Jones \&
  Jones}{2006}]{Jones2006}
Jones J.,  Jones W.,  2006, \mn@doi [Monthly Notices of the Royal Astronomical
  Society] {10.1111/j.1365-2966.2006.10025.x}, 367, 1050

\bibitem[\protect\citeauthoryear{{Jones}, {Brown}, {Ellis}, {Webster},
  {Campbell-Brown}, {Krzemenski}  \& {Weryk}}{{Jones} et~al.}{2005}]{Jones2005}
{Jones} J.,  {Brown} P.,  {Ellis} K.~J.,  {Webster} A.~R.,  {Campbell-Brown}
  M.,  {Krzemenski} Z.,   {Weryk} R.~J.,  2005, \mn@doi [Planetary and Space
  Science] {10.1016/j.pss.2004.11.002}, 53, 413

\bibitem[\protect\citeauthoryear{{Jopek}, {Rudawska}  \&
  {Pretka-Ziomek}}{{Jopek} et~al.}{2006}]{Jopek2006}
{Jopek} T.~J.,  {Rudawska} R.,   {Pretka-Ziomek} H.,  2006, \mn@doi [\mnras]
  {10.1111/j.1365-2966.2006.10770.x}, \href
  {https://ui.adsabs.harvard.edu/abs/2006MNRAS.371.1367J} {371, 1367}

\bibitem[\protect\citeauthoryear{Kareta, Reddy, Pearson, Sanchez  \&
  Harris}{Kareta et~al.}{2021}]{kareta2021}
Kareta T.,  Reddy V.,  Pearson N.,  Sanchez J.~A.,   Harris W.~M.,  2021, The
  Planetary Science Journal, 2, 190

\bibitem[\protect\citeauthoryear{Kashcheyev \& Lebedinets}{Kashcheyev \&
  Lebedinets}{1967}]{kashcheyev1967}
Kashcheyev B.,  Lebedinets V.,  1967, Smithsonian contributions to
  astrophysics, 11, 183

\bibitem[\protect\citeauthoryear{Kinoshita et~al.,}{Kinoshita
  et~al.}{2007}]{Kinoshita2007}
Kinoshita D.,  et~al., 2007, \mn@doi [Astronomy and Astrophysics]
  {10.1051/0004-6361:20066276}, 466, 1153

\bibitem[\protect\citeauthoryear{Licandro, Campins, Moth{\'{e}}-Diniz,
  Pinilla-Alonso  \& {De Le{\'{o}}n}}{Licandro et~al.}{2007}]{Licandro2007}
Licandro J.,  Campins H.,  Moth{\'{e}}-Diniz T.,  Pinilla-Alonso N.,   {De
  Le{\'{o}}n} J.,  2007, Astronomy \& Astrophysics, 461, 751

\bibitem[\protect\citeauthoryear{Lisse \& Steckloff}{Lisse \&
  Steckloff}{2022}]{lisse2022}
Lisse C.,  Steckloff J.,  2022, Icarus, p. 114995

\bibitem[\protect\citeauthoryear{MacLennan, Toliou  \& Granvik}{MacLennan
  et~al.}{2021}]{MacLennan2020}
MacLennan E.,  Toliou A.,   Granvik M.,  2021, Icarus, 366, 114535

\bibitem[\protect\citeauthoryear{Mazur, Pokorný, Brown, Weryk, Vida, Schult,
  Stober  \& Agrawal}{Mazur et~al.}{2020}]{Mazur2020}
Mazur M.,  Pokorný P.,  Brown P.,  Weryk R.~J.,  Vida D.,  Schult C.,  Stober
  G.,   Agrawal A.,  2020, \mn@doi [Radio Science] {10.1029/2019RS006987}, n/a,
  e2019RS006987

\bibitem[\protect\citeauthoryear{Moorhead, Egal, Brown, Moser  \&
  Cooke}{Moorhead et~al.}{2019}]{moorhead2019}
Moorhead A.~V.,  Egal A.,  Brown P.~G.,  Moser D.~E.,   Cooke W.~J.,  2019,
  Journal of Spacecraft and Rockets, 56, 1531

\bibitem[\protect\citeauthoryear{Neslu{\v{s}}an \&
  Hajdukov{\'{a}}}{Neslu{\v{s}}an \& Hajdukov{\'{a}}}{2017}]{Neslusan2017}
Neslu{\v{s}}an L.,  Hajdukov{\'{a}} M.,  2017, \mn@doi [Astronomy and
  Astrophysics] {10.1051/0004-6361/201629659}, 598, A40

\bibitem[\protect\citeauthoryear{Nilsson}{Nilsson}{1964}]{Nilsson1964}
Nilsson C.,  1964, \mn@doi [Australian Journal of Physics] {10.1071/PH640158},
  17, 158

\bibitem[\protect\citeauthoryear{Ohtsuka, Sekiguchi, Kinoshita, Watanabe, Ito,
  Arakida  \& Kasuga}{Ohtsuka et~al.}{2006}]{ohtsuka2006}
Ohtsuka K.,  Sekiguchi T.,  Kinoshita D.,  Watanabe J.-I.,  Ito T.,  Arakida
  H.,   Kasuga T.,  2006, Astronomy \& Astrophysics, 450, L25

\bibitem[\protect\citeauthoryear{{Pokorn\'y, P.} \& {Brown, P. G.}}{{Pokorn\'y,
  P.} \& {Brown, P. G.}}{2016}]{Pokorny2016}
{Pokorn\'y, P.} {Brown, P. G.} 2016, \mn@doi [Astronomy \& Astrophysics]
  {10.1051/0004-6361/201628134}, 592, A150

\bibitem[\protect\citeauthoryear{Pokorn{\'{y}}, Janches, Brown  \&
  Hormaechea}{Pokorn{\'{y}} et~al.}{2017}]{Pokorny2017}
Pokorn{\'{y}} P.,  Janches D.,  Brown P.~G.,   Hormaechea J.,  2017, \mn@doi
  [Icarus] {10.1016/j.icarus.2017.02.025}, 290, 162

\bibitem[\protect\citeauthoryear{Reddy, Kumar  \& Yellaiah}{Reddy
  et~al.}{2008}]{reddy2008}
Reddy K.~C.,  Kumar D. V.~P.,   Yellaiah G.,  2008, Planetary and Space
  Science, 56, 1014

\bibitem[\protect\citeauthoryear{Rendtel}{Rendtel}{2014}]{Rendtel2014}
Rendtel J.,  2014, {Meteor Shower Workbook 2014}.
International Meteor Organization

\bibitem[\protect\citeauthoryear{{Rudawska}, {Matlovi{\v{c}}}, {T{\'o}th}  \&
  {Korno{\v{s}}}}{{Rudawska} et~al.}{2015}]{Rudawaska2015}
{Rudawska} R.,  {Matlovi{\v{c}}} P.,  {T{\'o}th} J.,   {Korno{\v{s}}} L.,
  2015, \mn@doi [\planss] {10.1016/j.pss.2015.07.011}, \href
  {https://ui.adsabs.harvard.edu/abs/2015P&SS..118...38R} {118, 38}

\bibitem[\protect\citeauthoryear{Ryabova}{Ryabova}{2017}]{Ryabova2017}
Ryabova G.,  2017, \mn@doi [Planetary and Space Science]
  {10.1016/j.pss.2017.02.005}, 143, 125

\bibitem[\protect\citeauthoryear{Ryabova, Avdyushev  \& Williams}{Ryabova
  et~al.}{2019}]{Ryabova2019}
Ryabova G.~O.,  Avdyushev V.~A.,   Williams I.~P.,  2019, \mn@doi [Monthly
  Notices of the Royal Astronomical Society] {10.1093/mnras/stz658}, 485, 3378

\bibitem[\protect\citeauthoryear{{Sekanina}}{{Sekanina}}{1976}]{Sekanina1976}
{Sekanina} Z.,  1976, \mn@doi [\icarus] {10.1016/0019-1035(76)90009-9}, \href
  {https://ui.adsabs.harvard.edu/abs/1976Icar...27..265S} {27, 265}

\bibitem[\protect\citeauthoryear{Sonotaco}{Sonotaco}{2009}]{Sonotaco2009}
Sonotaco 2009, Wgn, Jimo, 37, 55

\bibitem[\protect\citeauthoryear{Spurn{\'{y}}}{Spurn{\'{y}}}{1993}]{Spurny1993}
Spurn{\'{y}} P.,  1993, in Meteoroids and their Parent Bodies. pp 193--196

\bibitem[\protect\citeauthoryear{Tabeshian, Wiegert, Ye, Hui, Gao  \&
  Tan}{Tabeshian et~al.}{2019}]{Tabeshian2019}
Tabeshian M.,  Wiegert P.,  Ye Q.,  Hui M.-T.,  Gao X.,   Tan H.,  2019,
  \mn@doi [The Astronomical Journal] {10.3847/1538-3881/ab245d}, 158, 30

\bibitem[\protect\citeauthoryear{Verniani}{Verniani}{1973}]{Verniani1973}
Verniani F.,  1973, Journal of Geophysical Research, 78, 8429

\bibitem[\protect\citeauthoryear{Vida, Mazur, {\v{S}}egon, Zubovi{\v{c}},
  Kuki{\v{c}}, Parag  \& Macan}{Vida et~al.}{2018}]{vida2018}
Vida D.,  Mazur M.,  {\v{S}}egon D.,  Zubovi{\v{c}} D.,  Kuki{\v{c}} P.,  Parag
  F.,   Macan A.,  2018, WGN, Journal of the International Meteor Organization,
  46, 71

\bibitem[\protect\citeauthoryear{Vida, Gural, Brown, Campbell-Brown  \&
  Wiegert}{Vida et~al.}{2020b}]{vida2020}
Vida D.,  Gural P.~S.,  Brown P.~G.,  Campbell-Brown M.,   Wiegert P.,  2020b,
  Monthly Notices of the Royal Astronomical Society, 491, 2688

\bibitem[\protect\citeauthoryear{Vida, Gural, Brown, Campbell-Brown  \&
  Wiegert}{Vida et~al.}{2020a}]{vida2020a}
Vida D.,  Gural P.~S.,  Brown P.~G.,  Campbell-Brown M.,   Wiegert P.,  2020a,
  \mn@doi [Monthly Notices of the Royal Astronomical Society]
  {10.1093/mnras/stz3160}, 491, 2688

\bibitem[\protect\citeauthoryear{Vida et~al.,}{Vida et~al.}{2021}]{Vida2021}
Vida D.,  et~al., 2021, Monthly Notices of the Royal Astronomical Society, 506,
  5046

\bibitem[\protect\citeauthoryear{Weiss}{Weiss}{1960}]{Weiss1960}
Weiss A.,  1960, \mn@doi [Monthly Notices of the Royal Astronomical Society]
  {10.1093/mnras/120.5.387}, 120, 387

\bibitem[\protect\citeauthoryear{Williams \& Wu}{Williams \&
  Wu}{1993}]{Williams1993}
Williams I.,  Wu Z.,  1993, Monthly Notices of the Royal Astronomical Society,
  262, 231

\bibitem[\protect\citeauthoryear{Williams, Jopek, Rudawska, Toth  \&
  Kornos}{Williams et~al.}{2019}]{Williams2019}
Williams I.~P.,  Jopek T.~J.,  Rudawska R.,  Toth J.,   Kornos L.,  2019, in
  Asher D.~J.,  Ryabova G.~O.,   Campbell-Brown M.~D.,  eds, Cambridge
  Planetary Science, Meteoroids: Sources of Meteors on Earth and Beyond.
Cambridge University Press, Cambridge, pp 210--234, \mn@doi{DOI:
  10.1017/9781108606462.016}, \url
  {https://www.cambridge.org/core/books/meteoroids/minor-meteor-showers-and-the-sporadic-background/81F0683515A682F6712191DBF2154C79}

\bibitem[\protect\citeauthoryear{Ye, Wiegert  \& Hui}{Ye et~al.}{2018}]{Ye2018}
Ye Q.,  Wiegert P.,   Hui M.-T.,  2018, \mn@doi [The Astrophysical Journal]
  {10.3847/2041-8213/aada46}, 864, L9

\bibitem[\protect\citeauthoryear{{Younger}, {Reid}, {Vincent}, {Holdsworth}  \&
  {Murphy}}{{Younger} et~al.}{2009}]{Younger2009}
{Younger} J.~P.,  {Reid} I.~M.,  {Vincent} R.~A.,  {Holdsworth} D.~A.,
  {Murphy} D.~J.,  2009, \mn@doi [\mnras] {10.1111/j.1365-2966.2009.15142.x},
  \href {https://ui.adsabs.harvard.edu/abs/2009MNRAS.398..350Y} {398, 350}

\makeatother
\end{thebibliography}

\clearpage

\bsp	
\label{lastpage}
\end{document}


\label{firstpage}
\pagerange{\pageref{firstpage}--\pageref{lastpage}}
\maketitle

\appendix
\section{Literature Comparison}\label{app:literature_appendix}
This appendix section contains two tables (Table \ref{literature_table} and Table \ref{literature_table_2}) that compare the radiant and orbital elements for the DSX in the literature with the values calculated in this study.
\begin{table*}
\begin{adjustbox}{angle=90}
\begin{tabular}{ |c|c|c|c|c|c|c|c|c|c| }
 \hline
  & \textbf{$\lambda_{max}$ (deg)} & \textbf{$\alpha_R$ (deg)} & \textbf{$\delta_R$ (deg)} & \textbf{$V_g$ (km/s)} & \textbf{$a$ (AU)} & \textbf{$e$} & \textbf{$i$ (deg)} & \textbf{$\omega$ (deg)} & \textbf{$\Omega$ (deg)} \\ 
 \hline
 \textbf{Weiss (1960)} & 187 & 155 $\pm$ 8 & 0 $\pm$ 10 & - & - & - & - & - & - \\ 
 \hline
 \textbf{Nilsson (1964)} & 183.6 & 151.7 $\pm$ 0.9 & -0.1 $\pm$ 1.5 & 32.2 $\pm$ & 0.89 $\pm$ 0.03 & 0.87 $\pm$ 0.01 & 21.8 $\pm$ 2.3 & 213.2 $\pm$ 2.1 & 3.6\\ 
 \hline
 \textbf{Sekanina (1976)} & 195 & 156.5 $\pm$ 0.9 & -8.3 $\pm$ 0.8 & 29.7 & 0.936 & 0.816 $\pm$ 0.011 & 31.1 $\pm$ 1.0 & 212.3 $\pm$ 1.0 & 15.1 $\pm$ 0.1 \\
 \hline
 \textbf{Jopek et al. (1999)} & 183 & 152 & 3 & 32 & - & 0.88 & 19 & 211 & 3 \\
 \hline
 \textbf{Galligan \& Baggaley (2002)} & 186.1 & 154.5 $\pm$ 2.7 & -1.5 $\pm$ 0.5 & 31.2 $\pm$ 1.6 & 1.04 $\pm$ 0.023 & 0.855 $\pm$ 0.023 & 23.1 $\pm$ 3.9 & 212.5 $\pm$ 3.0 & 6.1 $\pm$ 0.0 \\
 \hline
 \textbf{Brown et al. (2008)} & 187 & 154.6 & -1.4 & 31.84 & - & - & - & - & - \\
 \hline
 \textbf{Younger et al. (2009)} & 188.1 & 155.7 & -3.9 & 32.7 & 1.09 & 0.858 & 23.9 & 326.1 & 8.6 \\
 \hline 
 \textbf{SonotaCo (2009)} & 189.2 & 156.3 & -2.9 & 31.2 & - & - & - & - & - \\
 \hline 
 \textbf{Brown et al. (2010)} & 186 & 154.3 & -1 & 31.3 & 1.07 & 0.858 & 22.0 & 212.99 & 6.0 \\
 \hline
 \textbf{Rudawaska et al. (2015)} & 187.9 & 155.0 $\pm$ 1.5 & -1.4 $\pm$ 1.5 & 31.7 $\pm$ 1.2 & 1.0 & 0.9 & 23.4 & 211.4 & 7.9 \\
 \hline
  \textbf{Jenniskens et al} & 188 & 156.6  & -2.4  & 32.9  & 1.14 & 0.874 & 24.3 & 214.3 & 6.4 \\
 \hline
 \textbf{Pokorn{\'y} et al. (2017)} & 187 & 155.4 & -1.6 & 31.4 & 1.08 $\pm$ 0.08 & 0.858 $\pm$ 0.022 & 22.2 & 213.6 & 7 \\
 \hline
\makecell{\textbf{Bruzzone et al. (2020) } \\ \textbf{(CAMS)}} & 191 & 157.59 & -3.64 & 32.8 & 1.11 $\pm$ 0.02 & 0.878 $\pm$ 0.003 & 27 $\pm 1$ &  211.7 $\pm$ 0.9 & 11 \\
 \hline 
\makecell{\textbf{Bruzzone et al. (2020)} \\ \textbf{(SAAMER-OS)}} & 187 & 153.93 & -1.65 & 32.1 & 1.055 $\pm$ 0.009 & 0.872 $\pm$ 0.002 & 25.8 $\pm$ 0.5 & 210.8 $\pm$ 0.4 & 7 \\
 \hline 
 \textbf{Kipreos et al. (2022)} & 186 & 153.06 & -0.61 & 30.91 $\pm$ 2.33 & 0.98 $\pm$ 0.13 & 0.85 $\pm$ 0.03 & 22.57 $\pm$ 0.06 & 211.14 $\pm$ 0.05 & 6.36 $\pm$ 0.01 \\
 \hline 

\end{tabular}
\end{adjustbox}

\caption{Measurements of the Daytime Sextantids meteor shower made by previous groups, along with the calculations made in this study. The DSX measurements including in this table are the solar longitude at the peak of the shower ($\lambda_{max}$), right ascension ($\alpha_R$), declination ($\delta_R$), geocentric velocity ($V_g$), semi-major axis ($a$), eccentricity ($e$), inclination ($i$), argument of perihelion ($\omega$), and longitude of the ascending node ($\Omega$).}

\label{literature_table}
\end{table*}

\begin{table*}
\begin{adjustbox}{angle=90}

\begin{tabular}{ |c|c|c|c|c|c|c|c|c|c| } 
 \hline
  & \textbf{Year(s) of observation} & \textbf{Number of observations} & \textbf{Type of observations} & \textbf{Location} \\ 
 \hline
 \textbf{Weiss (1960)} & 1956 - 1956 & - & radar & - \\ 
 \hline
 \textbf{Nilsson (1964)} & 1961 & 9 & radar & Adelaide, Australia \\ 
 \hline
 \textbf{Sekanina (1976)} & 1968 - 1969 & 10 & radar & Illinois, USA \\
 \hline
 \textbf{Jopek et al. (1999)} & 1960 - 1961 and 1968 - 1969 & 14 & radar & Adelaide, Australia \\
 \hline
 \textbf{Galligan \& Baggaley (2002)} & 1995 - 1999 & 410 & radar & Adelaide, Austrailia \\
 \hline
 \textbf{Brown et al. (2008)} & 2001 - 2006 & - & radar & Tavistock, Ontario\\
 \hline
 \textbf{Younger et al. (2009)} & 2006 - 2007 & - & radar & Davis Station Antarctica and Darwin, Australia\\
 \hline 
 \textbf{SonotaCo (2009)} & 2007 - 2009 & 4 & optical & Japan (SonotaCo Network)\\
 \hline 
 \textbf{Brown et al. (2010)} & 2001 - 2008 & 1292 & radar & Tavistock, Ontario\\
 \hline
 \textbf{Rudawaska et al. (2015)} & 2001 - 2014 & 14 & optical & Europe (EDMOND database)\\
 \hline
  \textbf{Jenniskens et al. (2016)} & 2010 - 2013 & 14 & optical & Global (CAMS)\\
 \hline
 
 \textbf{Pokorn{\'y} et al. (2017)} & 2012 - 2015 & - & radar & Rio Grande, Argentina \\
 \hline
 \makecell{\textbf{Bruzzone et al. (2020)} \\ \textbf{(CAMS)}} & 2011 - 2017 & 25 & optical &  multiple\\
 \hline 
  \makecell{\textbf{Bruzzone et al. (2020)} \\\textbf{(SAAMER-OS)}} & 2012 - 2019 & 2255 & radar & Rio Grande, Argentina\\
 \hline 
 \textbf{Kipreos et al. (2022)} & 2002 - 2020 & 19,007 & radar & Tavistock, Ontario\\
 \hline 

\end{tabular}
\end{adjustbox}

\caption{Measurements of the Daytime Sextantids shower made by other research groups. The information contained in this table includes the years that the observations were taken, the number of total observations, the type of observation, and the location of the radar or camera system. The number of observations for our study is the number of total meteors located in the convex hull for the duration of the Daytime Sextantids meteor shower.}

\label{literature_table_2}
\end{table*}

\section{Convex Hull Results}\label{app:Convex Hull Results}
\begin{figure*}
\centering
\begin{tabular}{@{}c@{}}
    \includegraphics[width=0.45\textwidth]{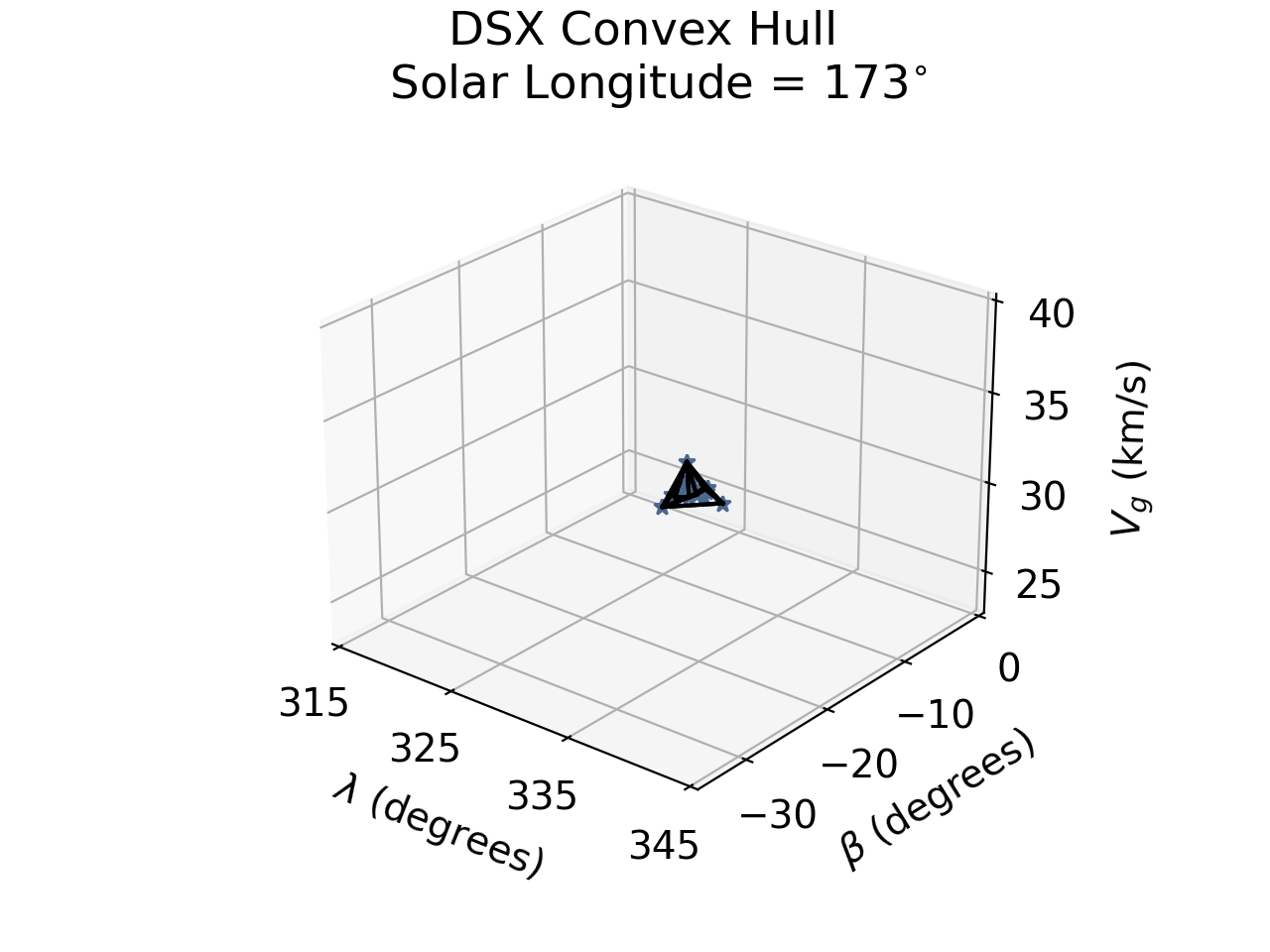}
\end{tabular}
\begin{tabular}{@{}c@{}}
    \includegraphics[width=0.45\textwidth]{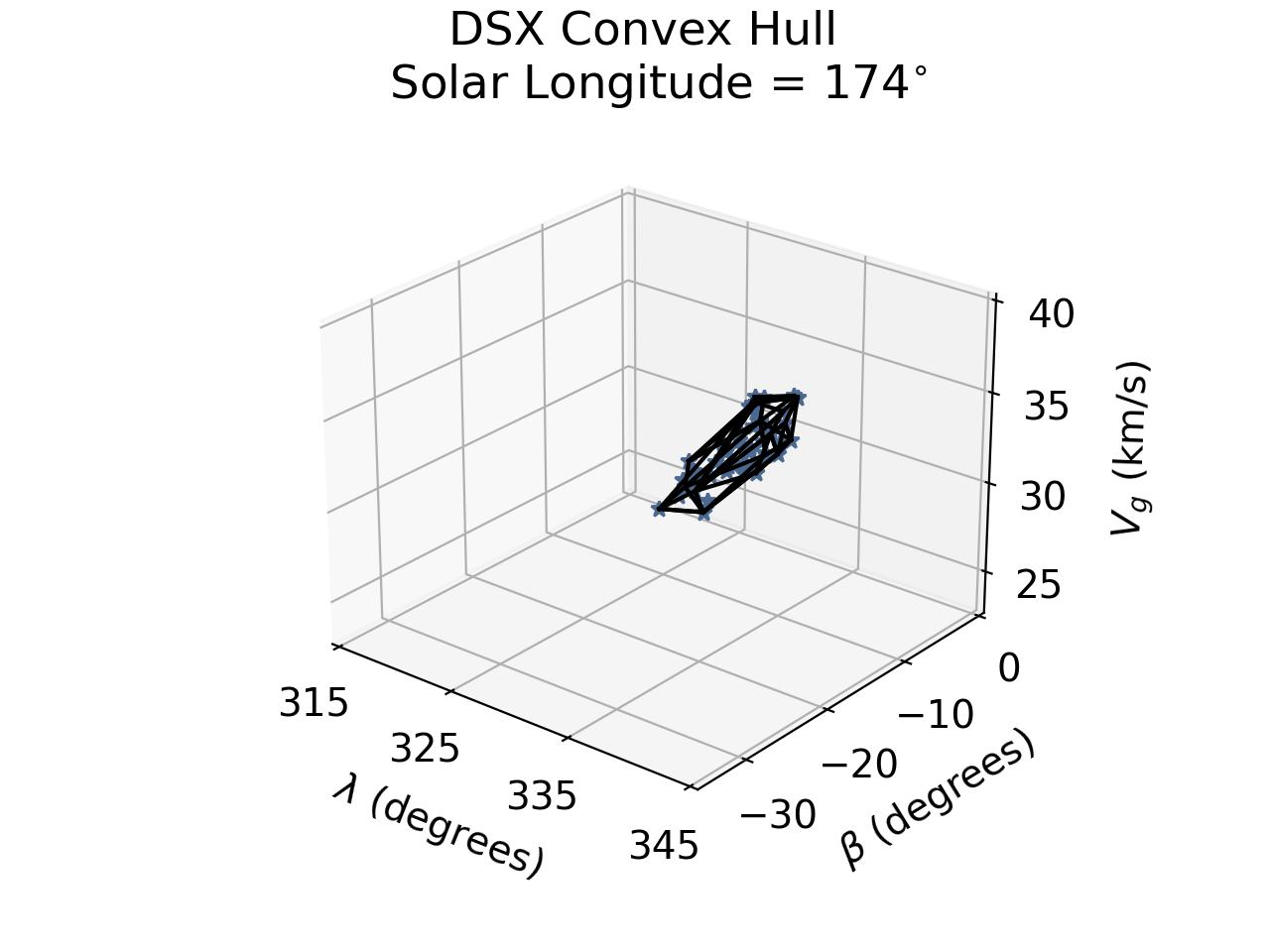}
\end{tabular}
\begin{tabular}{@{}c@{}}
    \includegraphics[width=0.45\textwidth]{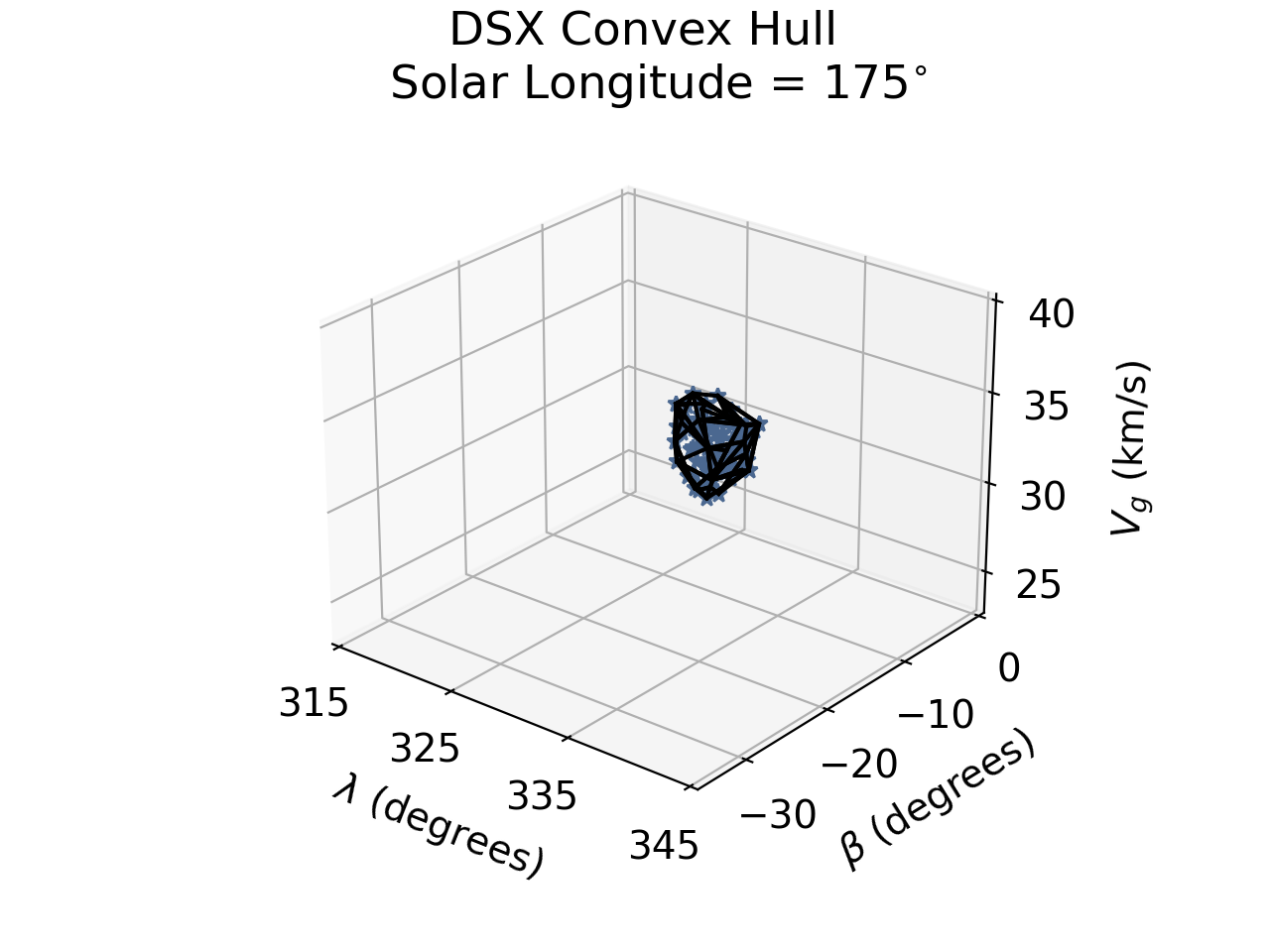}
\end{tabular}
\begin{tabular}{@{}c@{}}
    \includegraphics[width=0.45\textwidth]{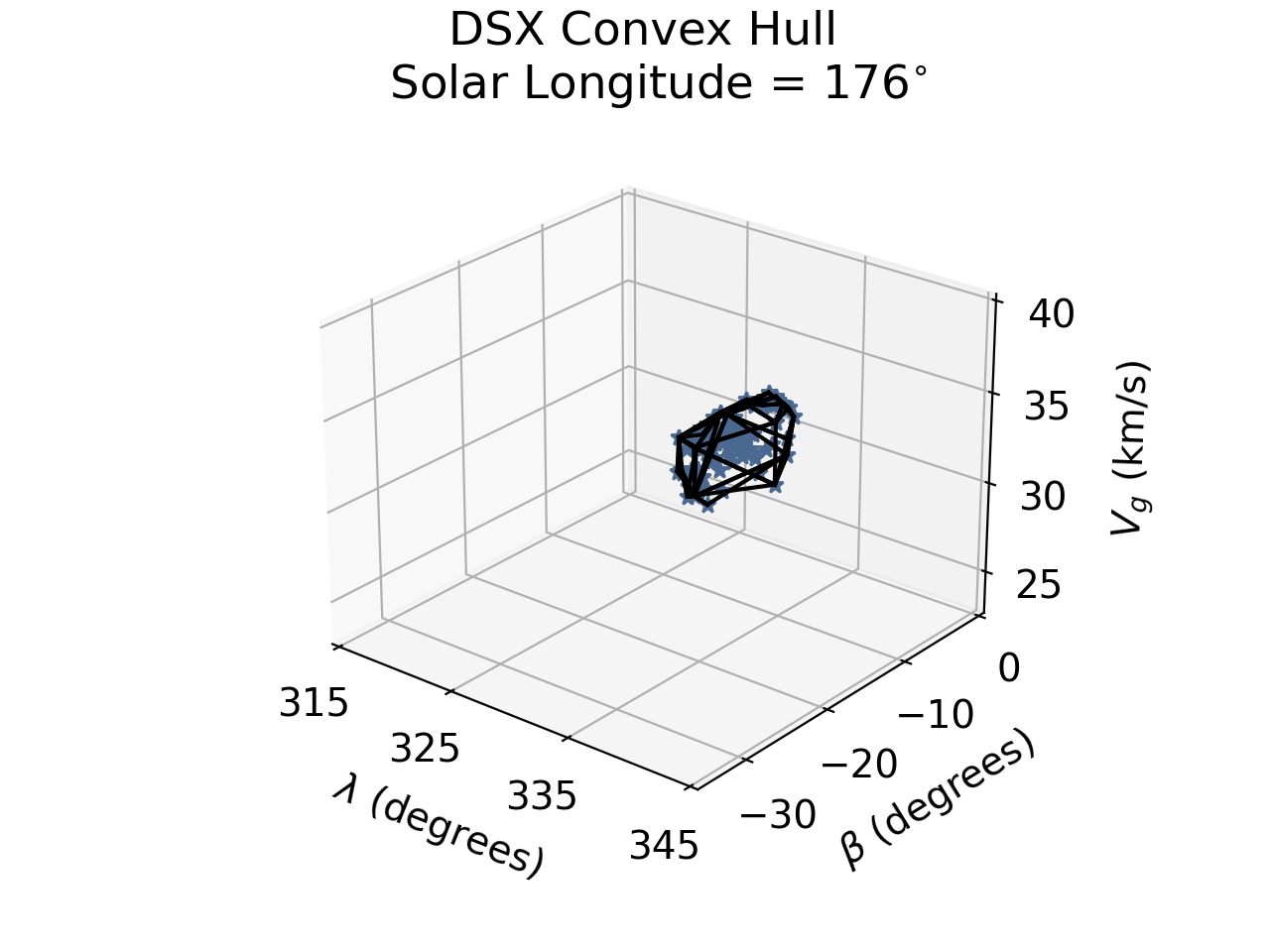}
\end{tabular}
\begin{tabular}{@{}c@{}}
    \includegraphics[width=0.45\textwidth]{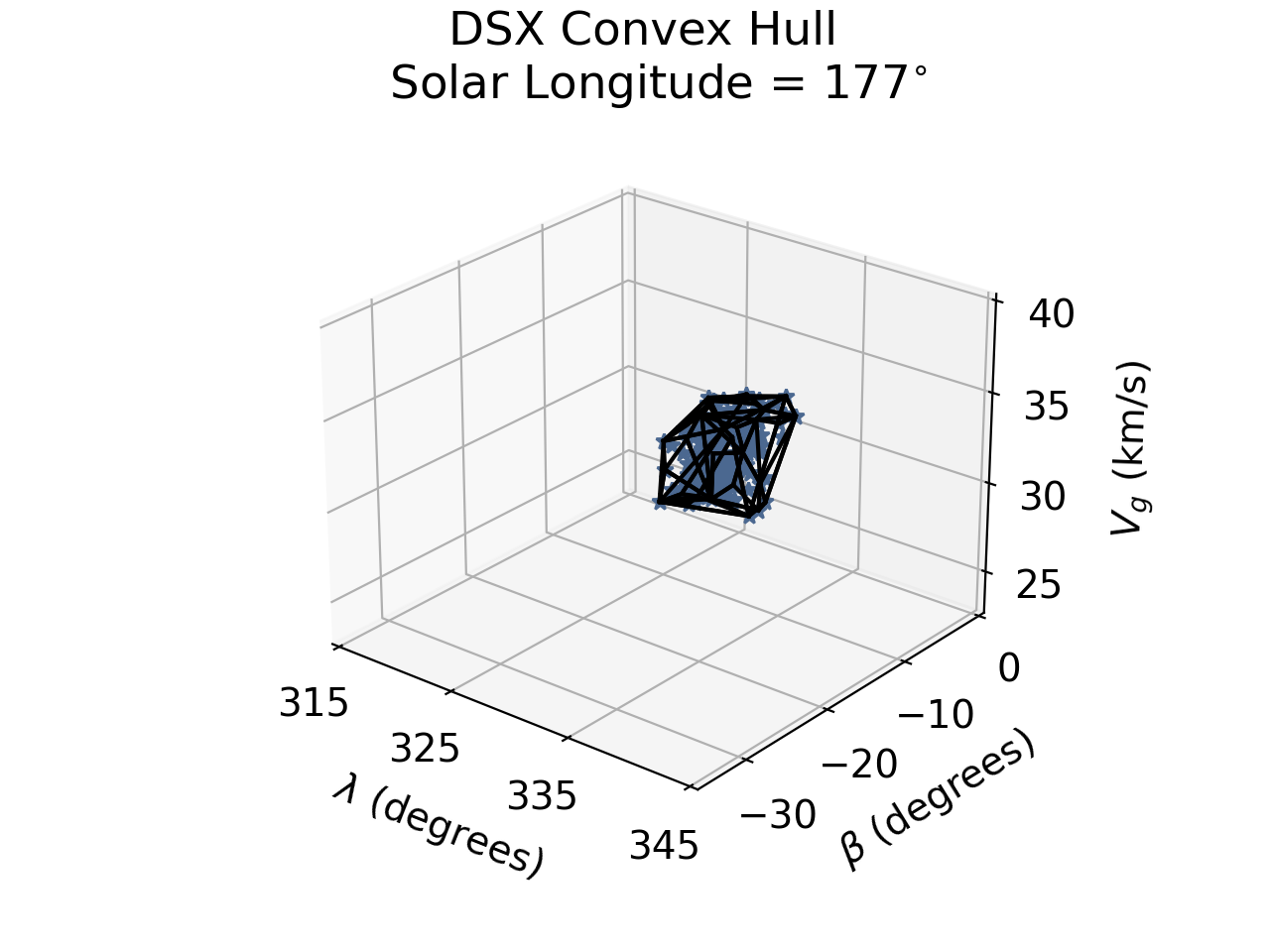}
\end{tabular}
\begin{tabular}{@{}c@{}}
    \includegraphics[width=0.45\textwidth]{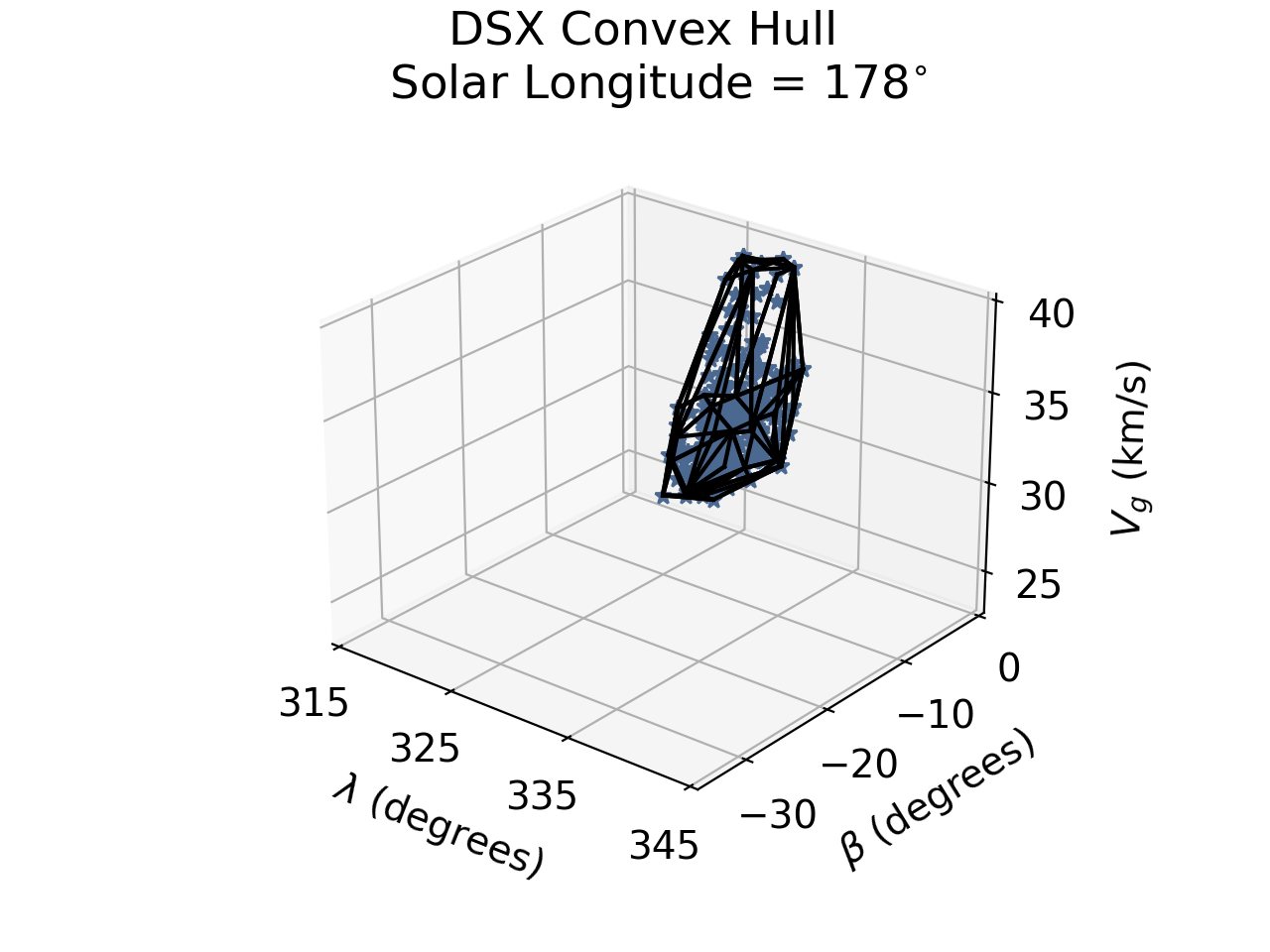}
\end{tabular}

\caption[short]{Convex hull results from solar longitude 173$^{\circ}$ to 178$^{\circ}$. The convex hull, described in Section 3.4.1 has been defined in such a way that any meteor with a radiant located within the hull is determined to be a member of the DSX shower, with a 95\% confidence level. These figures show the convex hull and all meteors that are located within it for a given solar longitude.}
\label{ch_1}
\end{figure*}

\begin{figure*}
\centering
\begin{tabular}{@{}c@{}}
    \includegraphics[width=0.45\textwidth]{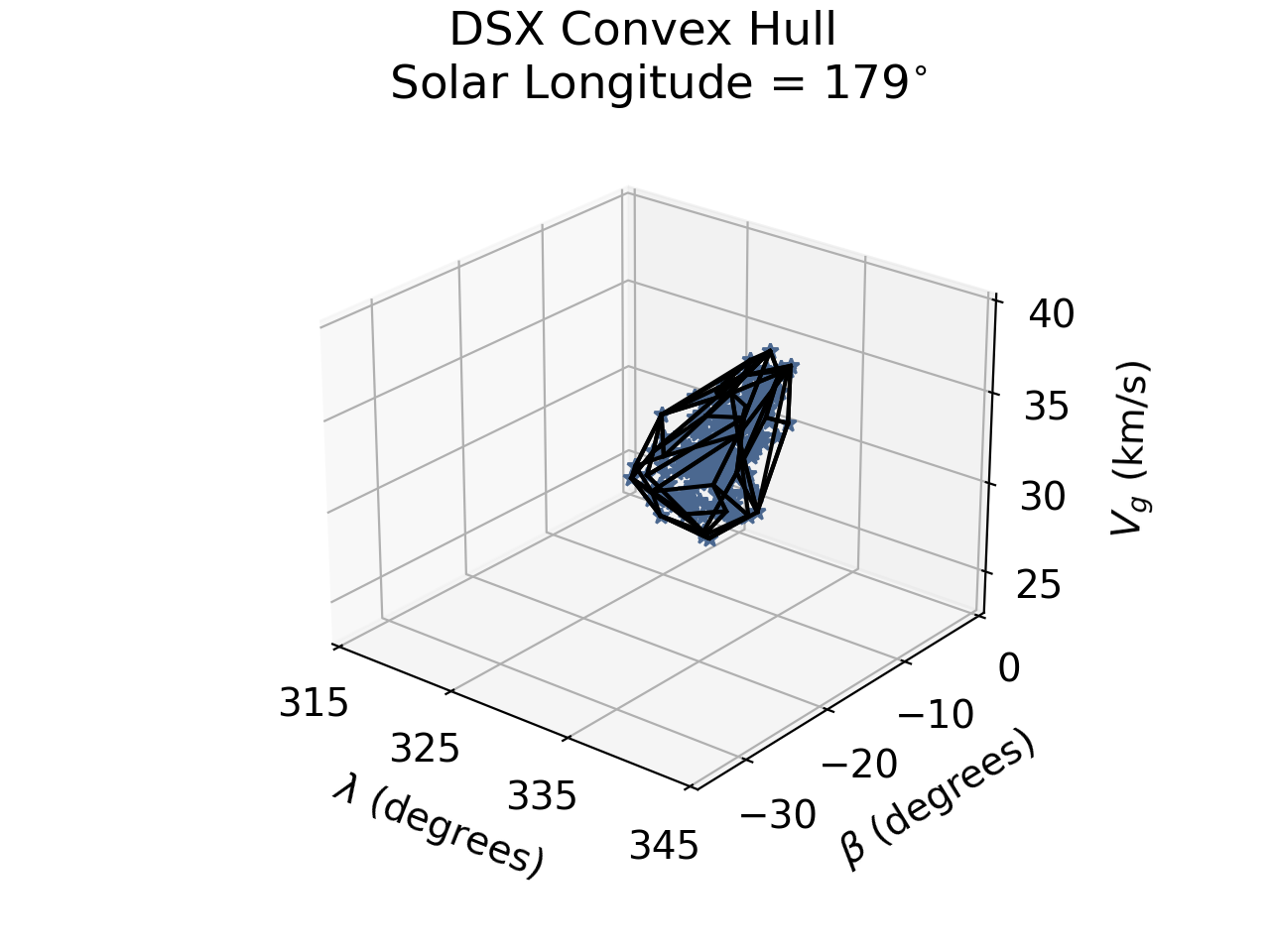}
\end{tabular}
\begin{tabular}{@{}c@{}}
    \includegraphics[width=0.45\textwidth]{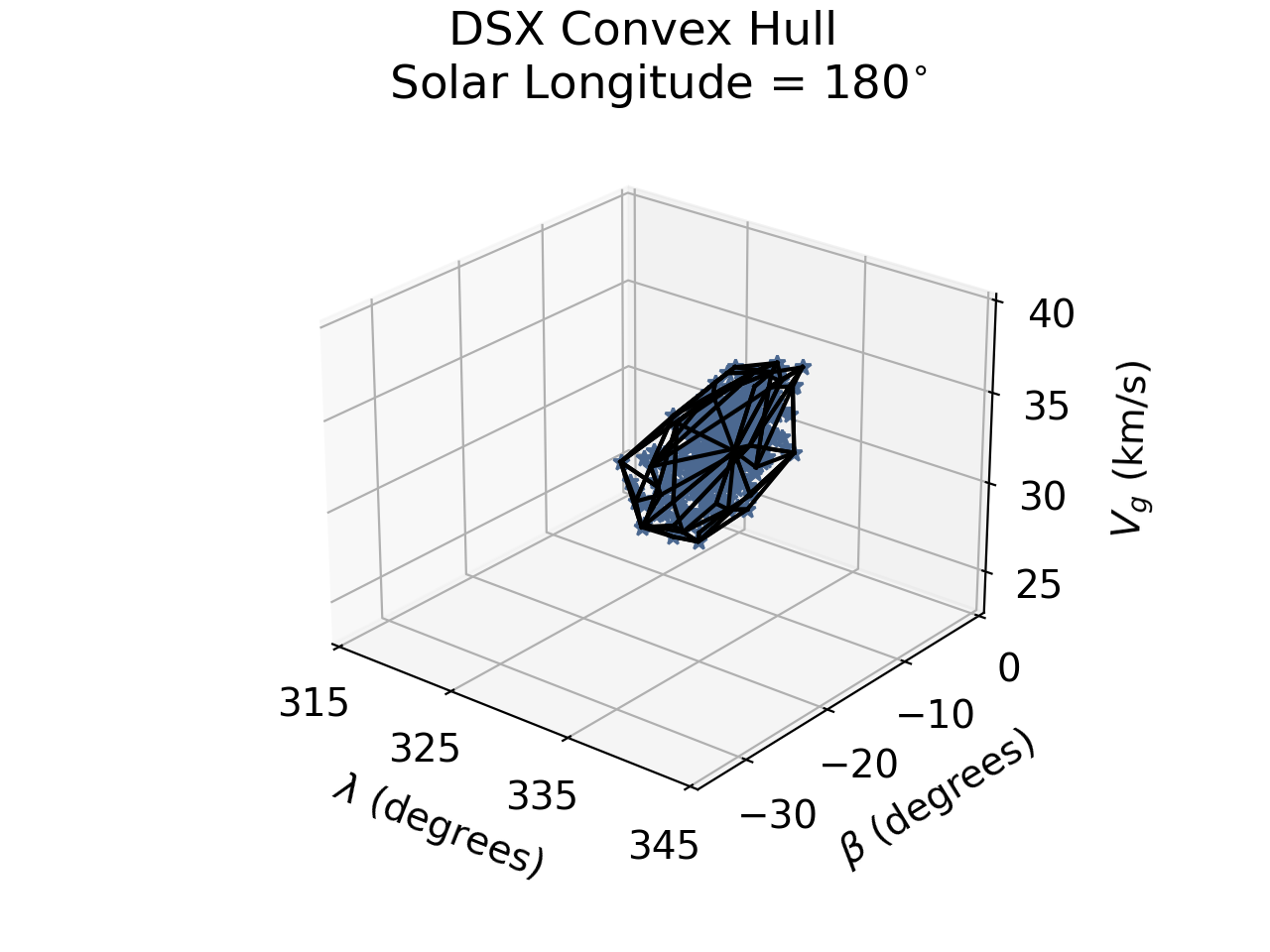}
\end{tabular}
\begin{tabular}{@{}c@{}}
    \includegraphics[width=0.45\textwidth]{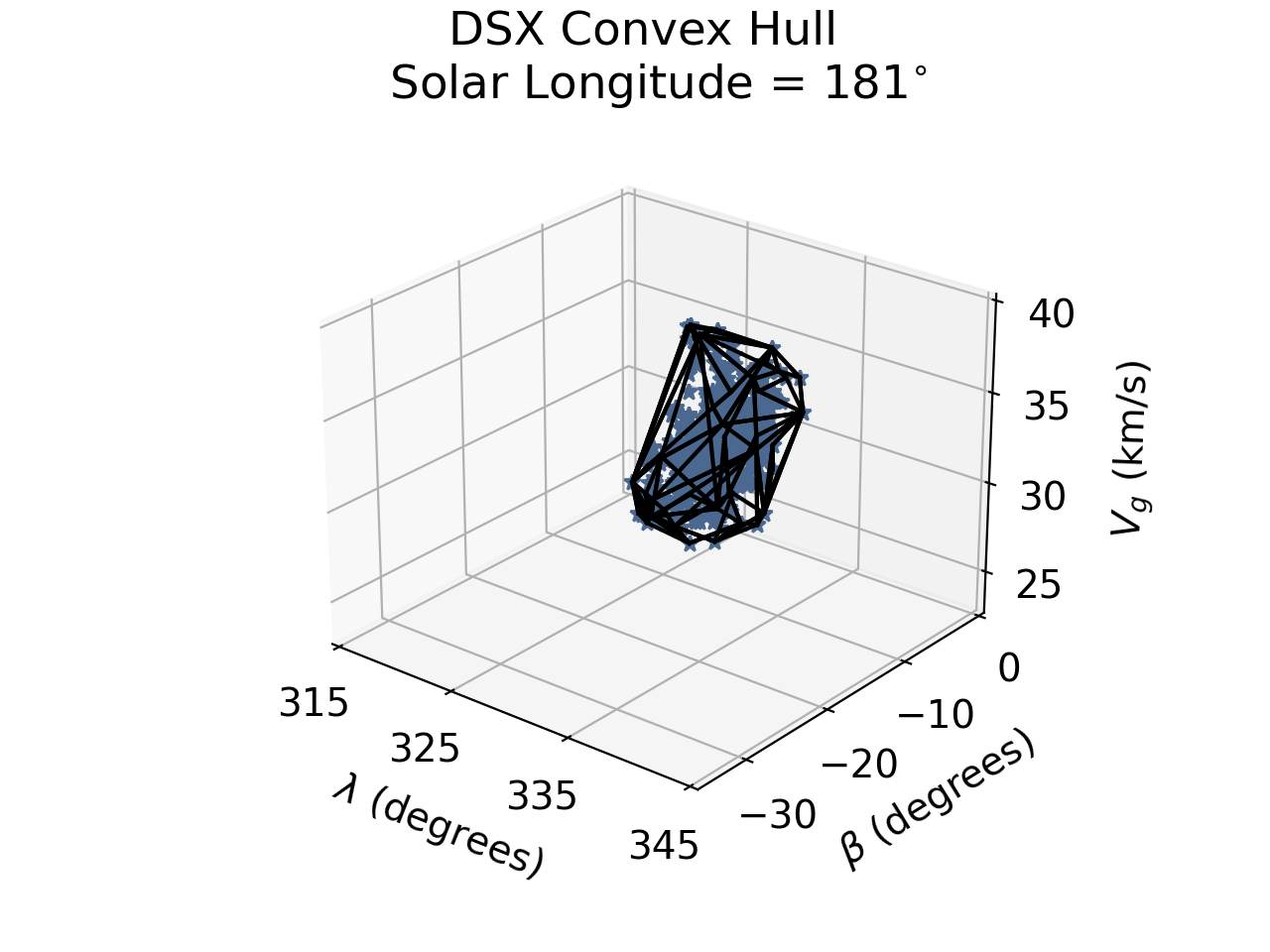}
\end{tabular}
\begin{tabular}{@{}c@{}}
    \includegraphics[width=0.45\textwidth]{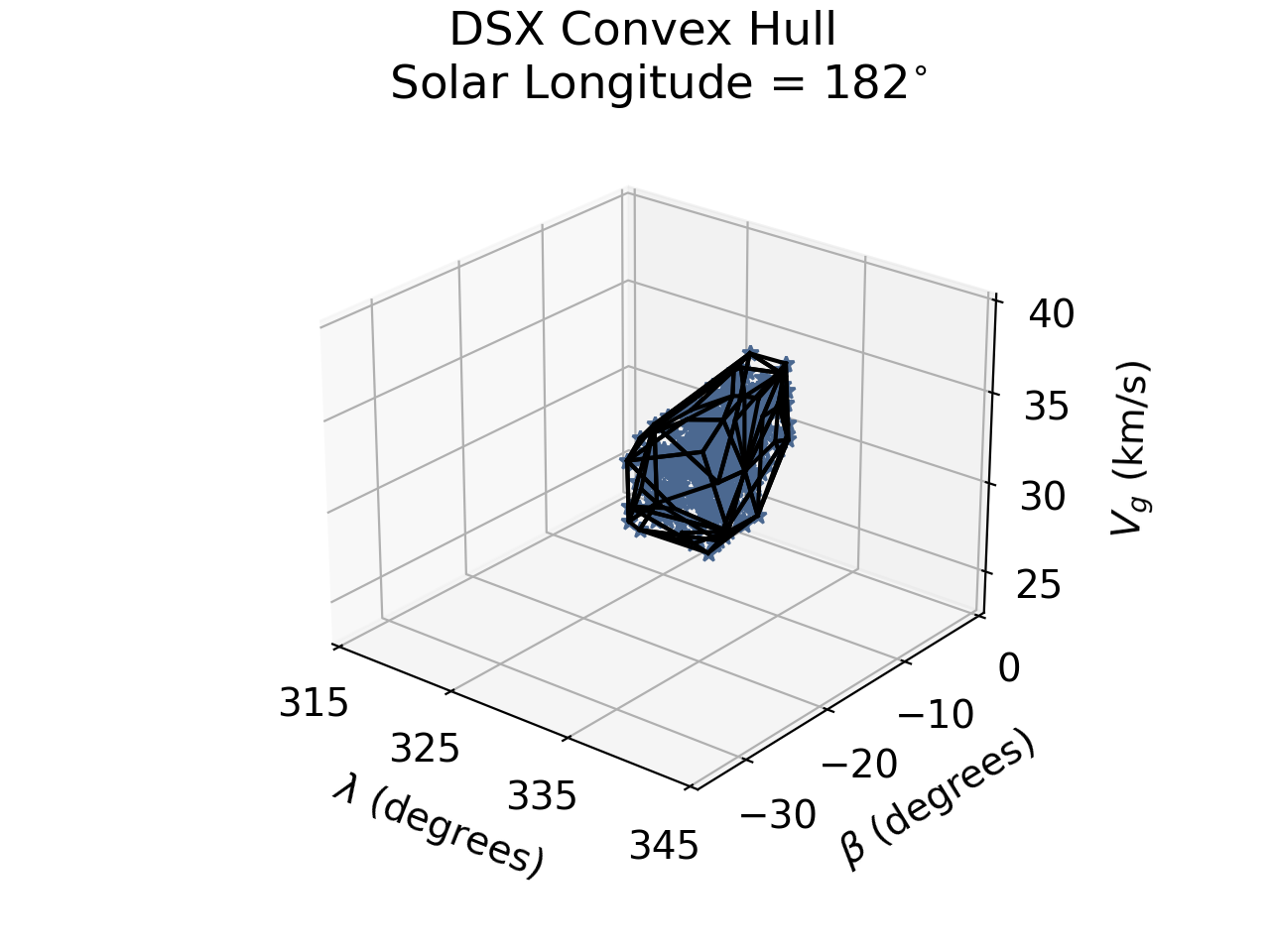}
\end{tabular}
\begin{tabular}{@{}c@{}}
    \includegraphics[width=0.45\textwidth]{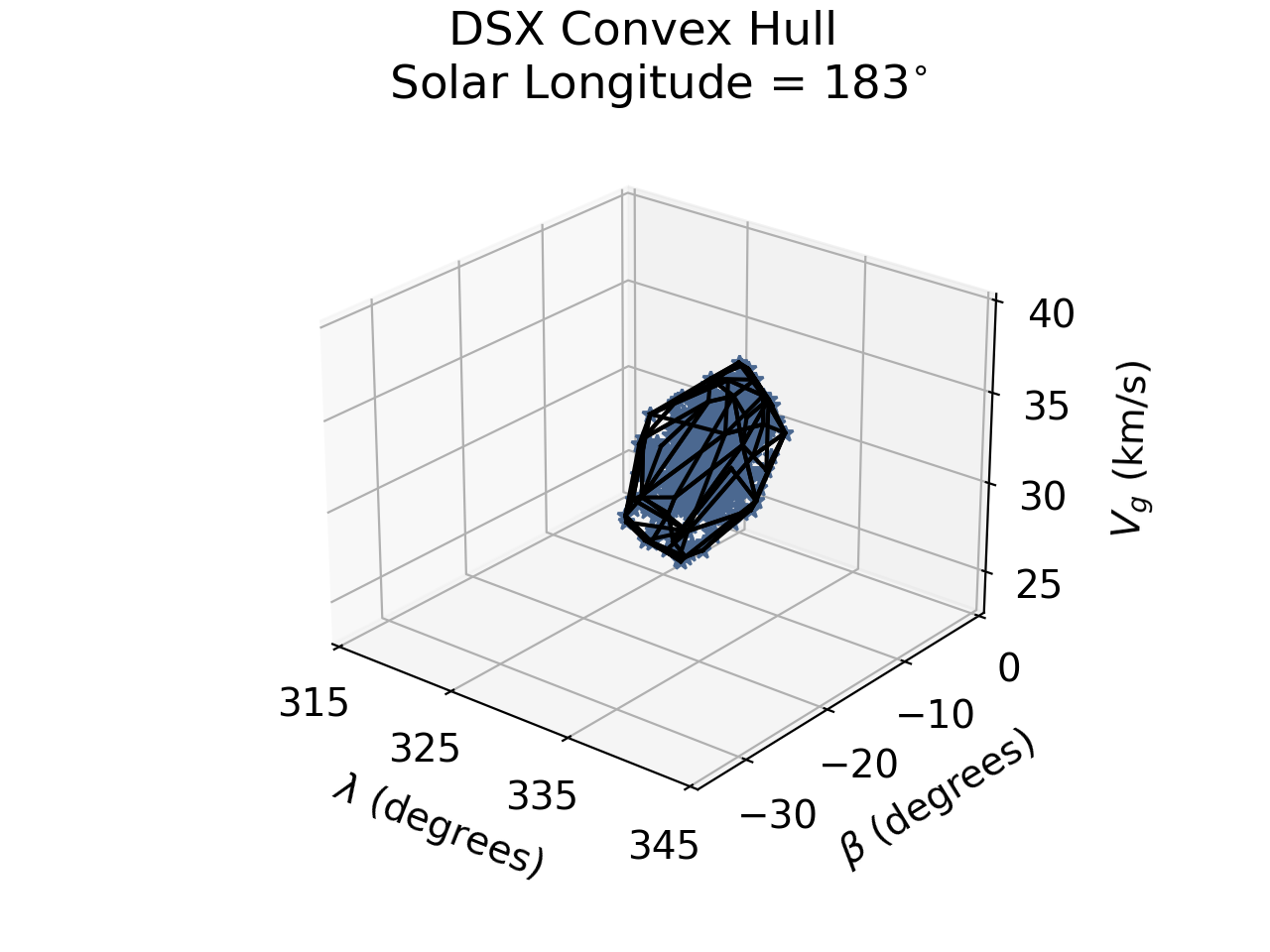}
\end{tabular}
\begin{tabular}{@{}c@{}}
    \includegraphics[width=0.45\textwidth]{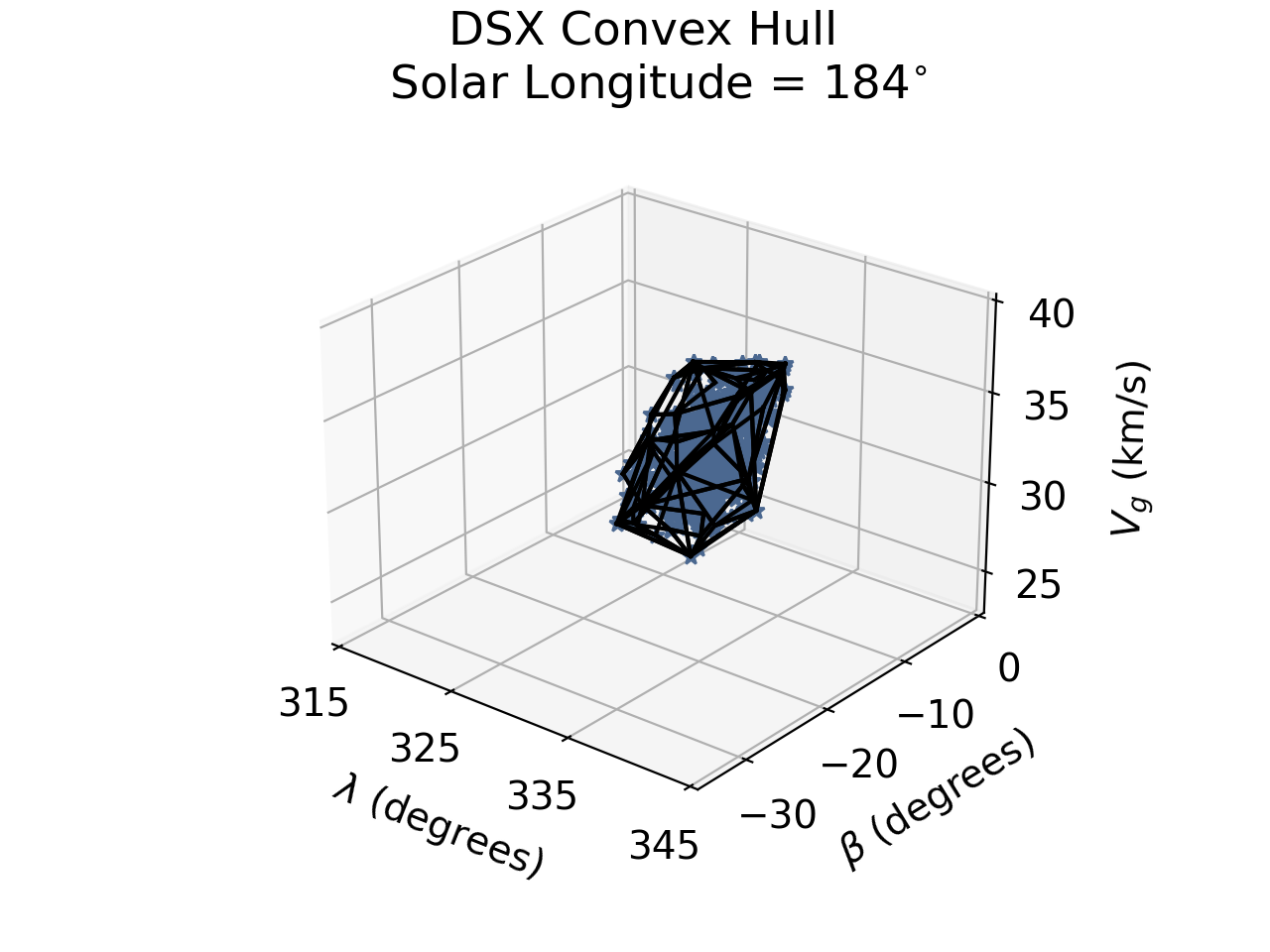}
\end{tabular}

\caption[short]{Convex hull results from solar longitude 179$^{\circ}$ to 184$^{\circ}$. The convex hull, described in Section 3.4.1 has been defined in such a way that any meteor with a radiant located within the hull is determined to be a member of the DSX shower, with a 95\% confidence level. These figures show the convex hull and all meteors that are located within it for a given solar longitude.}
\label{ch_2}
\end{figure*}

\begin{figure*}
\centering
\begin{tabular}{@{}c@{}}
    \includegraphics[width=0.45\textwidth]{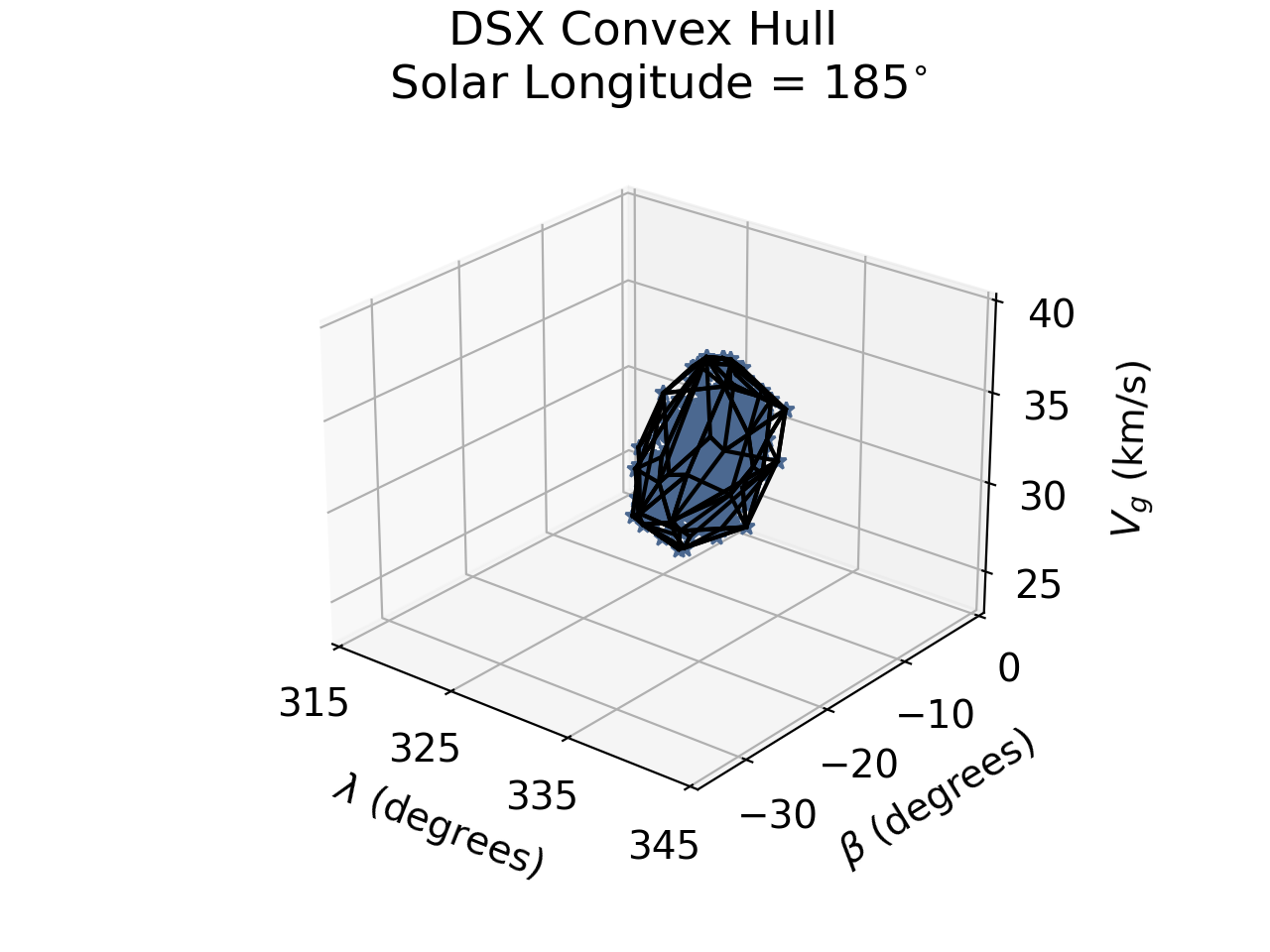}
\end{tabular}
\begin{tabular}{@{}c@{}}
    \includegraphics[width=0.45\textwidth]{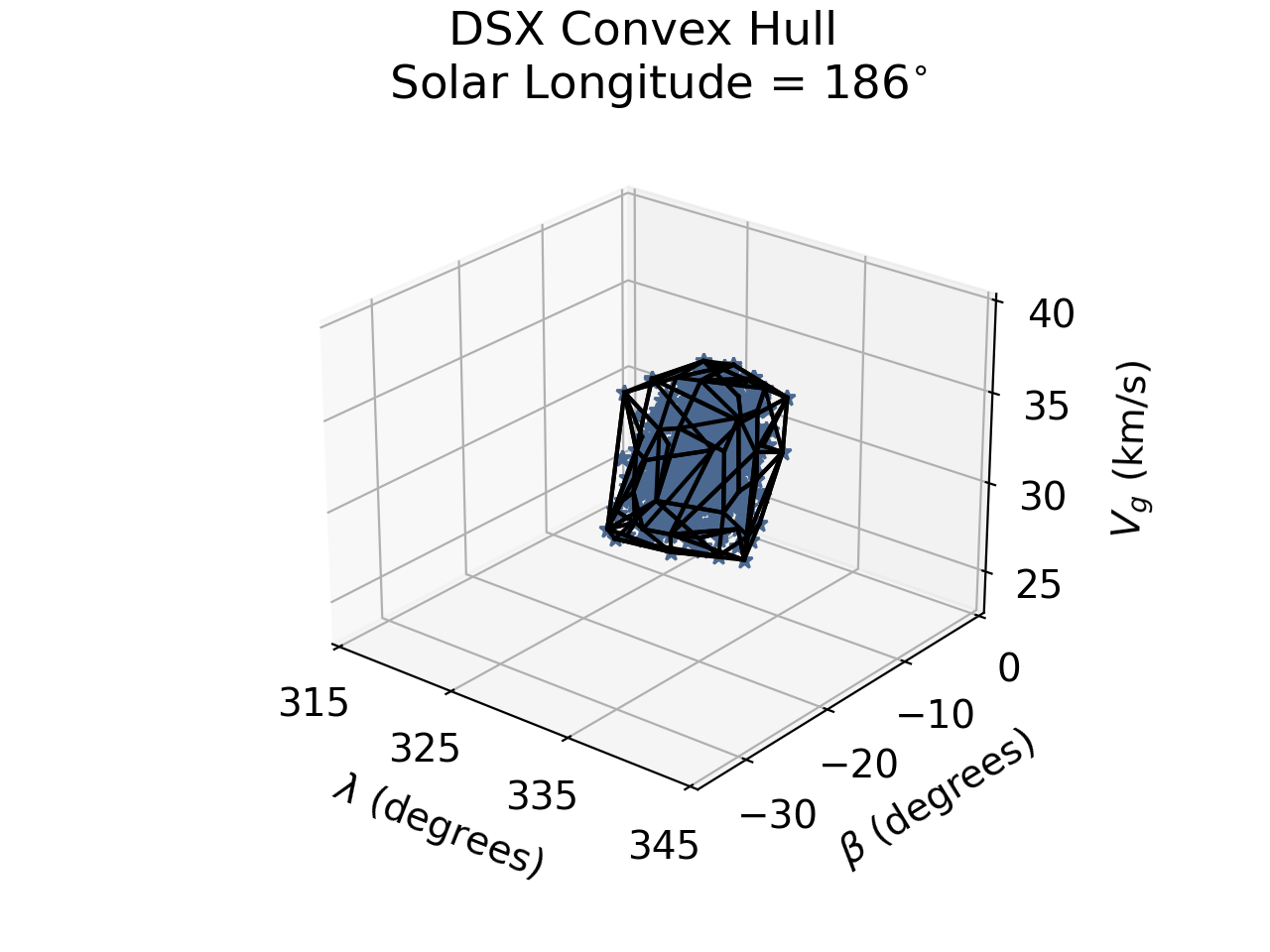}
\end{tabular}
\begin{tabular}{@{}c@{}}
    \includegraphics[width=0.45\textwidth]{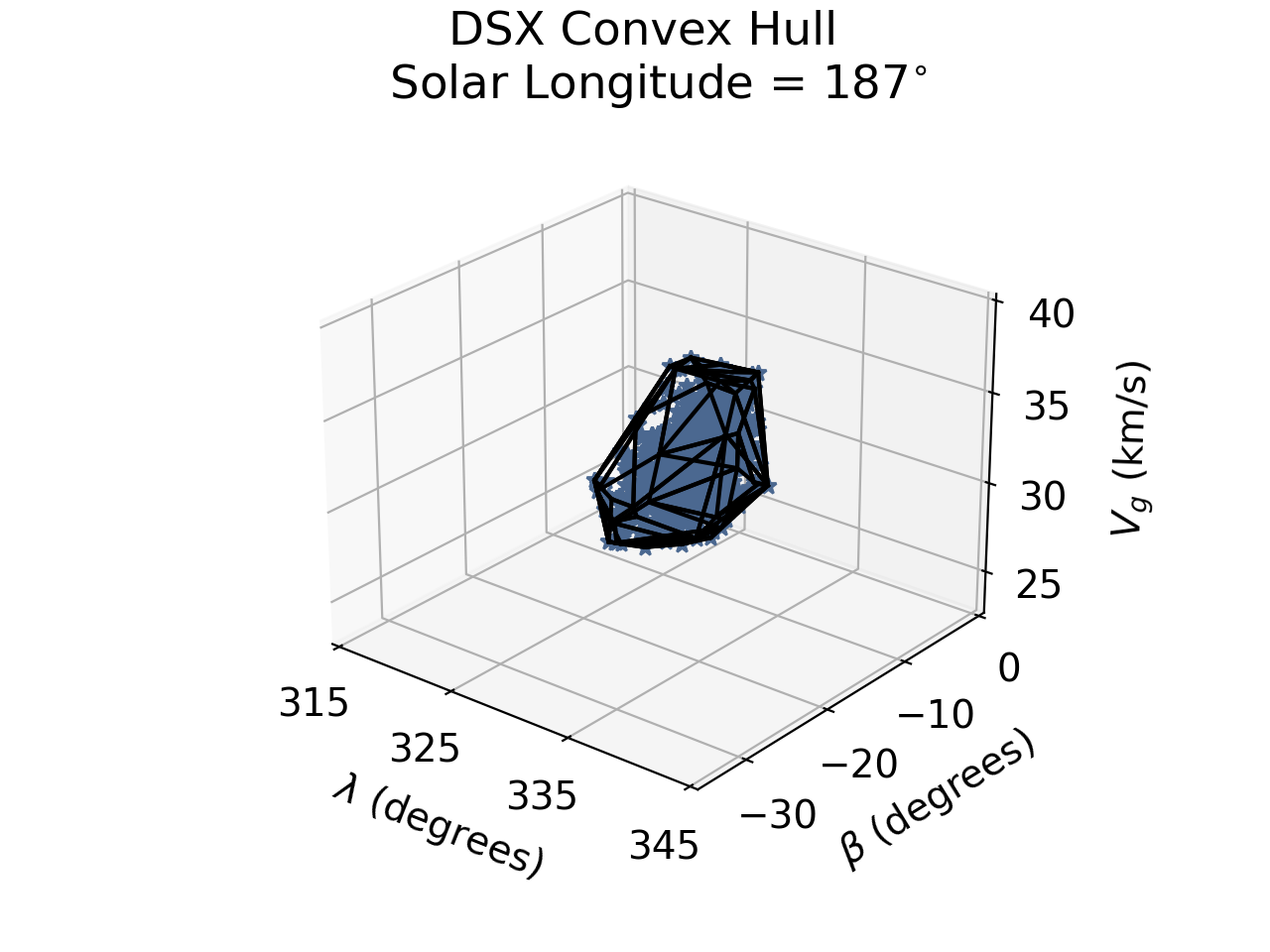}
\end{tabular}
\begin{tabular}{@{}c@{}}
    \includegraphics[width=0.45\textwidth]{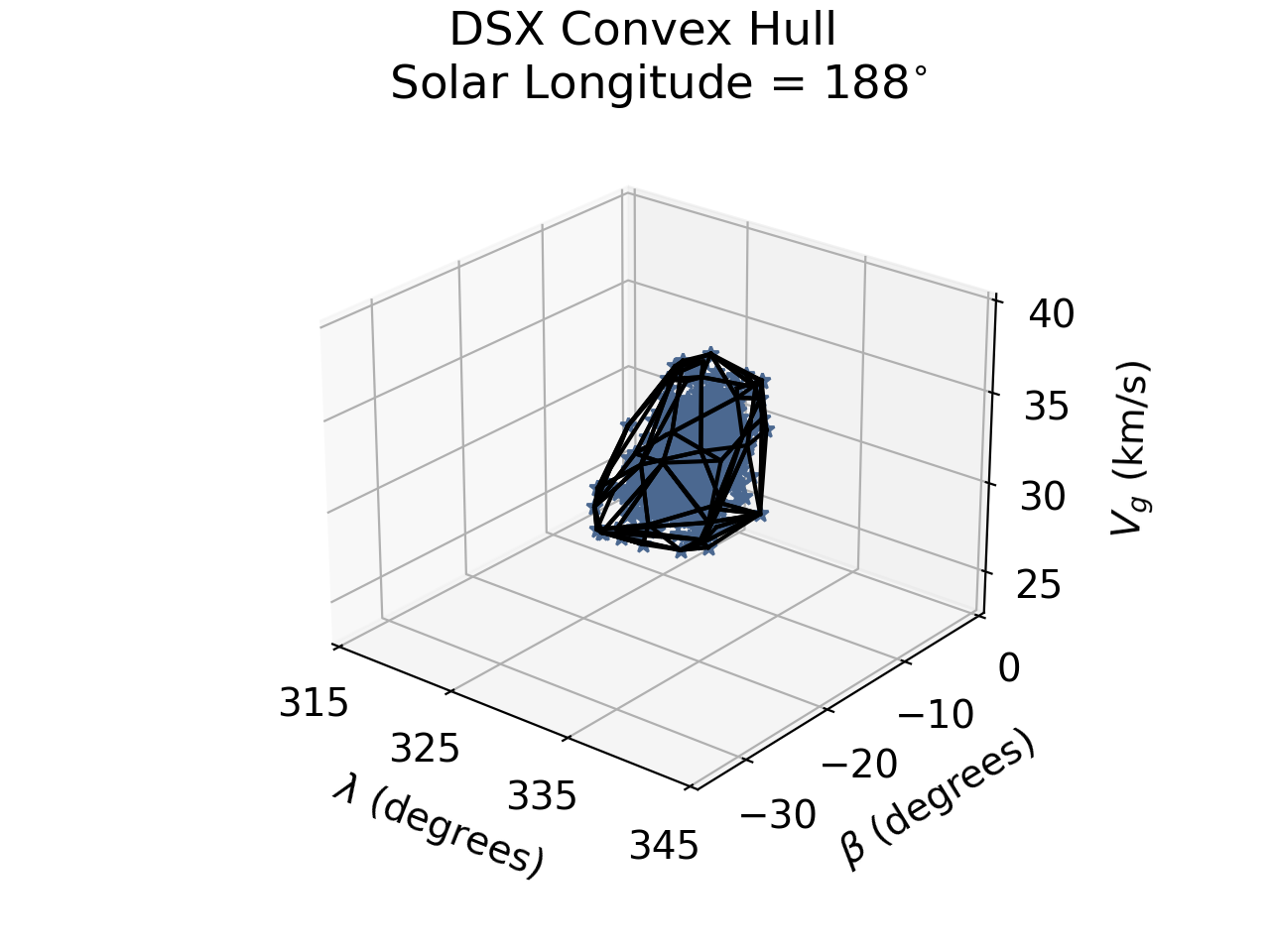}
\end{tabular}
\begin{tabular}{@{}c@{}}
    \includegraphics[width=0.45\textwidth]{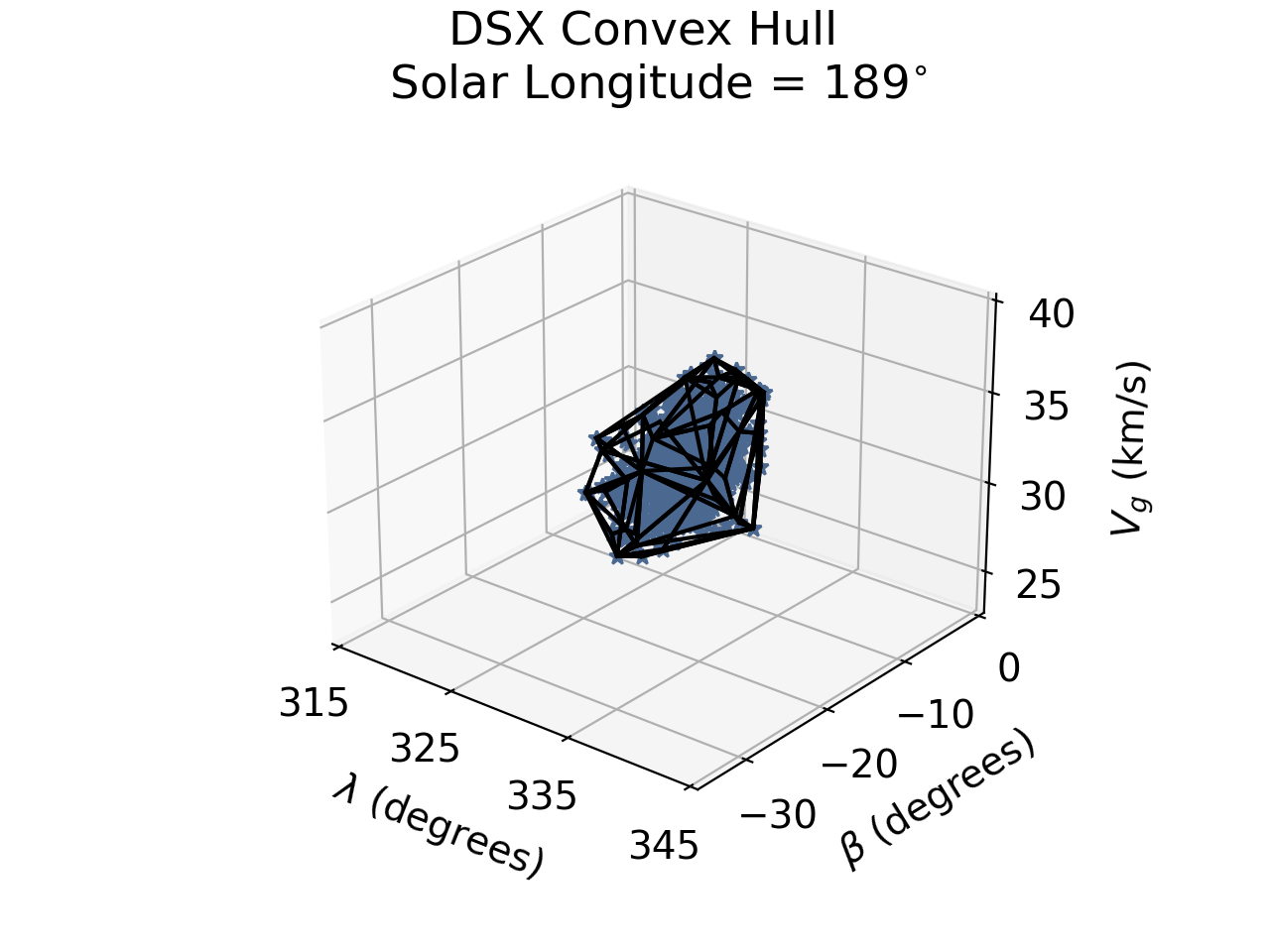}
\end{tabular}
\begin{tabular}{@{}c@{}}
    \includegraphics[width=0.45\textwidth]{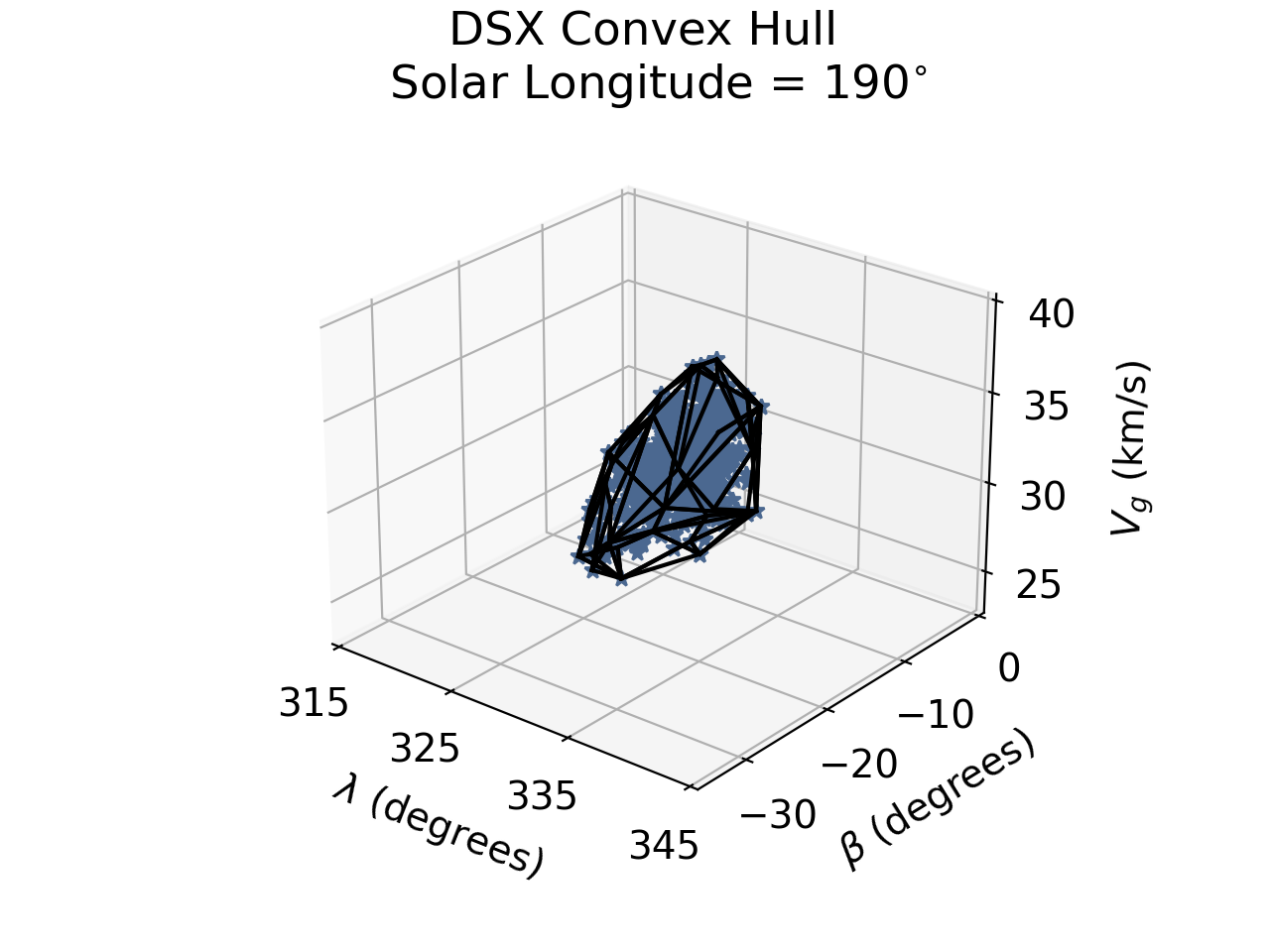}
\end{tabular}

\caption[short]{Convex hull results from solar longitude 185$^{\circ}$ to 190$^{\circ}$. The convex hull, described in Section 3.4.1 has been defined in such a way that any meteor with a radiant located within the hull is determined to be a member of the DSX shower, with a 95\% confidence level. These figures show the convex hull and all meteors that are located within it for a given solar longitude.}
\label{ch_3}
\end{figure*}

\begin{figure*}
\centering
\begin{tabular}{@{}c@{}}
    \includegraphics[width=0.45\textwidth]{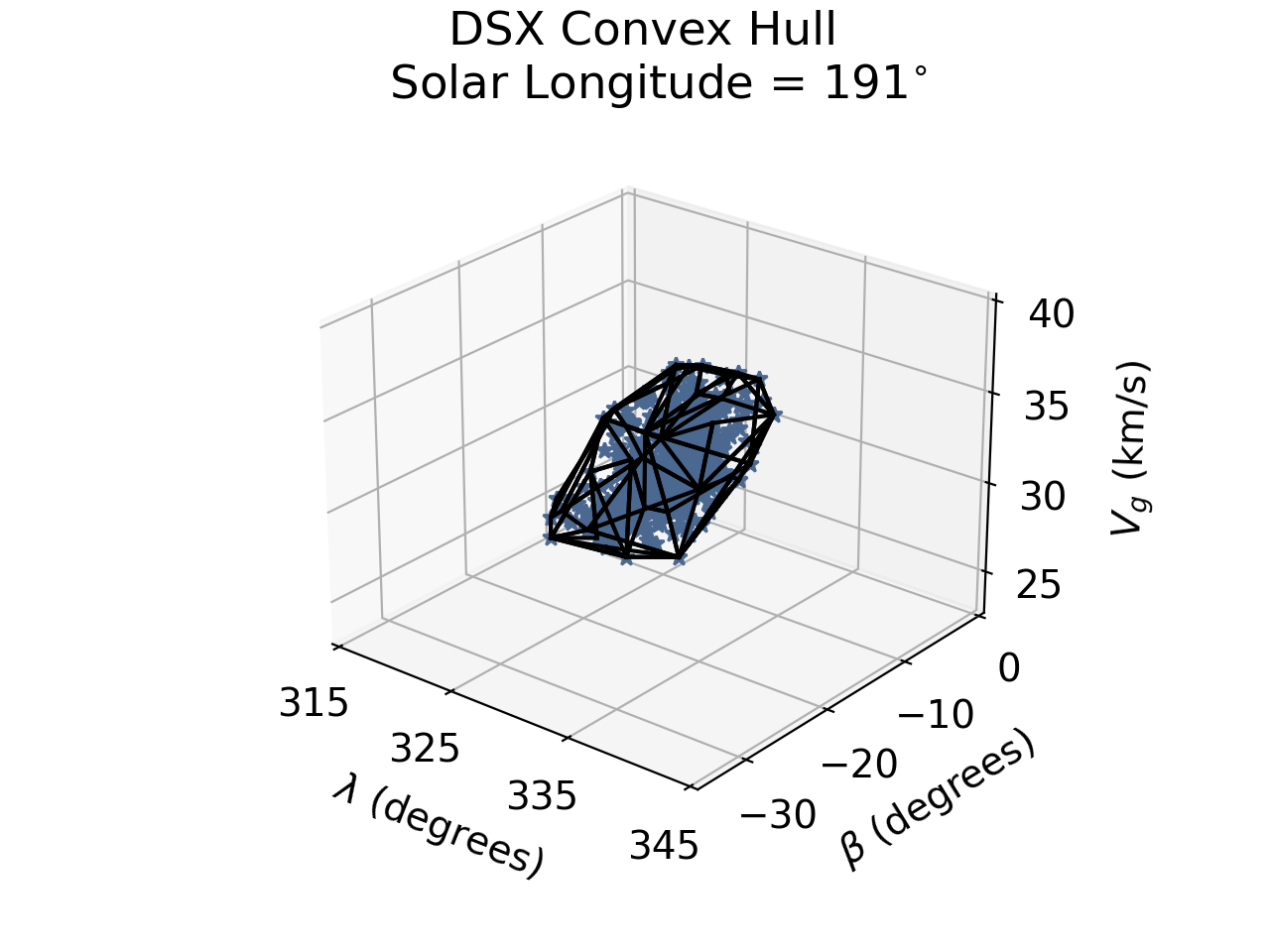}
\end{tabular}
\begin{tabular}{@{}c@{}}
    \includegraphics[width=0.45\textwidth]{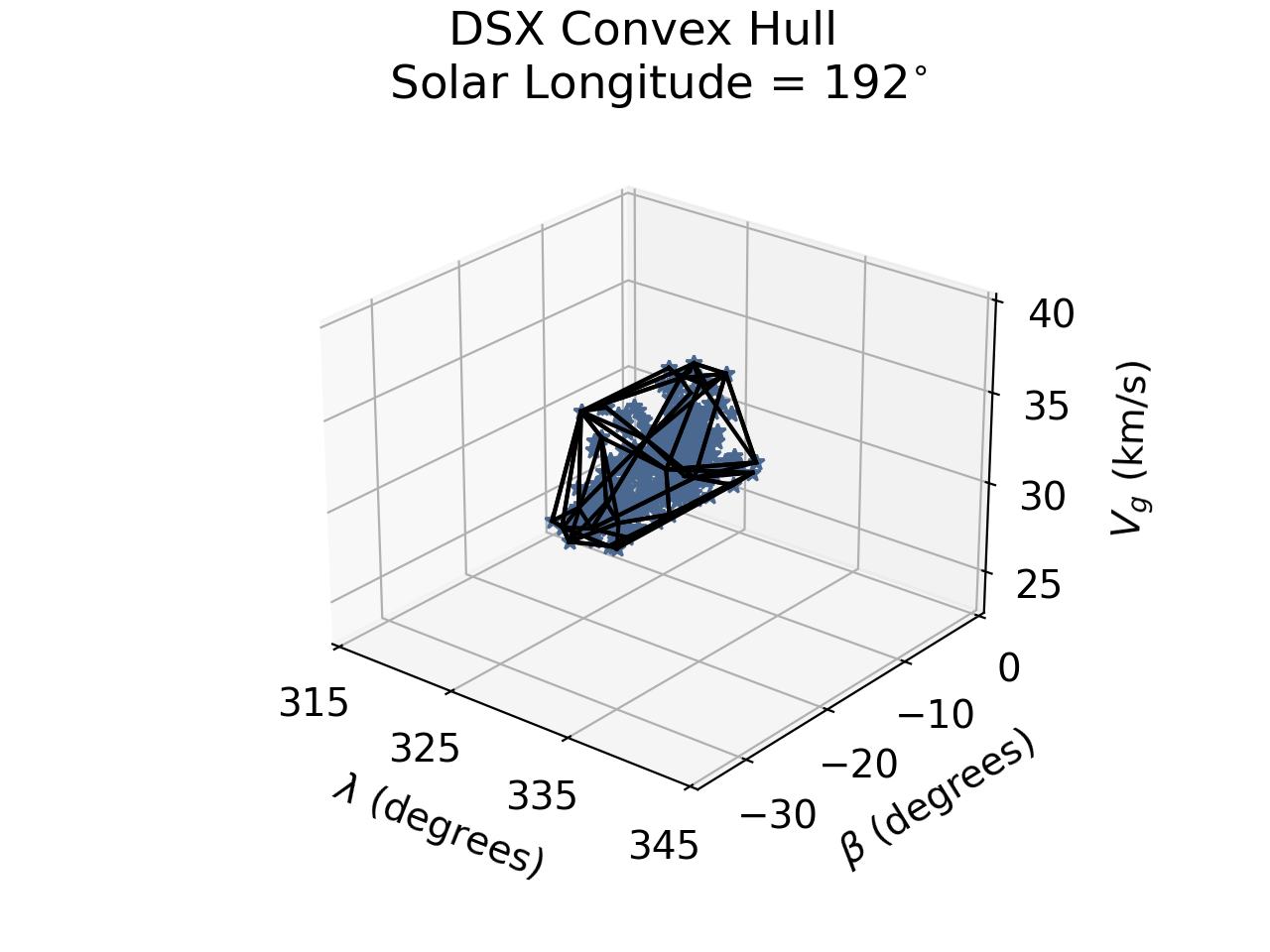}
\end{tabular}
\begin{tabular}{@{}c@{}}
    \includegraphics[width=0.45\textwidth]{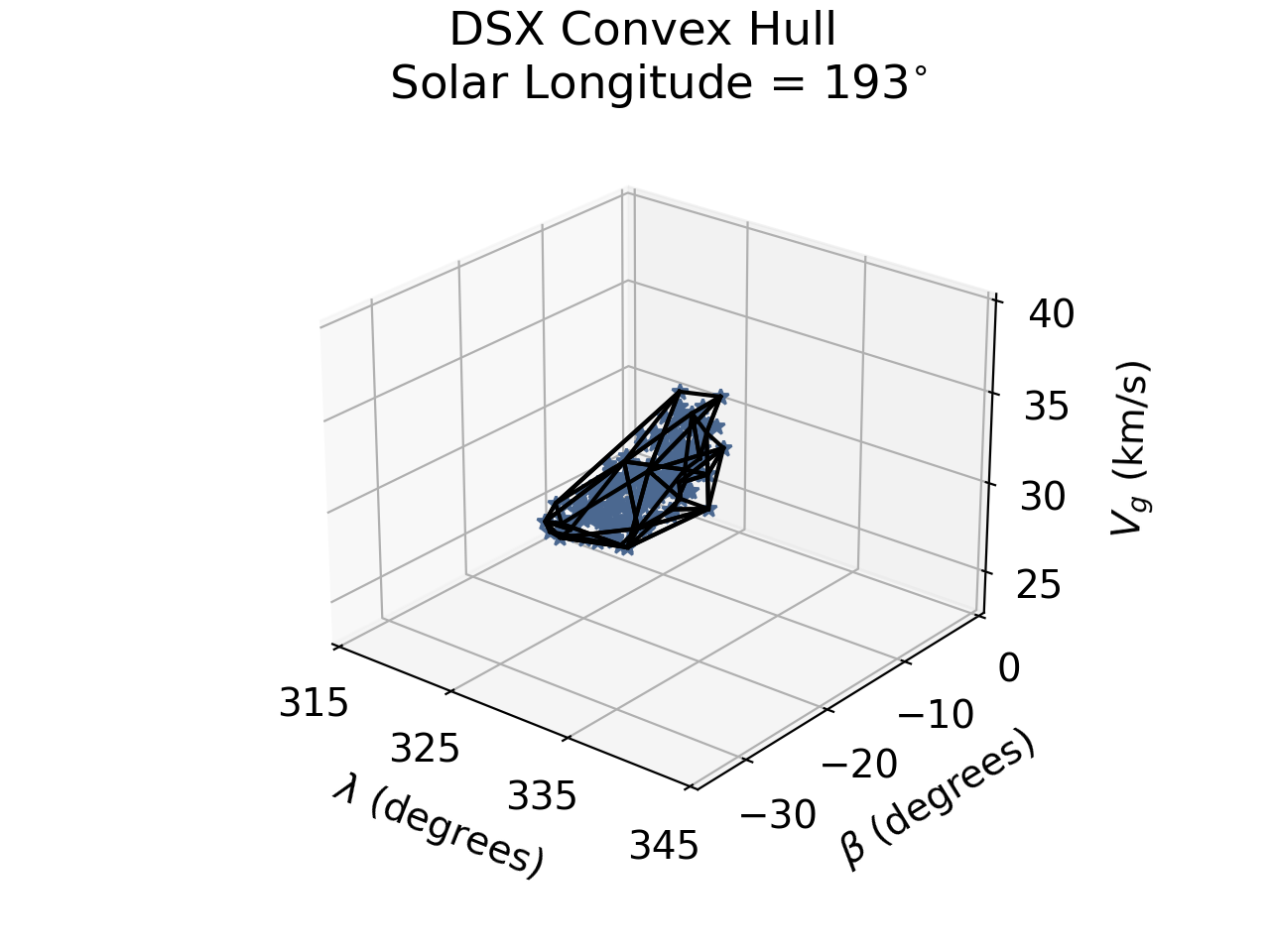}
\end{tabular}
\begin{tabular}{@{}c@{}}
    \includegraphics[width=0.45\textwidth]{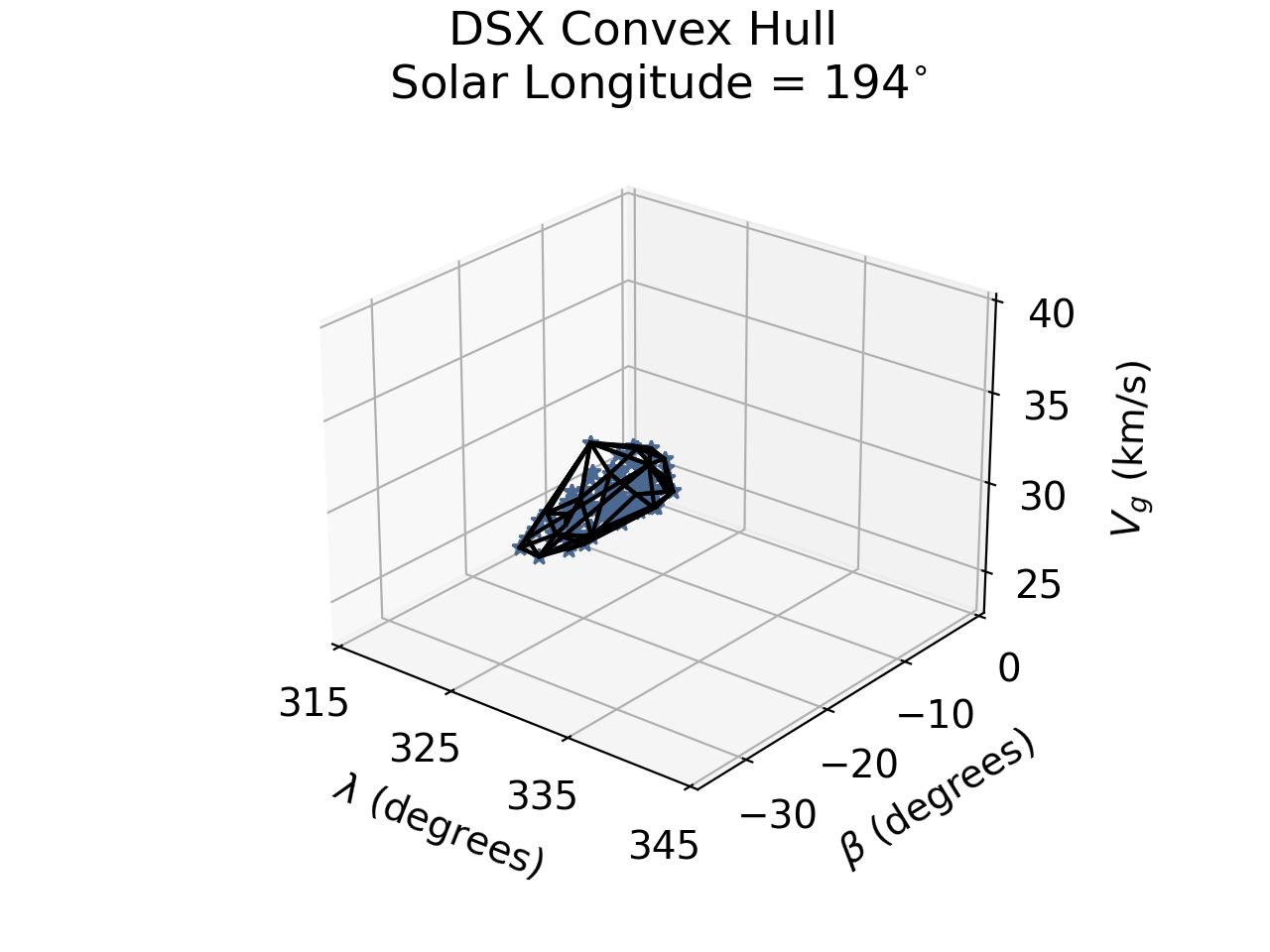}
\end{tabular}
\begin{tabular}{@{}c@{}}
    \includegraphics[width=0.45\textwidth]{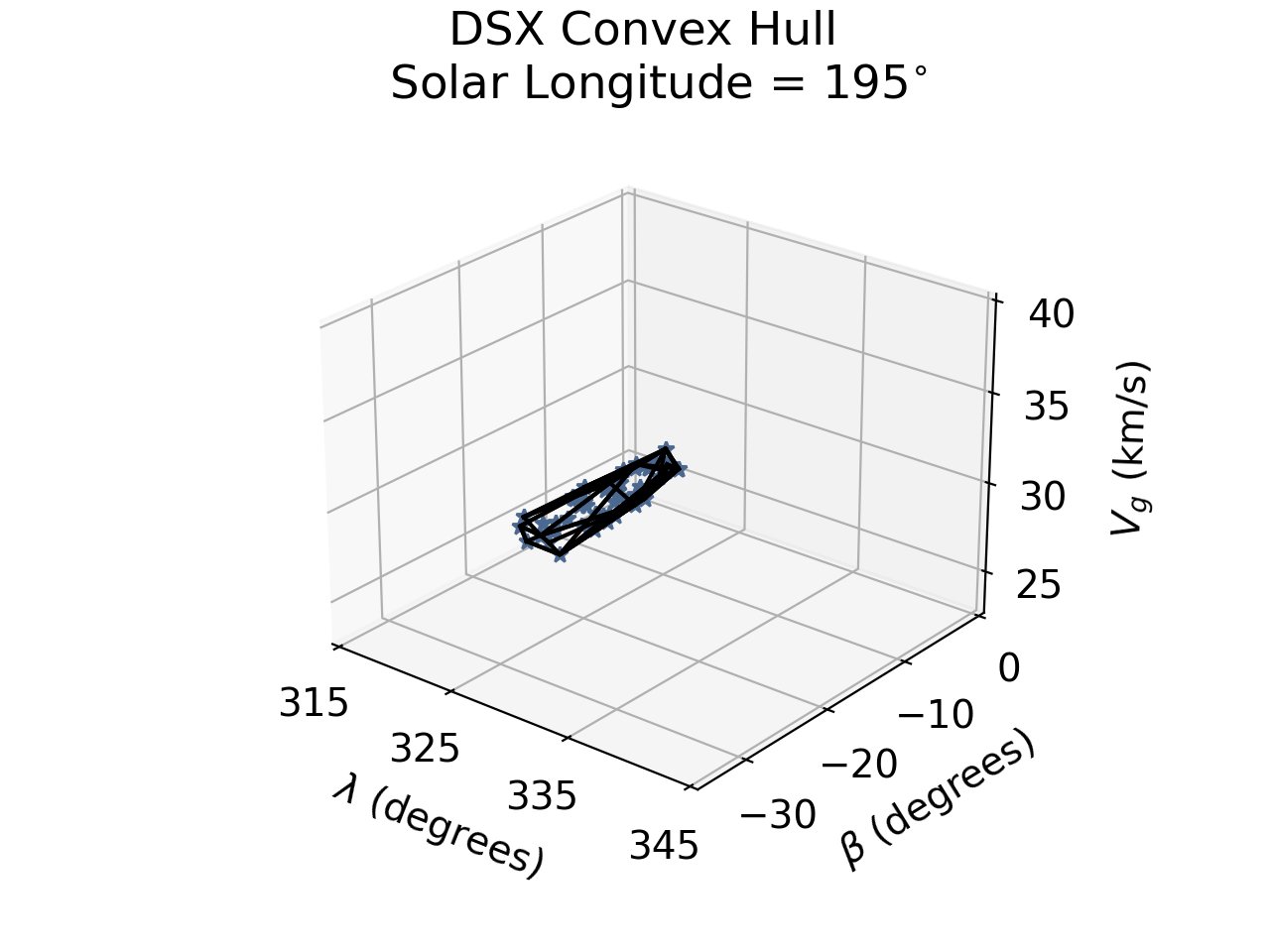}
\end{tabular}
\begin{tabular}{@{}c@{}}
    \includegraphics[width=0.45\textwidth]{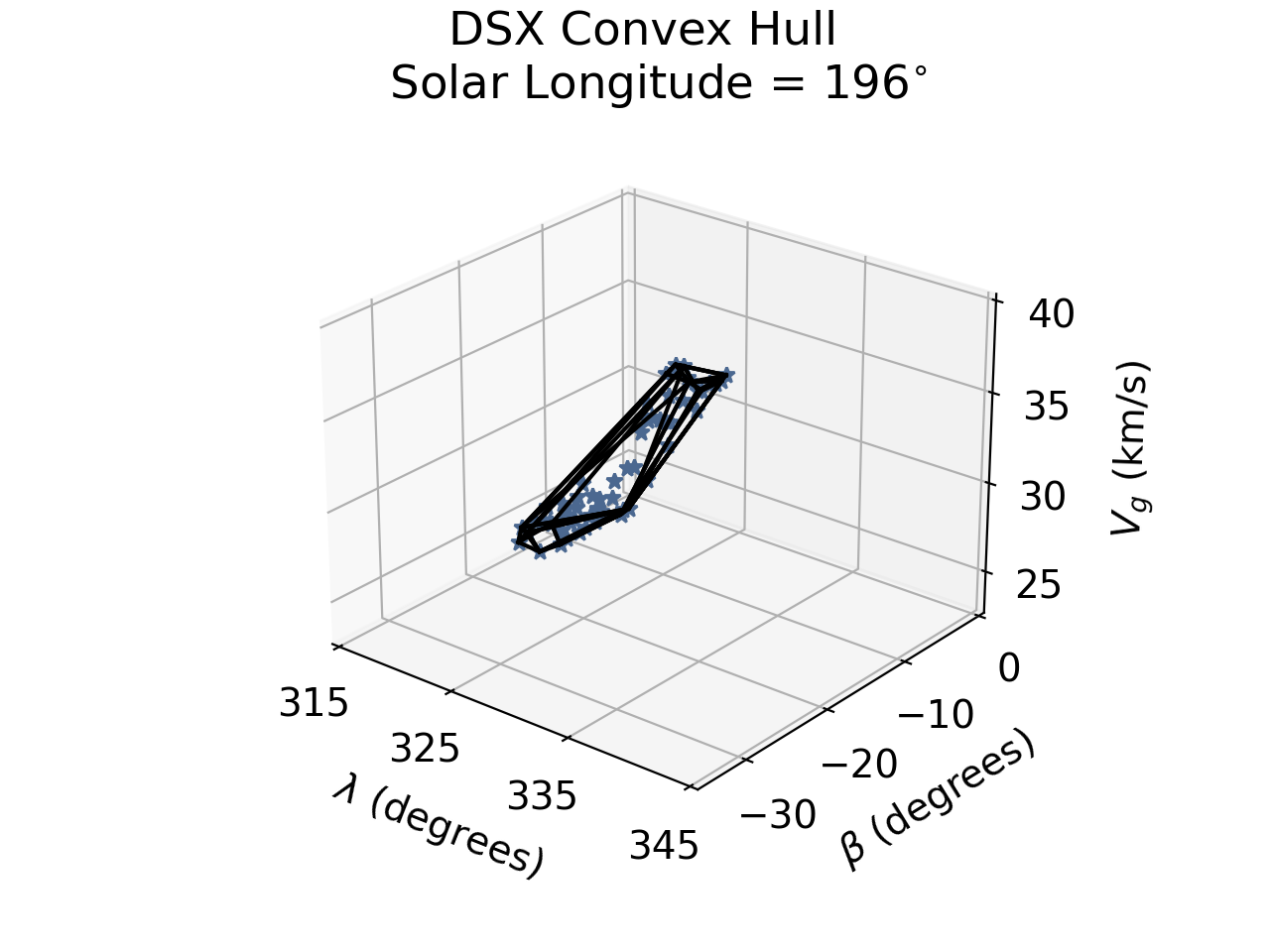}
\end{tabular}

\caption[short]{Convex hull results from solar longitude 191$^{\circ}$ to 196$^{\circ}$. The convex hull, described in Section 3.4.1 has been defined in such a way that any meteor with a radiant located within the hull is determined to be a member of the DSX shower, with a 95\% confidence level. These figures show the convex hull and all meteors that are located within it for a given solar longitude.}
\label{ch_4}
\end{figure*}

This table contains figures (Figures \ref{ch_1}, \ref{ch_2}, \ref{ch_3}, and \ref{ch_4}) of the convex hull results for each solar longitude of the DSX shower.

\section{An Alternative, More Robust Convex Hull Method}
\label{alt_convex_hull_method}

An assumption built into the Convex Hull meteor selection method, discussed in the main paper, is that the meteor radiants can be modeled as individual points in the radiant space. In reality, each meteor echo observed by CMOR has an uncertainty associated with its velocity and radiant measurement. Therefore each meteor echo is more realistically modeled by a three-dimensional Gaussian probability distribution in radiant space. This section explores whether this more computationally complex modeling method produces noticeably different results than the method discussed in Section 3.4.1 and whether it is a necessary modification. 

To model the meteor radiants as three-dimensional Gaussian probability distributions, we use three separate two-dimensional Gaussian probability distributions in each radiant space parameter: ($\lambda - \lambda_{\odot}$), $\beta$, and $V_g$. The center of each distribution is the value measured by CMOR, and measurement uncertainty represents one standard deviation from the mean per echo estimated using the Monte Carlo approach as described in \citet{WerykBrown2012}. We note that this approach ignores any non-diagonal covariance terms, however radiant covariances have been poorly explored so far \citep{vida2020}. 

The main difference in modeling the meteor radiants as three-dimensional Gaussian probability distributions instead of points in the radiant space is that the DSX and average background number density matrices must be calculated differently. The remaining steps in the Convex Hull meteor selection method remain the same. 

The extent of the Gaussian probability distribution of a meteor echo in three-dimensional radiant space can be very small, especially in the ecliptic longitude and latitude dimensions. For high-quality meteor echoes, the extent of this distribution is much smaller than the 8$\times$8$\times$8 voxels used in Section 3.4.1 to create the 3D number density matrix. Therefore more voxels are required to capture the scale of the 3D echo distributions. To increase the number of voxels, each 8$\times$8$\times$8 voxel is split into 100$\times$100$\times$100 sub-voxels, meaning that there are 800$\times$800$\times$800 sub-voxels in total.

Each meteor echo is modeled as a 3D Gaussian probability distribution. Instead of counting the number of whole meteors in each voxel, we calculate the probability that the meteor is located in each sub-voxel of radiant space. Each sub-voxel contains a small range of ($\lambda - \lambda_{\odot}$), $\beta$, and $V_g$ values. The Gaussian probability functions are used to determine the probability that a meteor is located within a given voxel, using the mean of each of the sub-voxel's ($\lambda - \lambda_{\odot}$), $\beta$, and $V_g$ ranges. A meteor's probability is set to zero if any of the parameter values are more than two standard deviations from the mean of a given distribution. After the probability calculations are completed in all applicable sub-voxels, the sum of the probabilities for a given meteor is normalized to one. 

In our application to the DSX, there are 512 million sub-voxels in total, so to make this process less computationally expensive, we reject any meteor with a maximum or minimum radiant value (using the CMOR measurement uncertainties) outside the ($\lambda - \lambda_{\odot}$), $\beta$, and $V_g$ radiant cuts. This rejection reduces the number of sub-voxels that need to be evaluated. For the DSX peak day, located at solar longitude 186$^{\circ}$, this rejection reduced the number of meteors from 1342 to 953.

After the 3D number density matrix is created, the 800$\times$800$\times$800 matrix is converted into an 8$\times$8$\times$8 matrix. This conversion is done by adding all values in the set of sub-voxels contained within each larger voxel. Once the number density matrix is recombined into an 8$\times$8$\times$8 size matrix, the rest of the analysis is identical to the process described in Section 3.4.1, except that this sub-voxel method is also used to create the average background density matrix. The complete set of convex hulls for the duration of the DSX shower is located in Appendix \ref{sec:Alt Convex Hull Results}.

\subsection{Calculating the DSX Orbital Elements from Radar Data} \label{orbital_element_section}

The convex hull results, calculated in the above sections, identify the set of individual DSX meteors with 95\% confidence for each solar longitude bin. Once the DSX meteor set has been isolated, the mean radiant and orbital elements are calculated using the method described in \citep{Jopek2006}. This method calculates the mean values using the least-squares method to average the heliocentric vectorial elements. 

\subsubsection{Comparing Results}
The alternate convex hull method is more rigorous, but much more computationally intensive, so we compare the results to determine whether the more complex method yields significantly different results.

\begin{figure*}
\centering
    \includegraphics[width=0.9\columnwidth]{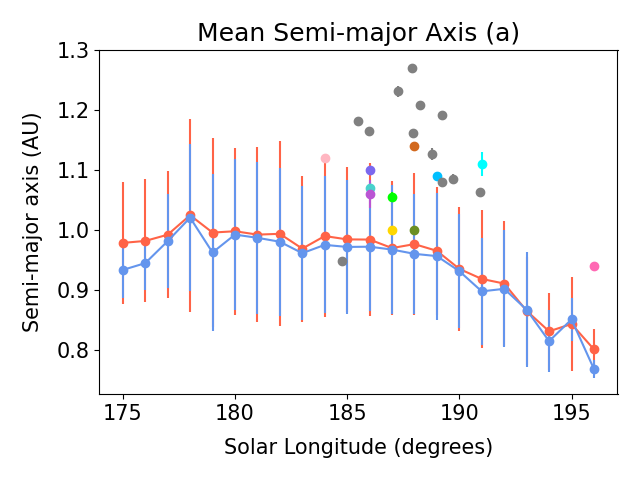}
    \includegraphics[width=0.9\columnwidth]{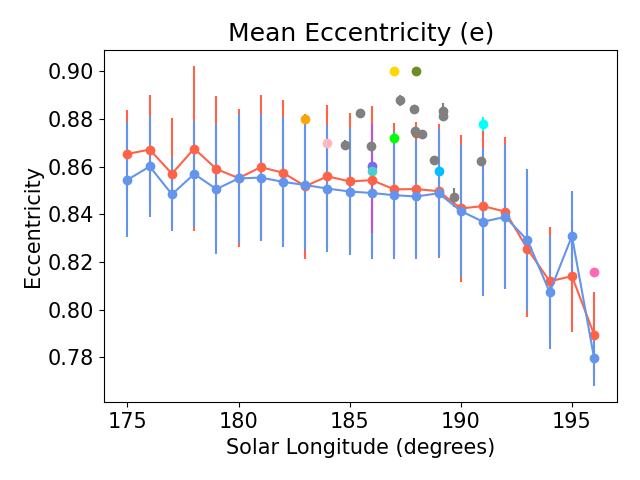}
    
    \includegraphics[width=0.9\columnwidth]{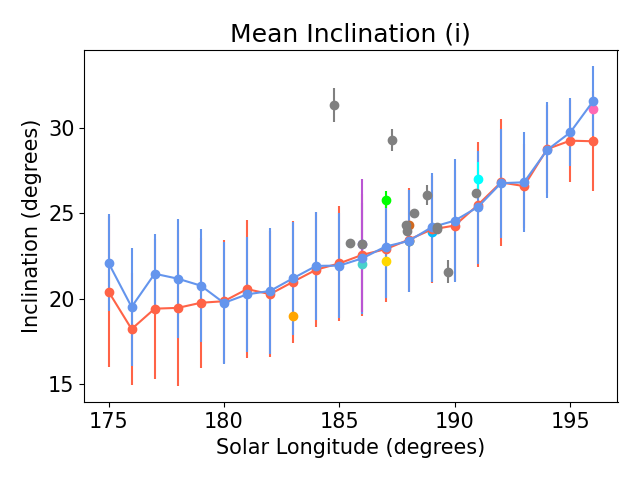}
    \includegraphics[width=0.9\columnwidth]{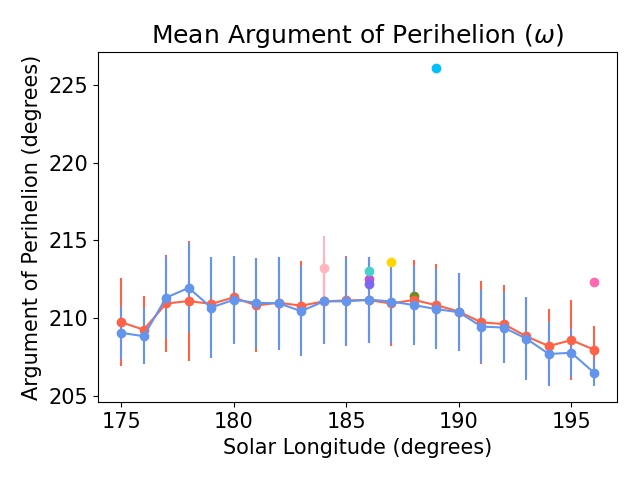}
    
    \includegraphics[width=0.9\columnwidth]{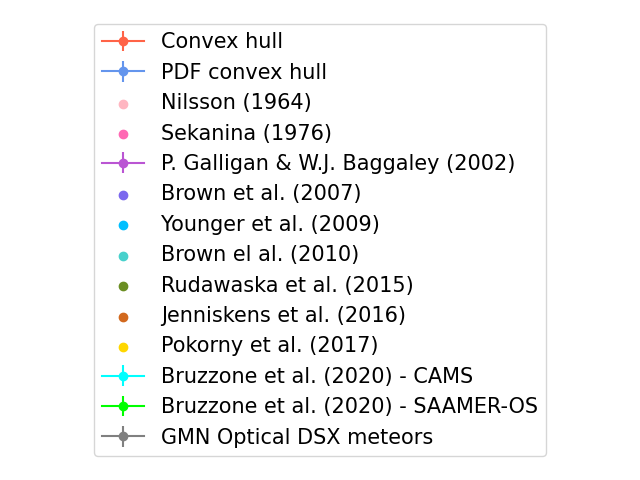}
\caption[short]{A comparison of the orbital elements over the duration of the DSX shower calculated using the both the alternate convex hull method, which models the meteor radiants as three-dimensional Gaussian probability distributions in radiant space instead of points, and of the computationally simple convex hull method, which models the meteors as points in radiant space. The convex hull is used to extract the set of DSX meteors with a confidence level of 95\%. The mean orbital elements have been calculated with the method described in \citet{Jopek2006}.The uncertainty bars represent one standard deviation of the DSX meteor set produced by each convex hull method. Results from literature, described in section \ref{app:literature_appendix}, are displayed. Note that the alternative convex hull method covers the solar longitude range from 175$^{\circ}$ to 196$^{\circ}$, whereas the computationally simple convex hull covers the solar longitude range from 173$^{\circ}$ to 196$^{\circ}$. }
\label{fig:orbital_elements_alt}
\end{figure*}

The mean orbital element and radiant values over the duration of the Daytime Sextantids shower are shown in Figures 11 and 12 for the computationally simple convex hull method described in Section 3.4.1. Figures \ref{fig:orbital_elements_alt} and \ref{fig:DSX radiant alt} contain the mean orbital elements and radiant results for the alternate method described in Section \ref{alt_convex_hull_method}. Note that the alternate convex hull creation method only detected the shower from a solar longitude range of 175$^{\circ}$ to 196$^{\circ}$. In contrast, the computationally simple convex hull method described in Section 3.4.1 detected the shower from 173$^{\circ}$ to 196$^{\circ}$.

\begin{figure*}
\centering
\begin{tabular}{@{}c@{}}
    \includegraphics[width=0.45\textwidth]{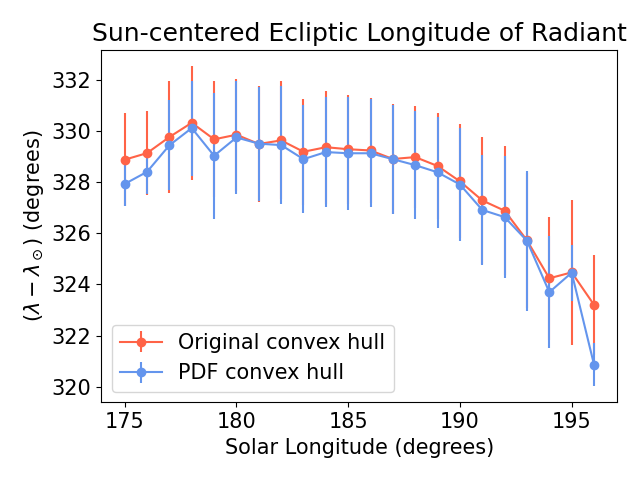}
\end{tabular}
\begin{tabular}{@{}c@{}}
    \includegraphics[width=0.45\textwidth]{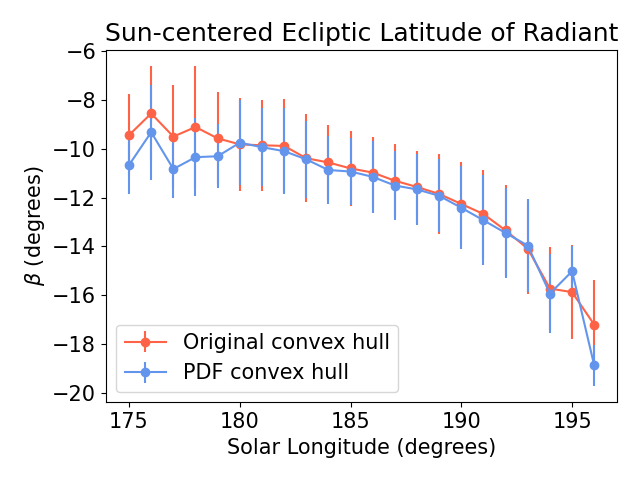}
\end{tabular}
\begin{tabular}{@{}c@{}}
    \includegraphics[width=0.45\textwidth]{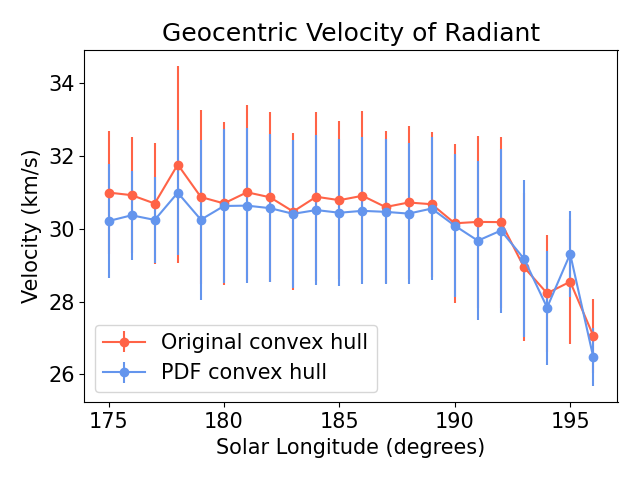}
\end{tabular}

\caption[short]{A comparison between the orbital elements of the DSX shower calculated using the computationally simple convex hull method, which models the meteor radiants as points in radiant space, and the alternative convex hull method, which models the radiants as 3D Gaussian PDFs. The convex hull is used to extract the set of DSX meteors with a confidence level of 95\%. The mean DSX radiant has been calculated with the method described in \citet{Jopek2006}. Note that these figures cover the solar longitude range from 175$^{\circ}$ to 196$^{\circ}$, which is the DSX duration calculated by the alternative convex hull method, whereas the computationally simple convex hull method calculated a duration of 173$^{\circ}$ to 196$^{\circ}$.}
\label{fig:DSX radiant}
\end{figure*}

\begin{figure*}
\centering
\begin{tabular}{@{}c@{}}
    \includegraphics[width=0.45\textwidth]{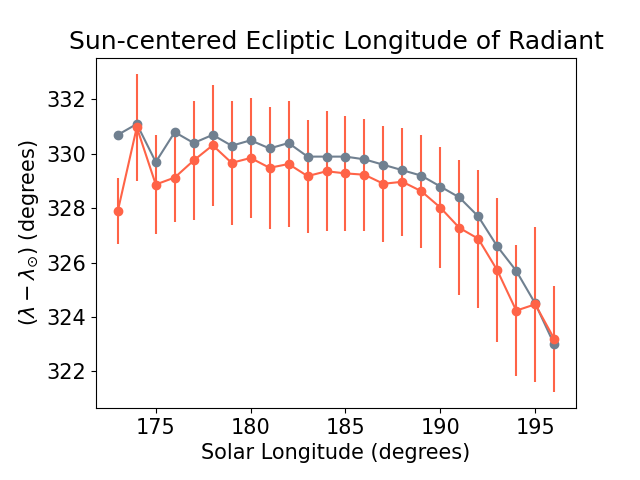}
\end{tabular}
\begin{tabular}{@{}c@{}}
    \includegraphics[width=0.45\textwidth]{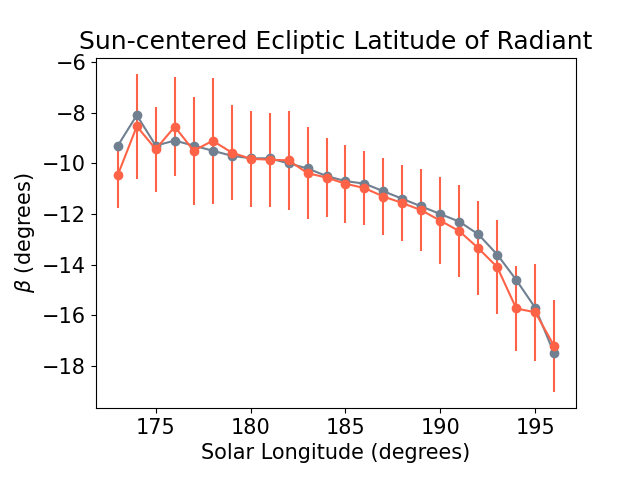}
\end{tabular}
\begin{tabular}{@{}c@{}}
    \includegraphics[width=0.45\textwidth]{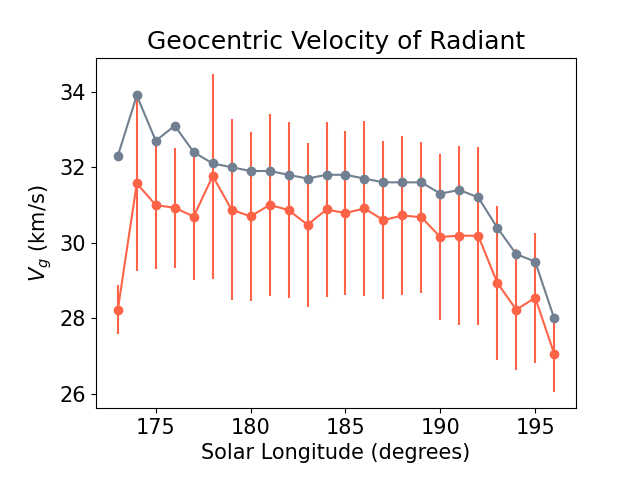}
\end{tabular}
\begin{tabular}{@{}c@{}}
    \includegraphics[width=0.45\textwidth]{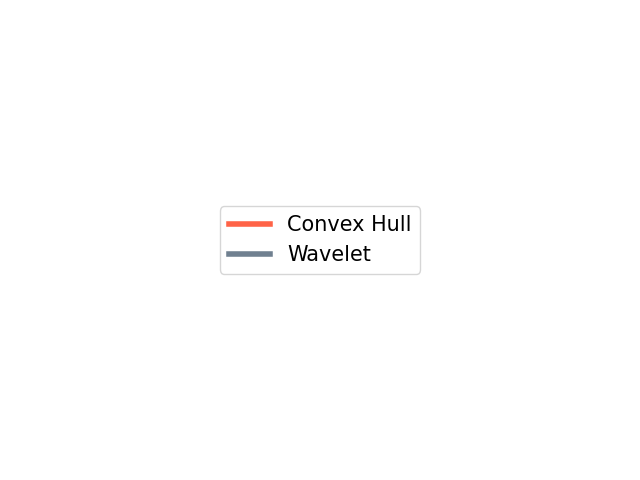}
\end{tabular}
\caption[short]{Radiant of the DSX shower as a function of solar longitude, calculated using the computationally simple convex hull method, compared with the wavelet-calculated radiant. The convex hull is used to extract the set of DSX meteors with a confidence level of 95\%. The mean DSX radiant has been calculated with the method described in \citet{Jopek2006}.}
\label{fig:DSX radiant alt}
\end{figure*}

\begin{figure*}
\centering
\begin{tabular}{@{}c@{}}
    
    \includegraphics[width=0.45\textwidth]{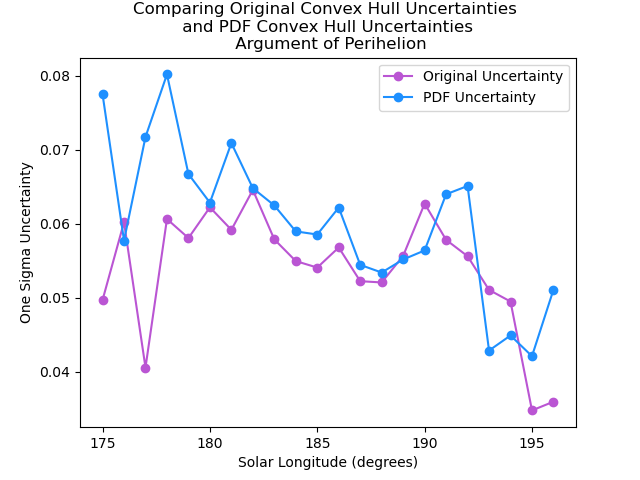}
\end{tabular}
\begin{tabular}{@{}c@{}}

    \includegraphics[width=0.45\textwidth]{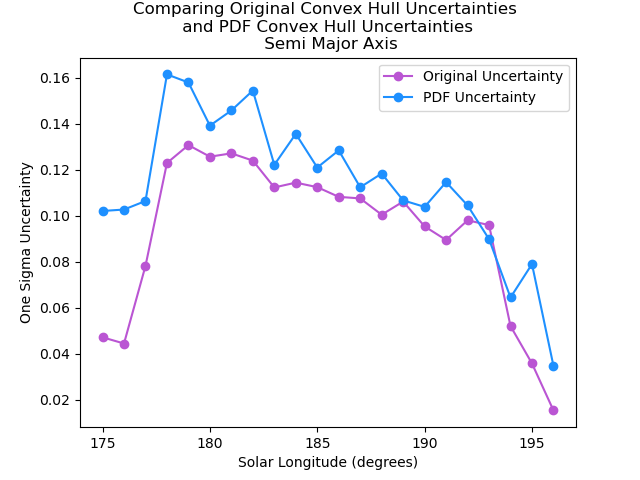}
\end{tabular}
\begin{tabular}{@{}c@{}}
    \includegraphics[width=0.45\textwidth]{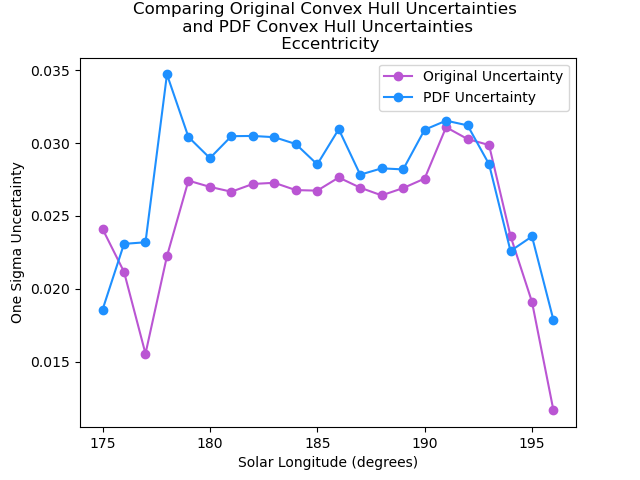}
\end{tabular}

\begin{tabular}{@{}c@{}}
    \includegraphics[width=0.45\textwidth]{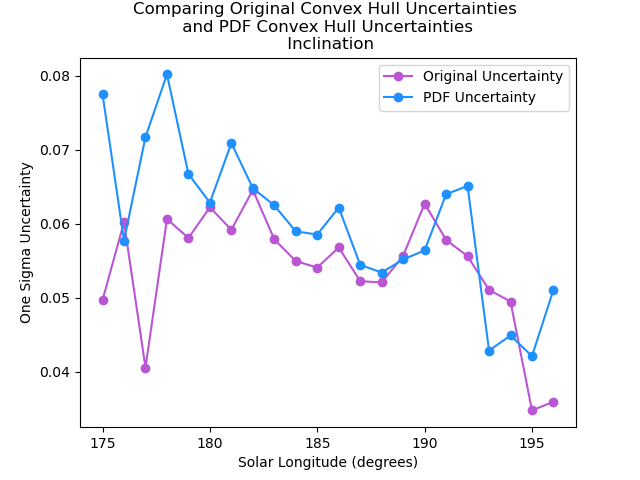}
\end{tabular}

\caption[short]{Orbital element uncertainties calculated using the \citet{Jopek2006} method for the computationally simple convex hull method, which models meteors as points in radiant space, and the alternate convex hull method, which models meteors as 3D Gaussian probability distributions. The alternate convex hull method produces larger uncertainties at the wings of the shower, but similar uncertainties near the shower's peak. This is likely due to lower number statistics at the wings from the rejection of meteors with large uncertainties performed for this method and not the computationally simple method.}
\label{fig:uncertainty orbital elements}
\end{figure*}

\begin{figure*}
\centering
\begin{tabular}{@{}c@{}}
    \includegraphics[width=0.45\textwidth]{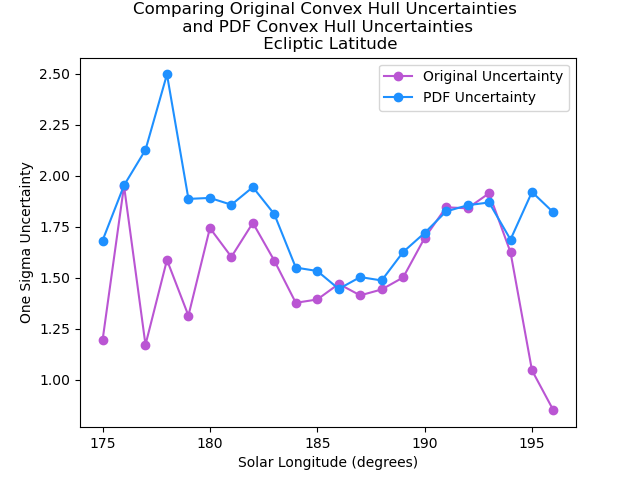}
\end{tabular}
\begin{tabular}{@{}c@{}}
    \includegraphics[width=0.45\textwidth]{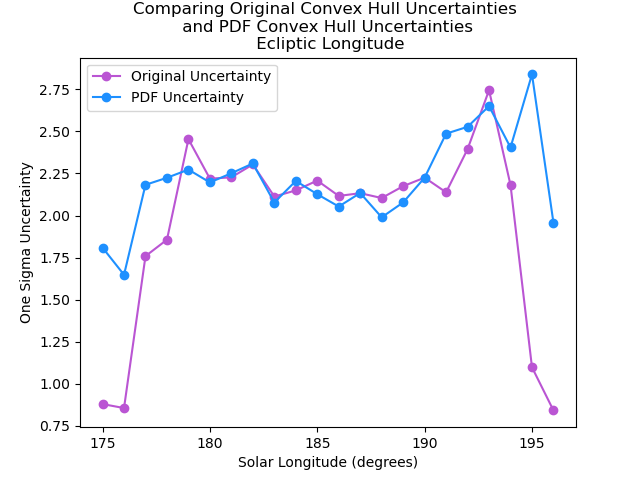}
\end{tabular}
\begin{tabular}{@{}c@{}}
    \includegraphics[width=0.45\textwidth]{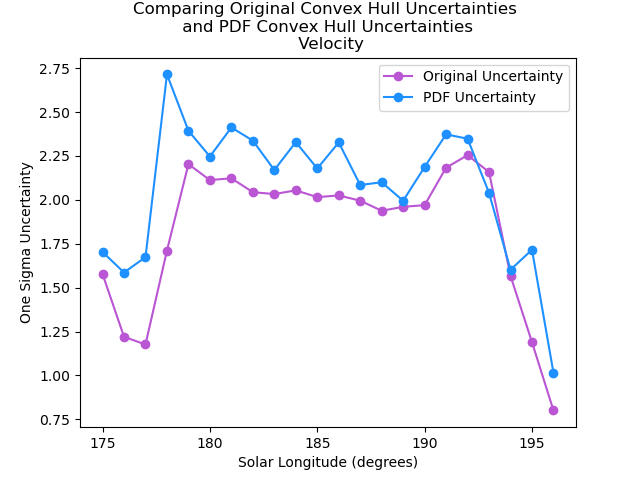}
\end{tabular}
\caption[short]{Radiant uncertainties calculated using the \citet{Jopek2006} method for the computationally simple convex hull method, which models meteors as points in radiant space, and the alternate convex hull method, which models meteors as 3D Gaussian probability distributions. The alternate convex hull method produces larger uncertainties at the wings of the shower, but similar uncertainties near the shower's peak. This is likely due to lower number statistics in the wings from the rejection of meteors with large uncertainties performed for this method and not the computationally simple method.}
\label{fig:uncertainty radiant}
\end{figure*}

Comparing the orbital element results in Figures \ref{fig:orbital_elements_alt}, and the radiant results in Figures \ref{fig:DSX radiant} and \ref{fig:DSX radiant alt}, we find that there is no significant difference in the radiant and orbital elements results. Figures \ref{fig:uncertainty orbital elements} and \ref{fig:uncertainty radiant} show the uncertainty in the results for each solar longitude for both convex hull methods. The uncertainties of the Jopek results per solar longitude are similar during the peak days for the two methods. The uncertainties produced by the alternate convex hull method are similar to the computationally simple method near the shower's peak but are larger around the wings of the shower. This effect is likely due to the lower number statistics in the alternate convex hull method due to the rejection of meteors with high uncertainties, which were not removed in the computationally simple method.

We have found that while the alternate convex hull method is a more robust method, the computationally simple convex hull method produces results similar enough that it is acceptable for the meteors to be modeled as points in radiant space instead of 3D Gaussian probability distributions.

Figures 11, \ref{fig:orbital_elements_alt}, \ref{fig:DSX radiant}, and \ref{fig:DSX radiant alt} compare our convex Hull and wavelet-based results of orbital element variations with solar longitude to those of previous work. Where past work measured orbits for a single solar longitude day of the shower their results are displayed on the corresponding solar longitude while if over a range of solar longitudes results are plotted at the reported DSX peak.  A detailed summary of these past results can be found in Appendix \ref{app:literature_appendix},  table \ref{literature_table} and Table \ref{literature_table_2}.

\section{Alternate Method Convex Hull Results}\label{sec:Alt Convex Hull Results}

\begin{figure*}
\centering
\begin{tabular}{@{}c@{}}
    \includegraphics[width=0.45\textwidth]{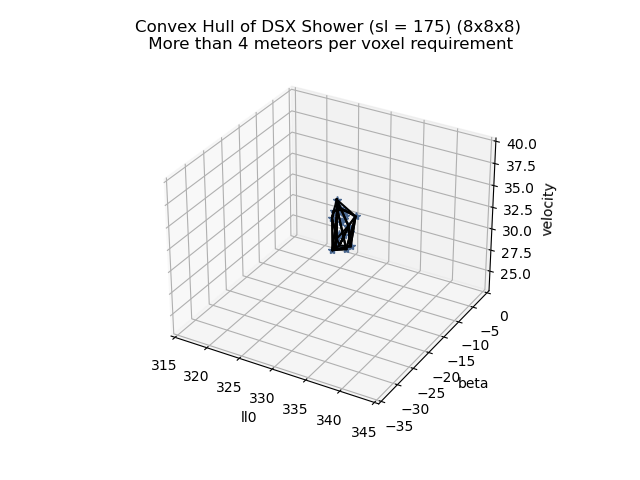}
\end{tabular}
\begin{tabular}{@{}c@{}}
    \includegraphics[width=0.45\textwidth]{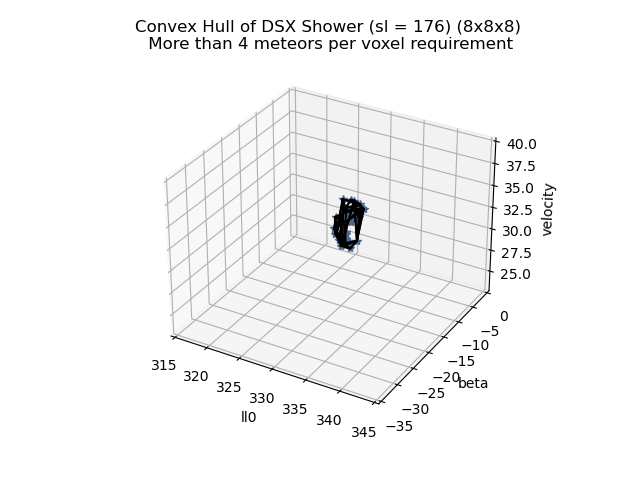}
\end{tabular}
\begin{tabular}{@{}c@{}}
    \includegraphics[width=0.45\textwidth]{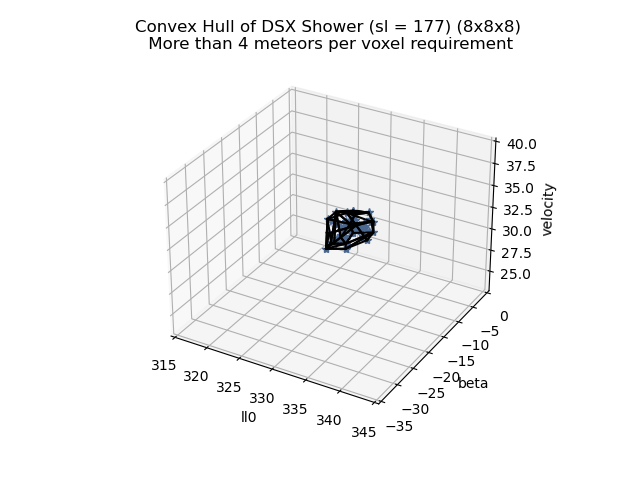}
\end{tabular}
\begin{tabular}{@{}c@{}}
    \includegraphics[width=0.45\textwidth]{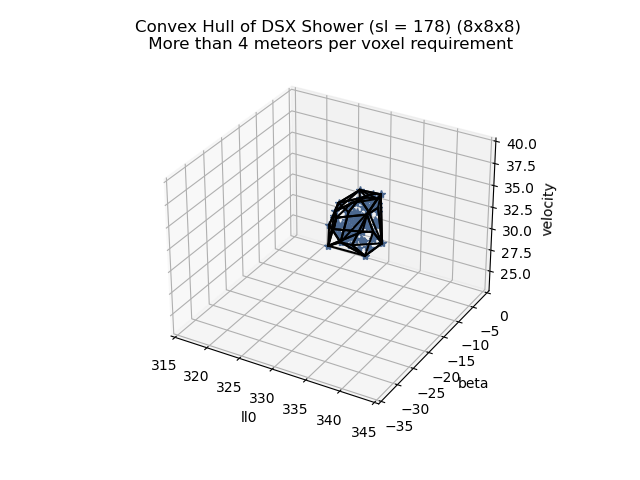}
\end{tabular}
\begin{tabular}{@{}c@{}}
    \includegraphics[width=0.45\textwidth]{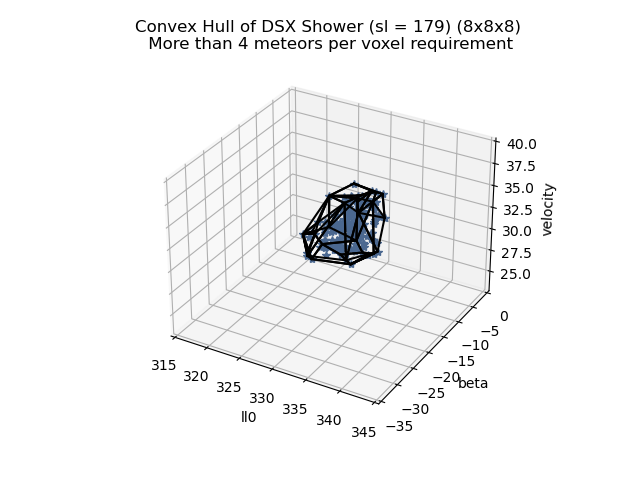}
\end{tabular}
\begin{tabular}{@{}c@{}}
    \includegraphics[width=0.45\textwidth]{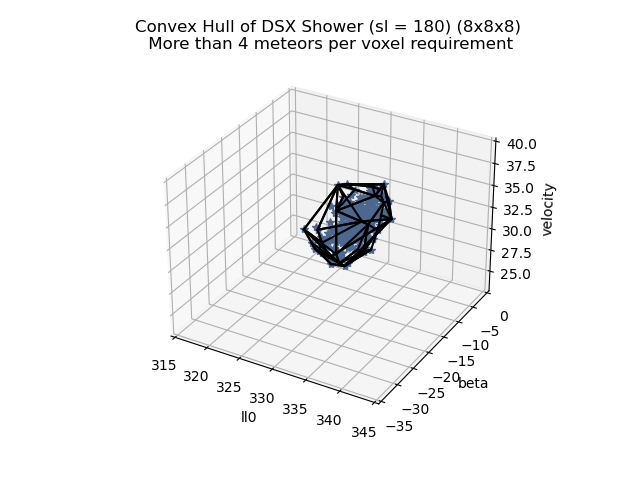}
\end{tabular}

\caption[short]{Convex hull results when the meteors are modelled as three-dimensional Gaussian probability distributions in radiant space, instead of being modelled as points. This sub-figure contains the results from solar longitude 175$^{\circ}$ to 180$^{\circ}$. The convex hull, described in Section 3.4.1 has been defined in such a way that any meteor with a radiant located within the hull is determined to be a member of the DSX shower, with a 95\% confidence level. These figures show the convex hull and all meteors that are located within it for a given solar longitude.}
\end{figure*}

\begin{figure*}
\centering
\begin{tabular}{@{}c@{}}
    \includegraphics[width=0.45\textwidth]{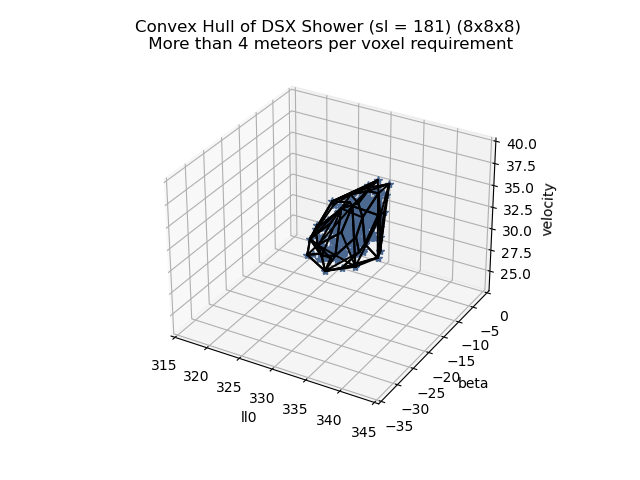}
\end{tabular}
\begin{tabular}{@{}c@{}}
    \includegraphics[width=0.45\textwidth]{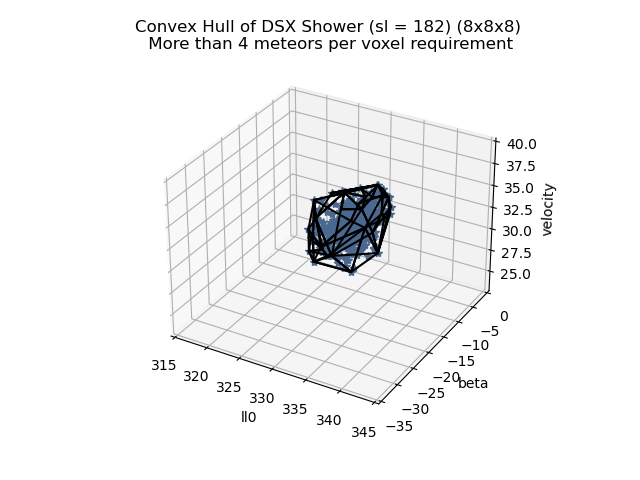}
\end{tabular}
\begin{tabular}{@{}c@{}}
    \includegraphics[width=0.45\textwidth]{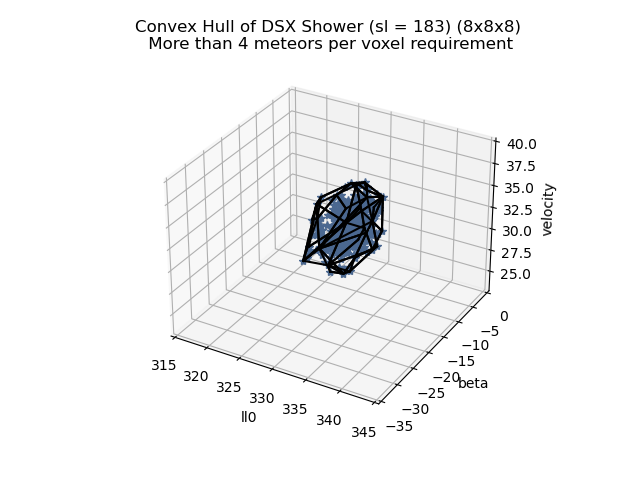}
\end{tabular}
\begin{tabular}{@{}c@{}}
    \includegraphics[width=0.45\textwidth]{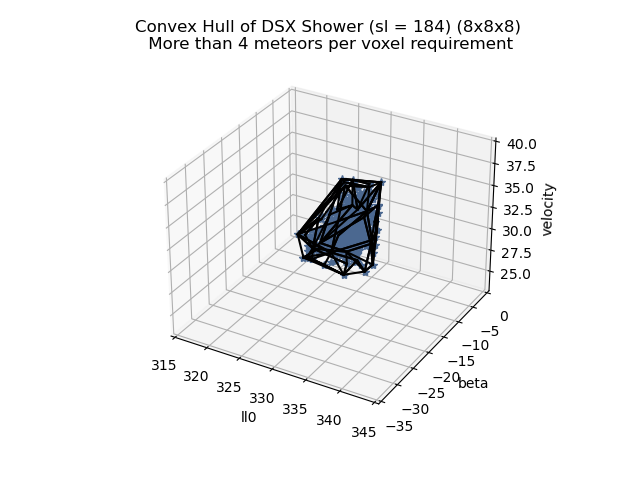}
\end{tabular}
\begin{tabular}{@{}c@{}}
    \includegraphics[width=0.45\textwidth]{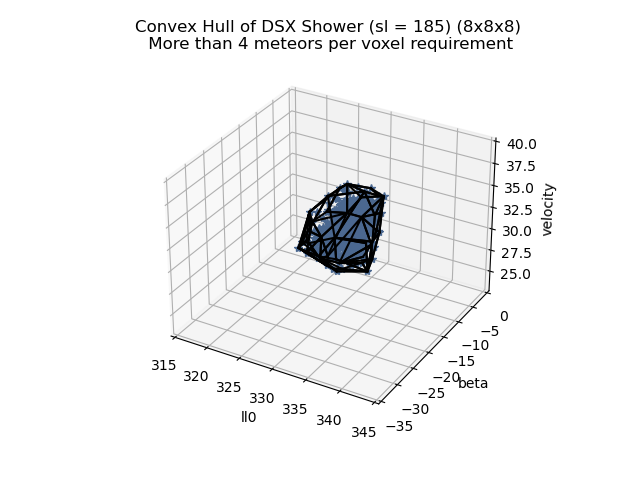}
\end{tabular}
\begin{tabular}{@{}c@{}}
    \includegraphics[width=0.45\textwidth]{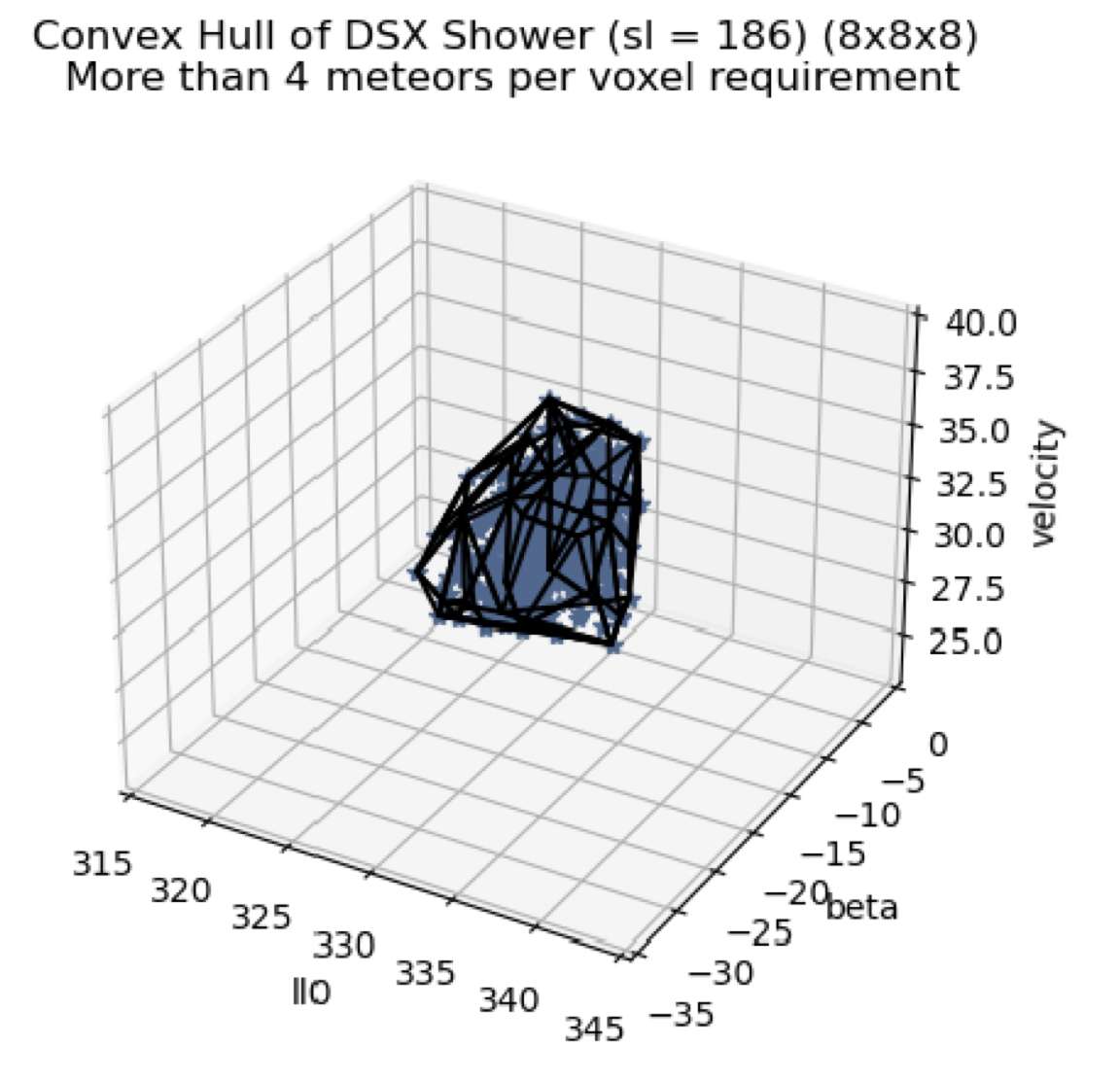}
\end{tabular}

\caption[short]{Convex hull results when the meteors are modelled as three-dimensional Gaussian probability distributions in radiant space, instead of being modelled as points. This sub-figure contains the results from solar longitude 181$^{\circ}$ to 186$^{\circ}$. The convex hull, described in Section 3.4.1 has been defined in such a way that any meteor with a radiant located within the hull is determined to be a member of the DSX shower, with a 95\% confidence level. These figures show the convex hull and all meteors that are located within it for a given solar longitude.}
\end{figure*}

\begin{figure*}
\centering
\begin{tabular}{@{}c@{}}
    \includegraphics[width=0.45\textwidth]{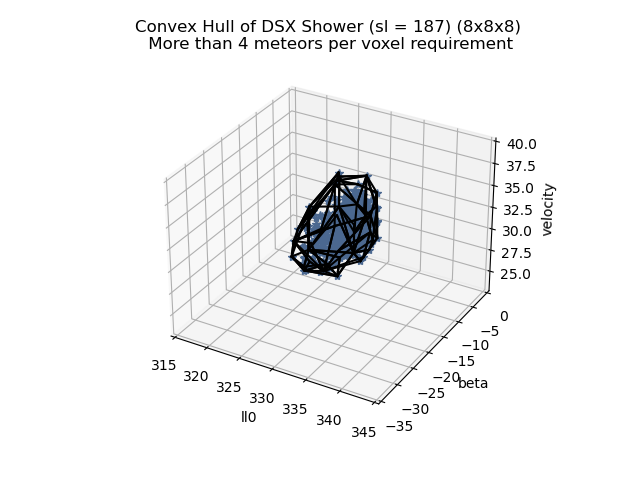}
\end{tabular}
\begin{tabular}{@{}c@{}}
    \includegraphics[width=0.45\textwidth]{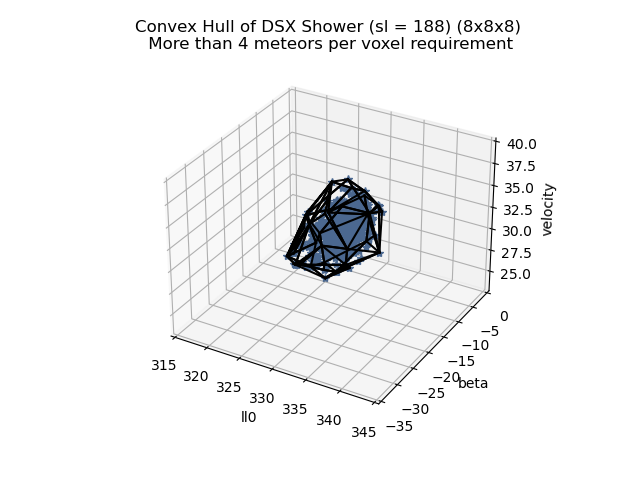}
\end{tabular}
\begin{tabular}{@{}c@{}}
    \includegraphics[width=0.45\textwidth]{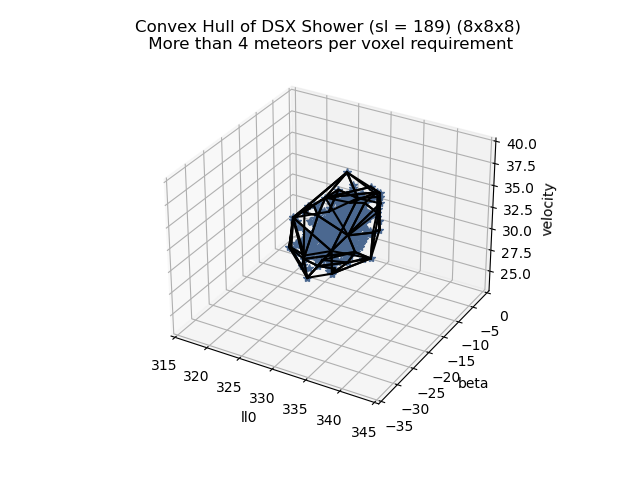}
\end{tabular}
\begin{tabular}{@{}c@{}}
    \includegraphics[width=0.45\textwidth]{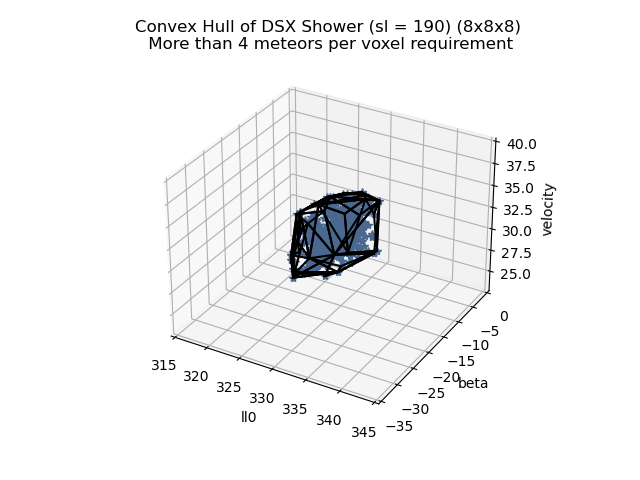}
\end{tabular}
\begin{tabular}{@{}c@{}}
    \includegraphics[width=0.45\textwidth]{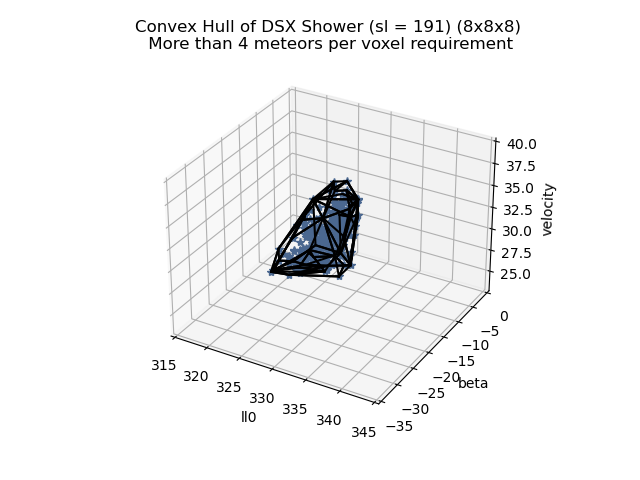}
\end{tabular}
\begin{tabular}{@{}c@{}}
    \includegraphics[width=0.45\textwidth]{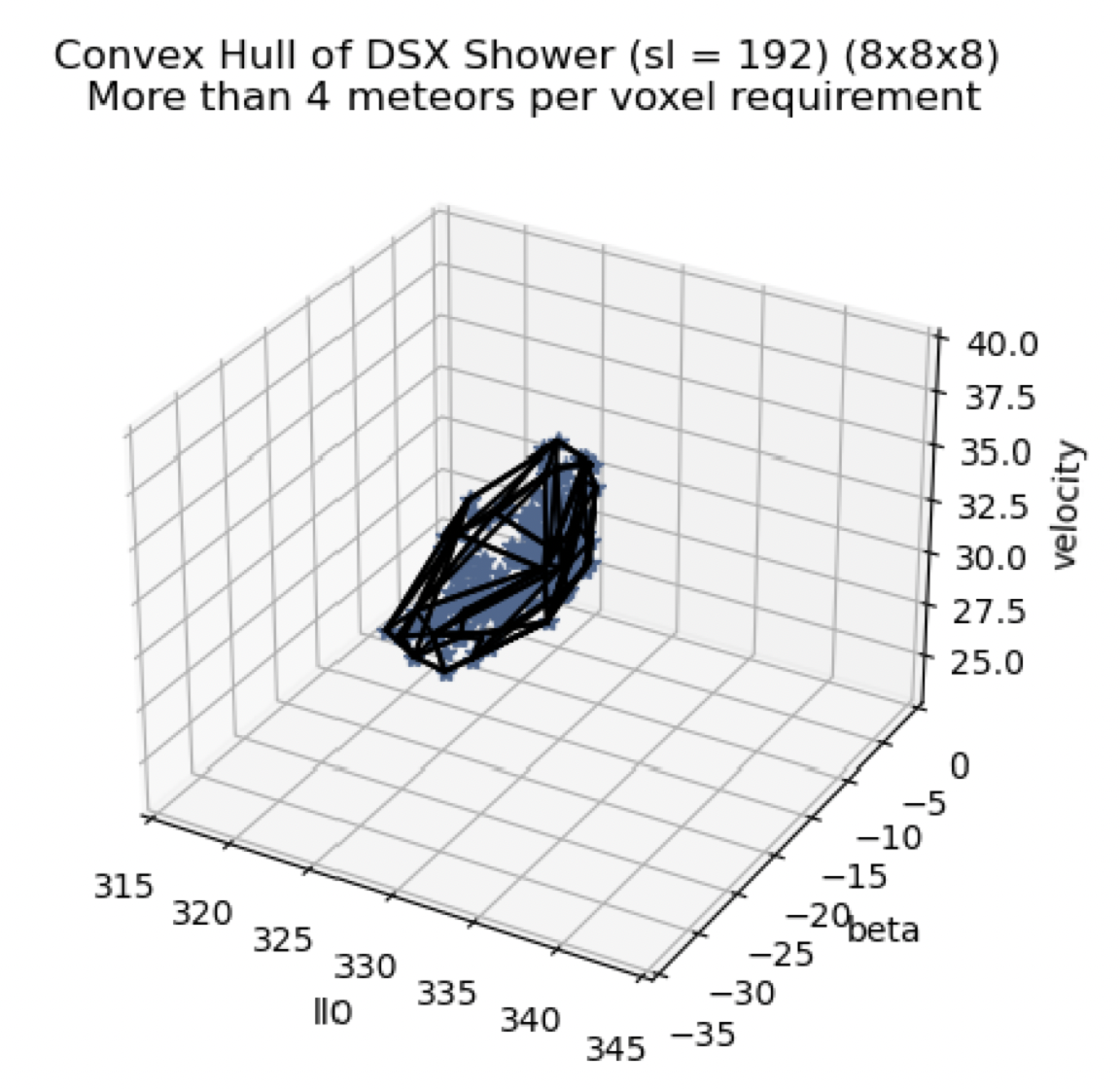}
\end{tabular}

\caption[short]{Convex hull results when the meteors are modelled as three-dimensional Gaussian probability distributions in radiant space, instead of being modelled as points. This sub-figure contains the results from solar longitude 187$^{\circ}$ to 192$^{\circ}$. The convex hull, described in Section 3.4.1 has been defined in such a way that any meteor with a radiant located within the hull is determined to be a member of the DSX shower, with a 95\% confidence level. These figures show the convex hull and all meteors that are located within it for a given solar longitude.}
\end{figure*}

\begin{figure*}
\centering
\begin{tabular}{@{}c@{}}
    \includegraphics[width=0.45\textwidth]{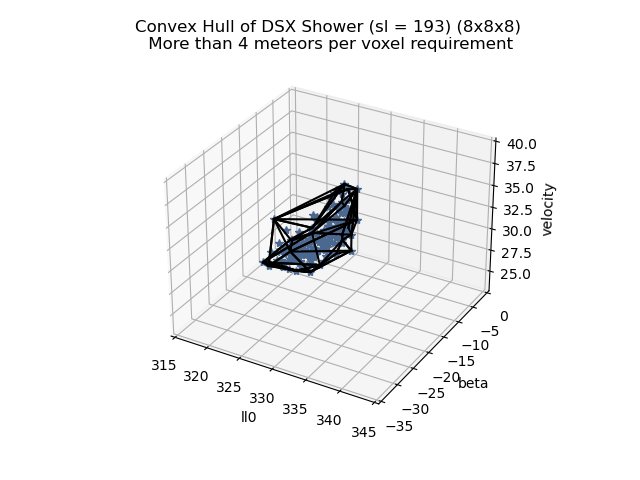}
\end{tabular}
\begin{tabular}{@{}c@{}}
    \includegraphics[width=0.45\textwidth]{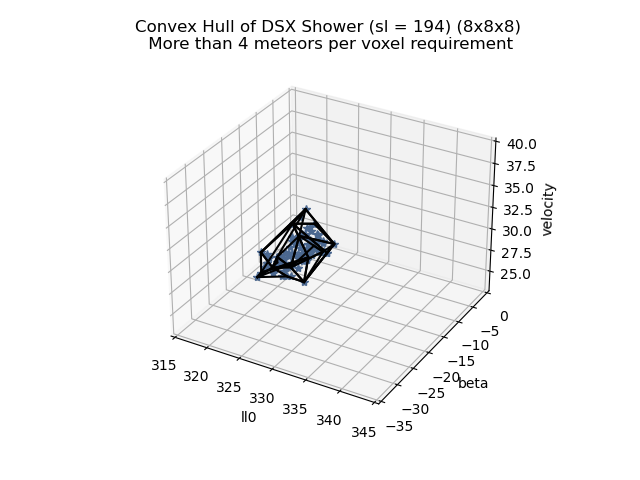}
\end{tabular}
\begin{tabular}{@{}c@{}}
    \includegraphics[width=0.45\textwidth]{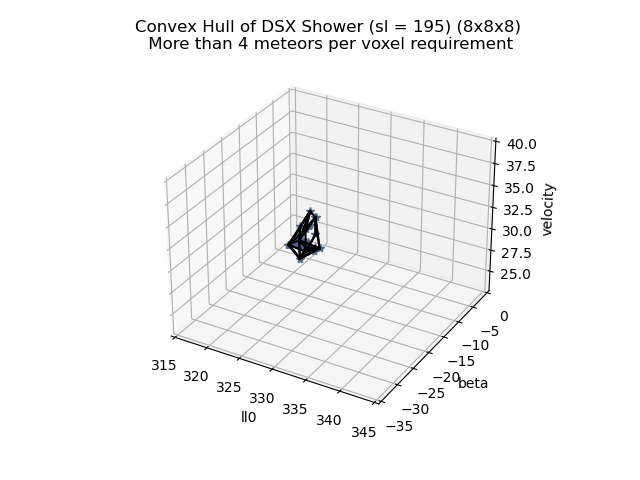}
\end{tabular}
\begin{tabular}{@{}c@{}}
    \includegraphics[width=0.45\textwidth]{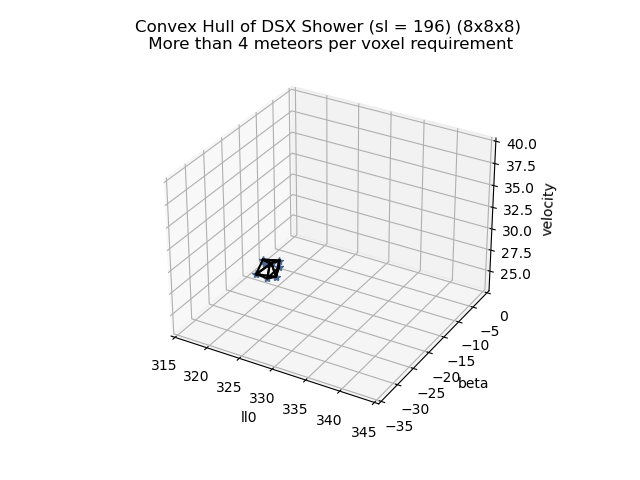}
\end{tabular}

\caption[short]{Convex hull results when the meteors are modelled as three-dimensional Gaussian probability distributions in radiant space, instead of being modelled as points. This sub-figure contains the results from solar longitude 193$^{\circ}$ to 196$^{\circ}$. The convex hull, described in Section 3.4.1 has been defined in such a way that any meteor with a radiant located within the hull is determined to be a member of the DSX shower, with a 95\% confidence level. These figures show the convex hull and all meteors that are located within it for a given solar longitude.}
\end{figure*}

\section{Additional Optical DSX Meteor Information}
\label{optical_appendix}

\begin{table*}
\begin{adjustbox}{angle=90}
\begin{tabular}{|c|c|c|c|c|c|c|c|c|}
\hline
\textbf{Time (UTC)} & \textbf{Shower} & \textbf{$\alpha_g$} & \textbf{$\delta_g$} & \textbf{$V_g$} & \textbf{$z_R$} & \textbf{$H_{\mathrm{b}}$} & \textbf{$\rho_b$} \\
\hline
\textbf{2019-09-28 11:14:28} & DSX &  $146.43 \pm 0.445$  &  $0.227 \pm 0.078$  &  $33.91 \pm 0.129$  &  $81.61 \pm 0.362$  &  $101.56 \pm 0.310$  &  $4.25 \times 10^{-10}$ \\
\hline
\textbf{2019-10-03 11:58:38} & DSX &  $155.37 \pm 0.294$  &  $-1.44 \pm 0.114$  &  $34.00 \pm 0.238$  &  $77.64 \pm 0.290$  &  $97.09 \pm  0.082$  &  $9.80 \times 10^{-10}$ \\
\hline
\textbf{2021-09-28 16:50:33} & DSX &  $153.010 \pm 0.047$  &  $-2.216 \pm 0.110$  &  $35.39 \pm 0.017$  &  $83.25 \pm 0.111$  &  $99.72 \pm 0.071$  &  $6.15 \times 10^{-10}$ \\
\hline
\textbf{2021-09-29 04:35:04} & DSX &  $153.77 \pm 0.017$  &  $0.62 \pm 0.020$  &  $34.51 \pm 0.010$  &  $86.74 \pm 0.023$  &  $102.35 \pm  0.016$  &  $3.72 \times 10^{-10}$ \\
\hline
\textbf{2021-09-30 12:22:06} & DSX &  $153.05 \pm 0.115$  &  $-1.93 \pm 0.057$  &  $36.94 \pm 0.207$  &  $79.38 \pm 0.095$  &  $101.94 \pm 0.118$  &  $3.87 \times 10^{-10}$ \\
\hline
\textbf{2021-10-01 03:48:48} & DSX &  $156.29 \pm 0.065$  &  $-0.235 \pm 0.125$  &  $35.99 \pm 0.043$  &  $81.14 \pm 0.084$  &  $99.84 \pm  0.033$  &  $5.80 \times 10^{-10}$ \\
\hline
\textbf{2021-10-01 04:41:16} & DSX &  $155.20 \pm 0.029$  &  $-0.603 \pm 0.015$  &  $35.00 \pm 0.006$  &  $80.68 \pm 0.029$  &  $102.88 \pm  0.025$  &  $3.25 \times 10^{-10}$ \\
\hline
\textbf{2021-10-01 12:16:03} & DSX &  $155.39 \pm 0.039$  &  $-1.37 \pm 0.069$  &  $35.16 \pm 0.024$  &  $79.38 \pm 0.071$  &  $102.23 \pm  0.036$  &  $3.65 \times 10^{-10}$ \\
\hline
\textbf{2021-10-02 01:41:41} & DSX &  $154.83 \pm 0.040$  &  $-0.80 \pm 0.352$  &  $34.28 \pm 0.078$  &  $83.80 \pm 0.240$  &  $101.34 \pm  0.054$  &  $4.43 \times 10^{-10}$ \\
\hline
\textbf{2021-10-02 11:43:31} & DSX &  $152.71 \pm 0.091$  &  $0.141 \pm 0.050$  &  $33.06 \pm 0.107$  &  $79.28 \pm 0.073$  &  $98.34 \pm  0.038$  &  $7.83 \times 10^{-10}$ \\
\hline
\textbf{2021-10-02 11:55:25} & DSX &  $155.32 \pm 0.114$  &  $-1.58 \pm 0.313$  &  $34.46 \pm 0.104$  &  $78.11 \pm 0.304$  &  $102.54 \pm 0.165$  &  $3.43 \times 10^{-10}$ \\
\hline
\textbf{2021-10-04 05:03:19} & DSX &  $156.10 \pm 0.027$  &  $-1.72 \pm 0.015$  &  $34.12 \pm 0.009$  &  $83.82 \pm 0.027$  &  $98.64 \pm 0.022$  &  $7.25 \times 10^{-10}$ \\
\hline
\textbf{2021-10-02 11:59:52} & DSX &  $156.41 \pm 0.019$  &  $-1.11 \pm 0.026$  &  $35.43 \pm 0.008$  &  $77.99 \pm 0.026$  &  $103.77 \pm 0.019$  &  $2.70 \times 10^{-10}$ \\

\Xhline{5\arrayrulewidth}

\textbf{2020-12-14 02:54:44} & GEM &  $109.95 \pm 0.263$  &  $32.75 \pm 0.074$  &  $36.20 \pm 0.011$  &  $77.40 \pm 0.16$  &  $101.57 \pm  0.096$  &  $3.97 \times 10^{-10}$ \\
\hline
\textbf{2019-12-14 17:50:47} & GEM &  $111.53 \pm 0.032$  &  $33.45 \pm 0.043$  &  $36.30 \pm 0.021$  &  $78.16 \pm 0.051$  &  $105.31 \pm  0.047$  &  $2.11 \times 10^{-10}$ \\
\hline
\textbf{2020-12-13 17:31:10} & GEM &  $111.94 \pm 0.046$  &  $33.90 \pm 0.024$  &  $35.90 \pm 0.023$  &  $83.84 \pm 0.038$  &  $103.32 \pm  0.027$  &  $3.05 \times 10^{-10}$ \\
\hline
\textbf{2020-12-14 02:33:39} & GEM &  $111.87 \pm 0.040$  &  $33.17 \pm 0.056$  &  $35.80 \pm 0.012$  &  $80.42 \pm 0.038$  &  $102.69 \pm  0.029$  &  $3.24 \times 10^{-10}$ \\
\hline
\textbf{2019-12-14 17:06:07} & GEM &  $111.85 \pm 0.024$  &  $33.97 \pm 0.033$  &  $35.76 \pm 0.011$  &  $82.90 \pm 0.030$  &  $102.33 \pm  0.023$  &  $3.68 \times 10^{-10}$ \\
\hline
\textbf{2019-12-14 18:08:00} & GEM &  $110.81 \pm 0.892$  &  $34.12 \pm 0.282$  &  $37.48 \pm 0.641$  &  $74.22 \pm 0.811$  &  $99.20 \pm 0.281$  &  $6.66 \times 10^{-10}$\\
\hline
\textbf{2019-12-15 02:20:12} & GEM &  $111.81 \pm 0.173$  &  $32.40 \pm 0.296$  &  $35.98 \pm 0.030$  &  $81.81 \pm 0.278$  &  $99.40 \pm 0.108$  &  $6.16 \times 10^{-10}$\\
\hline
\textbf{2020-12-12 02:29:53} & GEM &  $109.72 \pm 0.196$  &  $34.34 \pm 0.058$  &  $36.79 \pm 0.183$  &  $79.28 \pm 0.187$  &  $98.93 \pm 0.062$  &  $6.72 \times 10^{-10}$\\
\hline
\textbf{2020-12-14 02:25:50} & GEM &  $112.12 \pm 0.044$  &  $32.59 \pm 0.087$  &  $37.02 \pm 0.027$  &  $81.87 \pm 0.077$  &  $102.47 \pm 0.057$  &  $3.38 \times 10^{-10}$\\
\hline
\textbf{2019-12-14 17:41:47} & GEM &  $111.68 \pm 0.047$  &  $33.62 \pm 0.061$  &  $36.21 \pm 0.005$  &  $78.74 \pm 0.065$  &  $102.21 \pm 0.040$  &  $3.78 \times 10^{-10}$\\
\hline
\textbf{2020-12-11 17:46:30} & GEM &  $109.29 \pm 0.082$  &  $33.54 \pm 0.040$  &  $35.62 \pm 0.228$  &  $79.11 \pm 0.082$  &  $100.88 \pm 0.037$  &  $4.85 \times 10^{-10}$\\
\hline
\textbf{2020-12-13 17:45:54} & GEM &  $111.50 \pm 0.073$  &  $33.39 \pm 0.041$  &  $35.83 \pm 0.222$  &  $80.80 \pm 0.054$  &  $103.66 \pm 0.022$  &  $2.86 \times 10^{-10}$\\
\hline
\textbf{2020-12-14 02:58:56} & GEM &  $112.53 \pm 0.184$  &  $32.91 \pm 0.057$  &  $36.96 \pm 0.056$  &  $79.09 \pm 0.157$  &  $100.29 \pm 0.075$  &  $5.10 \times 10^{-10}$\\
\hline

\end{tabular}
\end{adjustbox}
\caption{Detailed atmospheric trajectory data for the 13 optical Daytime Sextantid  meteors analyzed in this paper and the 13 optical Geminid meteors used to compare their relative compositions. $\alpha_g$ (deg), $\delta_g$ (deg), and $V_g$ (km/s) are the geocentric radiant and velocity, $H_b$ (km) is the begin height, $\rho_b$ is the air mass density at the meteor begin point.}

\label{optical_table_extra}
\end{table*}

\begin{table*}
\begin{adjustbox}{angle=90}
\begin{tabular}{|c|c|c|c|c|c|}
\hline
\textbf{Time (UTC)} & \textbf{Shower}  & \textbf{$a$} & \textbf{$e$} & \textbf{$i$} & \textbf{$\omega$}\\
\hline
\textbf{2019-09-28 11:14:28} & DSX &  $0.95 \pm 0.006$ & $0.87 \pm 0.002$ & $29.97 \pm 1.002$ & $208.28 \pm 0.443$\\
\hline
\textbf{2019-10-03 11:58:37} & DSX & $1.07 \pm 0.008$ & $0.86 \pm 0.004$ & $24.12 \pm 0.678$ & $212.67 \pm 0.499$\\
\hline
\textbf{2021-09-28 16:50:33} & DSX & $1.18 \pm 0.002$ & $0.88 \pm 0.0004$ & $22.98 \pm 0.24$ & $212.90 \pm 0.091$\\
\hline
\textbf{2021-09-29 04:35:04} & DSX &  $1.16 \pm 0.001$ & $0.87 \pm 0.0001$ & $23.24 \pm 0.029$ & $214.50 \pm 0.039$\\
\hline
\textbf{2021-09-30 12:22:06} & DSX & $1.26 \pm 0.01$ & $0.89 \pm 0.002$ & $30.87 \pm 0.626$ & $213.26 \pm 0.133$\\
\hline
\textbf{2021-10-01 03:48:48} & DSX &  $1.27 \pm 0.003$ & $0.89 \pm 0.001$ & $23.70 \pm 0.193$ & $214.46 \pm 0.161$\\
\hline
\textbf{2021-10-01 04:41:16} & DSX & $1.17 \pm 0.002$ & $0.874 \pm 0.0001$ & $23.95 \pm 0.026$ & $213.96 \pm 0.050$\\
\hline
\textbf{2021-10-01 12:16:03} & DSX & $1.19 \pm 0.005$ & $0.87 \pm 0.0001$ & $24.73 \pm 0.097$ & $214.49 \pm 0.132$\\
\hline
\textbf{2021-10-02 01:41:41} & DSX & $1.10 \pm 0.010$ & $0.865 \pm 0.001$ & $24.51 \pm 0.590$ & $213.09 \pm 0.428$\\
\hline
\textbf{2021-10-02 11:43:31} & DSX & $0.96 \pm 0.004$ & $0.86 \pm 0.001$ & $23.55 \pm 0.220$ & $209.16 \pm 0.105$\\
\hline
\textbf{2021-10-02 11:55:25} & DSX & $1.11 \pm 0.006$ & $0.87 \pm 0.003$ & $24.81 \pm 0.150$ & $213.36 \pm 0.600$\\
\hline
\textbf{2021-10-04 05:03:19} & DSX & $1.06 \pm 0.001$ & $0.86 \pm 0.001$ & $26.13 \pm 0.016$ & $212.50 \pm 0.046$\\
\hline
\textbf{2021-10-02 11:59:52} & DSX & $1.19 \pm 0.002$ & $0.88 \pm 0.001$ & $24.18 \pm 0.042$ & $213.61 \pm 0.047$\\

\Xhline{5\arrayrulewidth}
\textbf{2020-12-14 02:54:44} & GEM & $1.44 \pm 0.016$ & $0.89 \pm 0.0002$ & $21.06 \pm 0.323$ & $321.89 \pm 0.318$\\
\hline
\textbf{2019-12-14 17:50:47} & GEM &$1.37 \pm 0.004$ & $0.89 \pm 0.0002$ & $22.77 \pm 0.070$ & $323.85 \pm 0.086$\\
\hline
\textbf{2020-12-13 17:31:10} & GEM & $1.29 \pm 0.001$ & $0.89 \pm 0.0004$ & $23.44 \pm 0.054$ & $324.82 \pm 0.080$\\
\hline
\textbf{2020-12-14 02:33:39} & GEM & $1.29 \pm 0.002$ & $0.89 \pm 0.0004$ & $22.72 \pm 0.093$ & $324.44 \pm 0.072$\\
\hline
\textbf{2019-12-14 18:08:00} & GEM & $1.57 \pm 0.115$ & $0.91 \pm 0.007$ & $25.16 \pm 0.745$ & $321.86 \pm 1.498$\\
\hline
\textbf{2019-12-14 17:06:07} & GEM & $1.30 \pm 0.002$ & $0.89 \pm 0.001$ & $22.89 \pm 0.067$ & $324.18 \pm 0.045$\\
\hline
\textbf{2019-12-15 02:20:12} & GEM & $1.30 \pm 0.011$ & $0.89 \pm 0.001$ & $21.28 \pm 0.422$ & $324.92 \pm 0.499$\\
\hline
\textbf{2020-12-12 02:29:53} & GEM & $1.39 \pm 0.028$ & $0.90 \pm 0.002$ & $25.84 \pm 0.184$ & $323.77 \pm 0.309$\\
\hline
\textbf{2020-12-14 02:25:50} & GEM & $1.38 \pm 0.003$ & $0.90 \pm 0.001$ & $23.52 \pm 0.112$ & $325.58 \pm 0.147$\\
\hline
\textbf{2019-12-14 17:41:47} & GEM & $1.35 \pm 0.003$ & $0.89 \pm 0.001$ & $23.04 \pm 0.087$ & $323.93 \pm 0.113$\\
\hline
\textbf{2020-12-11 17:46:30} & GEM & $1.28 \pm 0.022$ & $0.89 \pm 0.003$ & $21.72 \pm 0.315$ & $324.58 \pm 0.132$\\
\hline
\textbf{2020-12-13 17:45:54} & GEM & $1.30 \pm 0.006$ & $0.89 \pm 0.001$ & $22.18 \pm 0.079$ & $324.35 \pm 0.106$\\
\hline
\textbf{2020-12-14 02:58:56} & GEM & $1.36 \pm 0.006$ & $0.90 \pm 0.001$ & $24.53 \pm 0.151$ & $325.48 \pm 0.314$\\
\hline

\end{tabular}
\end{adjustbox}
\caption{Detailed orbital data for the 13 optical Daytime Sextantid  meteors and the 13 optical Geminid meteors used to compare their relative compositions. $a$ (AU) is the semi-major axis of the stream, $i$ (deg) is the inclination angle of the stream, $e$ is the eccentricity of the stream, and $\omega$ (deg) is the argument of the perihelion of the stream}

\label{optical_table_extra}
\end{table*}

\bibliographystyle{mnras}
\bibliography{bibliography}

\bsp	
\label{lastpage}